\documentclass[prb,aps,nofootinbib,amssymb,twocolumn,superscriptaddress,10pt]{revtex4-2}
\usepackage[T1]{fontenc}
\usepackage{graphicx}
\usepackage{changepage}
\usepackage{xcolor}
\usepackage{dcolumn}
\usepackage{comment}
\usepackage{amsmath,amssymb,bbold,bm}
\usepackage{hyperref}
\usepackage{amsfonts}
\usepackage{listings}
\usepackage{braket}
\usepackage{float,latexsym}
\usepackage{multirow}
\usepackage{tikz}
\usepackage{color}
\usepackage{xr}
\usepackage[normalem]{ulem}
\usepackage{bibunits}
\usepackage{cleveref}
\usepackage{physics}
\usepackage{blindtext}
\usepackage{comment}
\usepackage{tabularx}
\usepackage{ltablex}
\usepackage{array}
\usepackage[define-L-C-R]{nicematrix}
\usepackage{layouts}

\usepackage[caption=false]{subfig}
\usepackage{siunitx}
\usepackage{gensymb}
\usepackage{pifont}\usepackage{placeins}
\usepackage{pdfpages}
\newcommand{\cmark}{\ding{51}}\newcommand{\xmark}{\ding{55}}\keepXColumns

\crefname{appendix}{Appendix}{Appendices}
\crefname{equation}{Eq.}{Eqs.}
\crefname{figure}{Fig.}{Figs.}
\crefname{table}{Table}{Tables}
\crefname{section}{Section}{Sections}
\crefname{enumi}{case}{cases}
\creflabelformat{appendix}{[#2#1#3]}

\makeatletter
\AtBeginDocument{\let\LS@rot\@undefined}
\makeatother

\makeatletter     
\renewcommand\onecolumngrid{\do@columngrid{one}{\@ne}\def\set@footnotewidth{\onecolumngrid}\def\footnoterule{\kern-6pt\hrule width 1.5in\kern6pt}}

\let\vec\mathbf
\def\ie{{\it i.e.}\ }
\def\eg{{\it e.g.}\ }

\newlength{\nsubht}

\newsavebox{\nsubbox}
\allowdisplaybreaks

\newcommand{\citeBCS}{\cite{BRA17,VER17,ELC17,XU20,ELC20a}}
\newcommand{\titlePaper}{General Construction and Topological Classification of All Magnetic and Non-Magnetic Flat Bands}
\allowdisplaybreaks

\begin{document}
\title{\titlePaper}

\author{Dumitru C\u{a}lug\u{a}ru}\thanks{These authors contributed equally.}
\affiliation{Department of Physics, Princeton University, Princeton, New Jersey 08544, USA}

\author{Aaron Chew}\thanks{These authors contributed equally.}
\affiliation{Department of Physics, Princeton University, Princeton, New Jersey 08544, USA}

\author{Luis Elcoro}\thanks{These authors contributed equally.}
\affiliation{Department of Condensed Matter Physics, University of the Basque Country UPV/EHU, Apartado 644, 48080 Bilbao, Spain}

\author{Nicolas Regnault}\affiliation{Department of Physics, Princeton University, Princeton, New Jersey 08544, USA}
\affiliation{Laboratoire de Physique de l'Ecole normale superieure, ENS, Universit\'e PSL, CNRS, Sorbonne Universit\'e, Universit\'e Paris-Diderot, Sorbonne Paris Cit\'e, Paris, France}

\author{Zhi-Da Song}\affiliation{Department of Physics, Princeton University, Princeton, New Jersey 08544, USA}

\author{B. Andrei Bernevig}\email{bernevig@princeton.edu}
\affiliation{Department of Physics, Princeton University, Princeton, New Jersey 08544, USA}
\affiliation{Donostia International Physics Center, P. Manuel de Lardizabal 4, 20018 Donostia-San Sebastian, Spain}
\affiliation{IKERBASQUE, Basque Foundation for Science, Bilbao, Spain}

\begin{abstract}
Exotic phases of matter emerge from the interplay between strong electron interactions and non-trivial topology. Owing to their lack of dispersion at the single-particle level, systems harboring flat bands are excellent testbeds for strongly interacting physics, with twisted bilayer graphene serving as a prime example. On the other hand, existing theoretical models for obtaining flat bands in crystalline materials, such as the line-graph formalism, are often too restrictive for real-life material realizations. Here we present a generic technique for constructing perfectly flat bands from bipartite crystalline lattices. Our prescription encapsulates and generalizes the various flat band models in the literature, being applicable to systems with \emph{any} orbital content, \emph{with} or \emph{without} spin-orbit coupling. Using Topological Quantum Chemistry, we build a complete topological classification in terms of symmetry eigenvalues of all the gapped and gapless flat bands, for all 1651{} Magnetic Space Groups. In addition, we derive criteria for the existence of symmetry-protected band touching points between the flat and dispersive bands, and we identify the gapped flat bands as prime candidates for fragile topological phases. Finally, we show that the set of all (gapped and gapless) perfectly flat bands is finitely generated and construct the corresponding bases for all 1651{} Shubnikov Space Groups. 
\end{abstract}
\maketitle

\emph{Introduction.}~Under special conditions, translation-invariant systems harbor perfectly flat bands -- spectral bands whose energies are independent of crystal momentum. This extensive degeneracy of electron states at the single-particle level leads to a completely non-perturbative effect of arbitrary levels of interactions or disorder, making flat-band systems prime candidates for strongly-correlated phases of matter. Perhaps the most exciting recent developments in this direction concern twisted bilayer graphene (TBG), where the presence of almost flat bands~\cite{LOP07,SUA10,BIS11} has been linked to a plethora of magnetic and superconducting phases~\cite{CAO18,CAO18a,LU19,SHA19,XIE19,NUC20,WU21}.

Furthermore, consequent theoretical studies have highlighted the importance of the non-trivial topology of the flat bands, in addition to their dispersionless nature~\cite{AHN19,BER20b,BUL20,KAN19,LIA19,PEL18,PO18,PO19,SEO19,SON19,SON20b,WU18,XIE20,XIE20a,XU18,ZOU18}. For example, the Ginzburg-Landau theory predicts a vanishing superfluid weight for perfectly flat bands. However, a topologically non-trivial flat band has additional band-geometric contributions to the superfluid weight and can thus show superconductive behavior~\cite{HU19,XIE20,PER20,JUL20a}.

Flat bands which are not atomic in nature, but whose flatness arise from wave function interference, are a very rich playground for physical phenomena. In crystalline materials, the rich physics of interacting flat band systems \emph{predates} the advent of TBG, with theoretical proposals including Hubbard ferromagnetism~\cite{MIE91,MIE93}, Wigner crystallization~\cite{WU07}, supersolid formation~\cite{HUB10}, or Anderson transition~\cite{GOD06,CHA10}. On the other hand, the well-known theoretical constructions for engineering flat bands, such as the line-graph prescription~\cite{MIE91,MIE91a,MIE92,MIE92a,MIE93,KOL20}, are often restricted to toy-models comprised of $s$ orbitals and non-spin-orbit-coupled Hamiltonians with various geometric constraints, such as nearest-neighbor hoppings with the same sign. These simplifications have hindered the discovery of real-life crystalline materials with flat bands~\cite{HUD20,LIU20,MEI20}, which are usually obtained in optical lattices~\cite{BAB16}, or superconducting circuits~\cite{KOL19,KOL20,LEY18}. 

In this paper, we introduce a generic technique for constructing perfectly flat bands in a general class of systems that we term bipartite crystalline lattices (BCLs), where a lattice is divided into two sublattices with unequal numbers of atoms. The BCL construction can be applied to \emph{any} type of orbitals, \emph{with} or \emph{without} spin-orbit coupling, and in \emph{any} of the 1651{} Shubnikov Space Groups (SSGs), including the space groups with or without time-reversal symmetry and the magnetic space groups. Our prescription also encapsulates and generalizes the line-graph~\cite{MIE91,MIE91a,MIE92,MIE92a,MIE93,BER08,KOL19,KOL20,CHI20,MA20} and split-graph~\cite{LIE89,CHI20,MA20} formulations, as well as many of the models presented in literature~\cite{WEA71,TAS92,VID00,CRE01,DOU02,WU07,BER08,GRE10,HAT11,TAN11,CHE14,HAT15,DER15,MOR16,LEY18,ELS19,PO19,MIZ19a,MIZ19,KUN20,MIZ20,WEI20,GAO20,QI20,CHI20,LIU21b}. Applying the machinery of Magnetic Topological Quantum Chemistry (MTQC)~\cite{BRA17,ELC20a} and related theories~\cite{KRU17,PO17,WAT18} to our construction yields a complete symmetry eigenvalue-based classification for all BCL gapped and gapless flat bands in all SSGs. The main result of this work is that the BCL flat bands can be understood as formal differences of band representations (BRs). Firstly, this enables us to derive universal criteria for the existence of symmetry-protected band touching points (BTPs) between the flat and dispersive bands, which were previously only explained in an \textit{ad-hoc} manner~\cite{BER08,HWA21}: flat bands carry formal differences of irreducible (co)representations [(co)irreps], which can be exploited to diagnose protected BTPs. Moreover, gapped flat bands can realize any commensurate difference of BRs and therefore make flat bands prime candidates for fragile topological phases~\cite{BRA17,CAN18,PO18c,AHN19,ELS19,SON20,SON20c,CHI20}. Secondly, the relation between flat bands and BRs allows us to show that the set of all perfectly flat bands is finitely generated and construct the corresponding bases in all SSGs.

\emph{Model.}~We start by outlining the BCL construction (see \cref{app:sec:bipartite_crystalline_lattices:notation}). A BCL is a translation-invariant fermionic lattice which is partitioned into two different sublattices, $L$ and $\tilde{L}$. We assume that each BCL sublattice individually respects all the symmetries of the BCL's SSG. For each unit cell $\vec{R}$, we define fermionic annihilation operators $\hat{a}_{\vec{R},i}$ ($\hat{b}_{\vec{R},i}$) corresponding to each orbital $i$ from sublattice $L$ ($\tilde{L}$). In the case of spinful fermions, we consider the different spin states as distinct orbitals (with different spin states having different indices $i$). Within each sublattice $L$ and $\tilde{L}$, we place $N_L$ and $N_{\tilde{L}}$ orbitals per unit cell, respectively, and introduce  a unitary chiral operator $C$ acting differently on the two sublattices: $C \hat{a}^\dagger_{\vec{R},i} C^{-1} = \hat{a}^\dagger_{\vec{R},i}$ and $C \hat{b}^\dagger_{\vec{R},i} C^{-1} = -\hat{b}^\dagger_{\vec{R},i}$. We will first consider quadratic Hamiltonians $\mathcal{H}$ with chiral anti-commuting symmetry (\ie  $\{ C, \mathcal{H} \} = 0$) and show later how this constraint can be relaxed. Defining momentum space operators $\hat{c}_{\vec{k},i} = \frac{1}{\sqrt{N}}\sum_{\vec{R}}\hat{c}_{\vec{R},i} e^{i \vec{k} \cdot \left( \vec{R} + \vec{r}_i \right)}$, where $\vec{r}_i$ denotes the position of the $i$-th orbital relative to the unit cell origin and $\hat{c}_{}=\hat{a}_{},\hat{b}_{}$ for the two sublattices, the Hamiltonian (which includes only generic hoppings between the $L$ and $\tilde{L}$ sublattice) can be written as $\mathcal{H}=\sum_{\vec{k}} \hat{\Psi}^\dagger_{\vec{k}} H_{\vec{k}} \hat{\Psi}_{\vec{k}}$, where the first-quantized Hamiltonian matrix is
\begin{equation}
	\label{eqn:bip_ham}
	H_{\vec{k}} = \begin{pmatrix}
		\mathbb{0} & S_\vec{k} \\
		S_\vec{k}^\dagger & \mathbb{0}
\end{pmatrix}
\end{equation}
and $\hat{\Psi}_{\vec{k}}^T = \left( \hat{a}_{\vec{k},1}, \dots, \hat{a}_{\vec{k},N_L}, \hat{b}_{\vec{k},1}, \dots, \hat{b}_{\vec{k},N_{\tilde{L}}} \right)$ is an $ \left(N_{L} + N_{\tilde{L}} \right)$-dimensional spinor. In \cref{eqn:bip_ham}, the presence of chiral symmetry $C$ forbids any intra-sublattice hopping terms, while $S_{\vec{k}}$ denotes the $N_L \times N_{\tilde{L}}$ hopping matrix between the orbitals belonging to different sublattices~\cite{LIE89}. If $S_{\vec{k}}$ is rectangular then its rank $r_{\vec{k}}$ is bounded by the smaller of its dimensions. Taking $N_L > N_{\tilde{L}}$, it follows (see \cref{app:sec:bipartite_crystalline_lattices:chiral}) that $H_{\vec{k}}$ contains at least $N_L - N_{\tilde{L}}$ zero modes for all $\vec{k}$, giving rise to $N_L - N_{\tilde{L}}$ perfectly flat bands, as well as $2N_{\tilde{L}}$ dispersive bands coming in pairs related by chiral symmetry. Generically, $S_{\vec{k}}$ has maximal rank $r_{\vec{k}}$ for all $\vec{k}$ with the exception of BTPs between the flat and the dispersive bands. 

The Hamiltonian $H_{\vec{k}}$ can be diagonalized using the singular value decomposition $S_{\vec{k}} = W_{\vec{k}} \Sigma_{\vec{k}} V_{\vec{k}}^\dagger$, where $W_{\vec{k}}$ ($V_{\vec{k}}$) denotes an $N_L \times N_L$ ($N_{\tilde{L}} \times N_{\tilde{L}}$) unitary matrix, while $\Sigma_{\vec{k}}$ is a diagonal $N_{L} \times N_{\tilde{L}}$ matrix of singular values (listed in descending order). We define $\psi_{\vec{k},\alpha}$ ($\phi_{\vec{k},\alpha}$) to be the vector formed by the $\alpha$-th column of $V_{\vec{k}}$ ($W_{\vec{k}}$). For each $\alpha \leq r_{\vec{k}}$, the eigenvectors of $H_{\vec{k}}$ corresponding to the dispersive bands are
\begin{equation}
	\label{eqn:dispersive_eigs}
	\Psi^\pm_{\vec{k},\alpha} = \frac{1}{\sqrt{2}}\begin{pmatrix}
		\pm \phi_{\vec{k},\alpha} \\
		\psi_{\vec{k},\alpha}
	\end{pmatrix},
\end{equation}
with energies $\pm\epsilon_{\vec{k},\alpha}$, where $\epsilon_{\vec{k},\alpha}$ is the $\alpha$-th singular value of $S_{\vec{k}}$ (\ie the $\alpha$-th nonzero diagonal entry of $\Sigma_{\vec{k}}$). The zero modes of $H_{\vec{k}}$ are $\Psi^{+T}_{\vec{k},\alpha} = \left( \phi_{\vec{k},\alpha}^T,0 \right)$ for $r_{\vec{k}} < \alpha \leq N_L$ and $\Psi^{-T}_{\vec{k},\alpha} = \left(0, \psi_{\vec{k},\alpha}^T \right)$ for $r_{\vec{k}} < \alpha \leq N_{\tilde{L}}$, with $S_{\vec{k}} \psi_{\vec{k},\alpha} = 0$ and $S^{\dagger}_{\vec{k,\alpha}} \phi_{\vec{k},\alpha} = 0$, respectively. These include both the flat band eigenstates (for which $S^{\dagger}_{\vec{k}} \phi_{\vec{k},\alpha} = 0$ is true for all $\vec{k}$) and the zero modes of the dispersive bands at the BTPs (where $r_{\vec{k}}<N_{\tilde{L}}$). 

It is instructive to consider ``integrating'' out the degrees of freedom on the smaller sublattice $\tilde{L}$. This is equivalent to adding a large chemical potential term for the orbitals in sublattice $\tilde{L}$ and then including their effects on sublattice $L$ using degenerate second order perturbation theory (see \cref{app:sec:bipartite_crystalline_lattices:effective:def}). Up to multiplicative factors and constant offsets, the resulting effective Hamiltonian is $T_{\vec{k}} = S_{\vec{k}} S_{\vec{k}}^\dagger$. The eigenvectors of $T_{\vec{k}}$ are simply $\phi_{\vec{k},\alpha}$ ($1 \leq \alpha \leq N_{L})$ and thus include the flat band modes of the original BCL (being in the kernel of $S_{\vec{k}}^\dagger$, they are also zero modes of $T_{\vec{k}}$). Alternatively, one may integrate the other sublattice yielding the Hamiltonian $\tilde{T}_{\vec{k}} = S_{\vec{k}}^\dagger S_{\vec{k}}$, whose eigenstates $\psi_{\vec{k},\alpha}$ ($1 \leq \alpha \leq N_{\tilde{L}}$) do not include the flat band modes, but possesses the same nonzero eigenvalues as $T_{\vec{k}}$: $\epsilon_{\vec{k},\alpha}^2$ ($1 \leq \alpha \leq r_{\vec{k}}$). There is also a direct mapping between the non-zero eigenstates of the two Hamiltonians: $\phi_{\vec{k},\alpha} = \frac{1}{\epsilon_{\vec{k},\alpha}} S_{\vec{k}} \psi_{\vec{k},\alpha}$. Because $H_{\vec{k}}$ and $T_{\vec{k}}$ posses identical flat band wave functions and share the same SSG, they offer identical information on the topology of the flat band and can be used interchangeably to derive the properties of the flat bands. Additionally, this formal integration procedure is reminiscent of the construction of a line-graph Hamiltonian ($T_{\vec{k}}$) from a Hamiltonian defined on a ``root'' euclidean graph ($\tilde{T}_{\vec{k}}$) with the aid of the incidence matrix of the ``root'' graph ($S_{\vec{k}}$)~\cite{MIE91,KOL20}. This connection is discussed in more detail in \cref{app:sec:bipartite_crystalline_lattices:effective:lin_split}. It is worth noting that unlike the line-graph construction, in the BCL construction, $S_{\vec{k}}$ can denote \emph{any} type of inter-sublattice hopping matrix between \emph{any} orbitals (with or without spin-orbit coupling) and is not restricted to binary incidence matrices in spinless systems of $s$ orbitals. We present several examples in \cref{app:sec:example}.

The chiral BCL Hamiltonian $\mathcal{H}$ can be generalized by including generic intra-sublattice hopping terms between the orbitals of the $\tilde{L}$ sublattice. While the chiral symmetry no longer holds, a similar argument for the existence of flat bands remains. To see this, consider the Hamiltonian
\begin{equation}
	\label{eqn:bip_ham_gen}
	H_{\vec{k}} = \begin{pmatrix}
		A_{\vec{k}} & S_\vec{k} \\
		S_\vec{k}^\dagger & B_{\vec{k}} 
\end{pmatrix},
\end{equation}
where $A_{\vec{k}}$ ($B_{\vec{k}}$) is an $N_{L} \times N_{L}$ ($N_{\tilde{L}} \times N_{\tilde{L}}$) Hermitian matrix denoting the intra-sublattice hopping inside the $L$ ($\tilde{L}$) sublattice. We assume that $A_{\vec{k}}$ has a momentum-independent eigenvalue $a$ with degeneracy $n_a$. If $N_{\tilde{L}} < n_a \leq N_{L}$, then the Hamiltonian in \cref{eqn:bip_ham_gen} has at least $n_a-N_{\tilde{L}}$ flat bands of energy $a$ irrespective of $B_{\vec{k}}$. The proof of this statement is relegated in \cref{app:sec:breaking_chiral}. The simplest case is to consider $A_{\vec{k}}$ to be proportional to the identity matrix (\ie $A_{\vec{k}} = a \mathbb{1}$), while placing no constraints on $B_{\vec{k}}$, a construction which we term a \emph{generalized} BCL. One can also imagine a more general possibility in which $A_{\vec{k}}$ \emph{itself} is a BCL Hamiltonian with $n_a$ perfectly flat bands. As such, we can consider $A_{\vec{k}}$ to be a generalized BCL Hamiltonian comprised of sublattices $L'$ and $\tilde{L}'$ with $n_a = N_{L'}-N_{\tilde{L}'}$. However, it can be shown (see \cref{app:sec:breaking_chiral}) that by redefining $\tilde{L} \leftarrow \tilde{L}'$ and $L \leftarrow L \oplus L' $, such a Hamiltonian can always be brought in the form of \cref{eqn:bip_ham_gen} with $A_{\vec{k}}$ proportional to identity. We will henceforth consider $A_{\vec{k}} = a \mathbb{1}$. Moreover, because they were essentially defined from the kernel of $S^{\dagger}_{\vec{k}}$ and have support only on the $L$ sublattice, the chiral BCL flat band eigenstates $\Psi^{+}_{\vec{k},\alpha}$ (and corresponding BTPs) for $r_{\vec{k}} < \alpha \leq N_{L}$ will remain eigenvectors of $H_{\vec{k}}$, but with eigenvalue $a$. The corresponding flat band and BTP wave functions will not be affected by the introduction of the intra-sublattice hopping matrices $A_{\vec{k}} = a \mathbb{1}$ and $B_{\vec{k}}$. On the other hand the BTPs corresponding to $\Psi^{-}_{\vec{k},\alpha}$ for $r_{\vec{k}} < \alpha \leq N_{\tilde{L}}$ will generically be gapped. This implies that the topological properties of the flat bands, as well as the corresponding BTPs with the dispersive bands in a generalized BCL can be inferred from the zero modes of the effective Hamiltonian $T_{\vec{k}} = S_\vec{k}S^\dagger_\vec{k}$. The zero modes of $\tilde{T}_{\vec{k}} = S^\dagger_\vec{k} S_\vec{k}$ will not correspond to BTPs in the generalized BCL Hamiltonian from \cref{eqn:bip_ham_gen}.

{\emph{Symmetries in a BCL.}}~According to MTQC, the (co)irreps of an electronic band at high-symmetry momentum points in the Brillouin zone can be used to diagnose its topology~\cite{BRA17,PO17,WAT18,ELC20a}. Therefore MTQC is a natural starting point for discussing the topology of the BCL flat bands. We assume that the BCL Hamiltonian from \cref{eqn:bip_ham}, as well as each of the two sublattices \emph{individually} are invariant under a certain SSG $\mathcal{G}$ (in principle, each sublattice might be invariant under a supergroup of $\mathcal{G}$, the consequences of which will not be considered in this paper). At a given high-symmetry momentum point $\vec{K}$, the first-quantized Hamiltonian $H_{\vec{K}}$ must be invariant under the symmetry transformations belonging to $\mathcal{G}_{\vec{K}}$, the little-group corresponding to $\vec{K}$. Since the partitioning of the BCL obeys the symmetries of $\mathcal{G}$, every unitary or anti-unitary symmetry operation $g \in \mathcal{G}_{\vec{K}}$ is implemented individually in sublattice $L$ ($\tilde{L}$) by a unitary matrix $U(g)$ [$\tilde{U}(g)$] such that $\hat{g} \hat{a}^\dagger_{i,\vec{K}} \hat{g}^{-1} = \sum_{j=1}^{N_{L}} U_{ji}(g)  \hat{a}^\dagger_{j,\vec{K}}$ [$ \hat{g} \hat{b}^\dagger_{i,\vec{K}} \hat{g}^{-1} = \sum_{j=1}^{N_{\tilde{L}}} \tilde{U}_{ji}(g)  \hat{b}^\dagger_{j,\vec{K}}$]. Consequently (see \cref{app:sec:unitary_sym}), the inter-sublattice hopping matrix is invariant under the symmetry $g$, \ie $U(g) S^{(*)}_{\vec{K}} \tilde{U}^{\dagger} (g) = S_{\vec{K}}$, where ${}^{(*)}$ denotes complex conjugation when $g$ is anti-unitary. Next, we consider two sets of eigenstates $\phi_{\vec{K},\beta}$ and $\psi_{\vec{K},\beta}$ (labeled by $\beta$) corresponding to the two effective Hamiltonians $T_{\vec{K}}$ and $\tilde{T}_{\vec{K}}$ with identical eigenvalues $\epsilon^2_{\vec{K},\beta} = \mathcal{E}^2$ (with $\mathcal{E}>0$). Under the symmetry $g$, the eigenstates $\psi_{\vec{K},\beta}$ will transform under a certain (co)irrep of the little-group $\mathcal{G}_{\vec{K}}$, \ie
\begin{equation}
	\label{eqn:symmetry_rep_small}
	\tilde{U}(g) \psi^{(*)}_{\vec{K},\beta} = \sum_{\alpha} \left[\mathcal{R}^{\mathcal{E}}_{\vec{K}}(g) \right]_{\beta \alpha} \psi_{\vec{K},\alpha},
\end{equation}
where the sum runs over the states $\alpha$ for which $\epsilon_{\alpha, \vec{K}} = \mathcal{E}$. Left-multiplying \cref{eqn:symmetry_rep_small} by
$\frac{1}{\mathcal{E}} S_{\vec{K}}$ and employing the invariance of $S_{\vec{K}}$ under the group $\mathcal{G}_{\vec{K}}$ as well as the mapping between the nonzero eigenstates of $\tilde{T}_{\vec{K}}$ and $T_{\vec{K}}$, we find that
\begin{equation}
	\label{eqn:symmetry_rep_big}
	\frac{1}{\mathcal{E}} S_{\vec{K}} \tilde{U}(g) \psi^{(*)}_{\vec{K},\beta} = U(g) \phi^{(*)}_{\vec{K},\beta} = \sum_{\alpha} \left[\mathcal{R}^{\mathcal{E}}_{\vec{K}}(g) \right]_{\beta \alpha} \phi_{\vec{K},\alpha}.
\end{equation}
\Cref{eqn:symmetry_rep_big} implies that the set of eigenstates $\phi_{\vec{K},\beta}$ of the Hamiltonian $T_{\vec{K}}$ will transform according to the same (co)irrep $\mathcal{R}^{\mathcal{E}}_{\vec{K}}$ of the little group $\mathcal{G}_{\vec{K}}$. We conclude that the dispersive bands of the two effective Hamiltonians $T_{\vec{k}}$ and $\tilde{T}_{\vec{k}}$ are not only identical in energies, but also share the identical (co)irreps at the high symmetry momentum points. The only exceptions are the zero modes of the two effective Hamiltonians for which there is no direct mapping between the eigenstate. For the zero modes of $T_{\vec{k}}$ and $\tilde{T}_{\vec{k}}$ an indirect mapping between the (co)irreps will be derived below.

{\emph{Flat band (co)irreps.}}~We now derive the formula for the (co)irreps of the perfectly flat bands in $T_{\vec{k}}$, and correspondingly the flat bands of the generalized BCL Hamiltonian from \cref{eqn:bip_ham_gen} with $A_{\vec{k}} = a \mathbb{1}$, which are identical to the former. The proof outlined here relies on the effective sublattice Hamiltonians, but an alternative one which does not is presented in \cref{app:sec:unitary_sym}. Since $T_{\vec{k}}$ is defined on the $L$ sublattice, the BR corresponding to all the bands in $T_{\vec{k}}$ ($\mathcal{BR}_{L}$) can be found straight-forwardly: it is just the sum of all the elementary band representations (EBRs) induced from all the orbitals of the $L$ sublattice. Similarly, the BR of all the bands (including the flat bands) of $\tilde{T}_{\vec{k}}$ ($\mathcal{BR}_{\tilde{L}}$) is just the sum of EBRs induced from all the orbitals of the $\tilde{L}$ sublattice. Furthermore, we know that $T_{\vec{k}}$ and $\tilde{T}_{\vec{k}}$ share the same dispersive bands with the same (co)irreps at the high symmetry momentum points, except at the zero modes in $\tilde{T}_{\vec{k}}$ or at the BTPs with the flat bands in $T_{\vec{k}}$. We conclude the (co)irreps of the perfectly flat bands in $T_{\vec{k}}$ (which we term $\mathcal{B}_{\mathrm{FB}}$) are independent of the inter-sublattice hoppping matrix $S_{\vec{k}}$ and \emph{must} be given by those (co)irreps of $\mathcal{BR}_{L}$ which are not in $\mathcal{BR}_{\tilde{L}}$.

In what follows, it will be useful to extend the (co)irreps of the perfectly flat bands to include formal differences of (co)irreps at given momentum points by writing
\begin{equation}
	\label{eqn:br_subtraction}
	\mathcal{B}_{\mathrm{FB}} = \mathcal{BR}_{L} \boxminus \mathcal{BR}_{\tilde{L}}.
\end{equation}
This can be understood as formally introducing an identity element $\emptyset$ for the direct sum operation of (co)irreps ($\oplus$) followed by assigning an ``inverse'' $\boxminus \Xi$ to each (co)irrep $\Xi$, such that $\Xi \oplus \left( \boxminus \Xi \right) = \emptyset$. As flat bands are differences of BRs, they form a perfect playground for realizing fragile topological phases of matter, which also emerge as differences of BRs~\cite{BRA17,CAN18,PO18c,AHN19,ELS19,SON20,SON20c,CHI20}. When evaluating differences of (co)irreps, we will generically encounter expressions such as $\left( \Gamma_2 \oplus \Gamma_3\right) \boxminus \left( \Gamma_1 \oplus \Gamma_3\right) = \Gamma_2 \boxminus \Gamma_1$ which cannot be simplified further to (co)representation. Nevertheless, they will prove instrumental in diagnosing and understanding the BTPs between the flat and the dispersive bands arising in a generalized BCL (see \cref{app:sec:gapless}).

\begin{figure*}[!t]
\captionsetup[subfloat]{farskip=0pt}\sbox\nsubbox{
		\resizebox{\textwidth}{!}
		{\includegraphics[height=6cm]{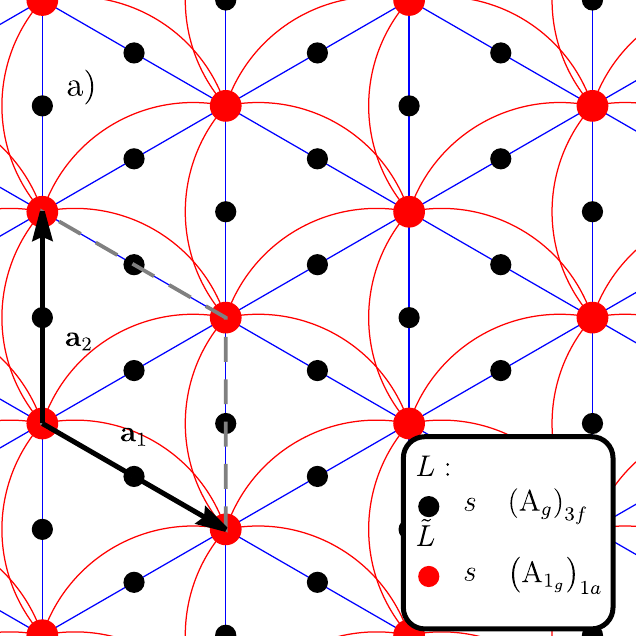}\includegraphics[height=6cm]{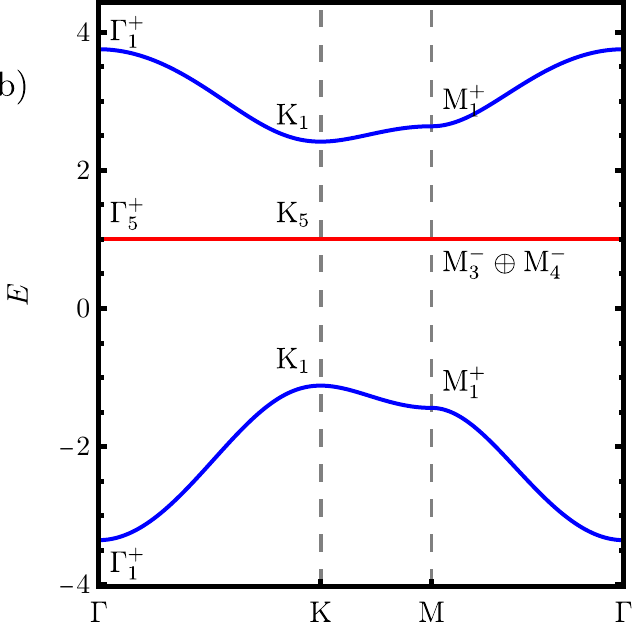}\includegraphics[height=6cm]{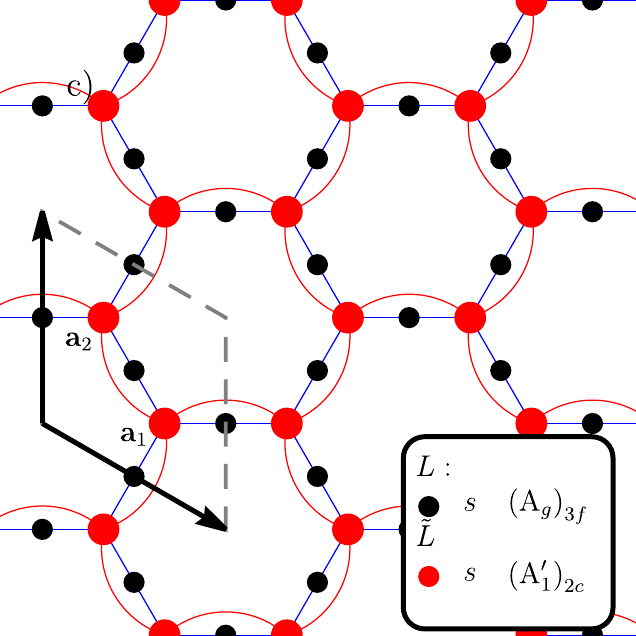}\includegraphics[height=6cm]{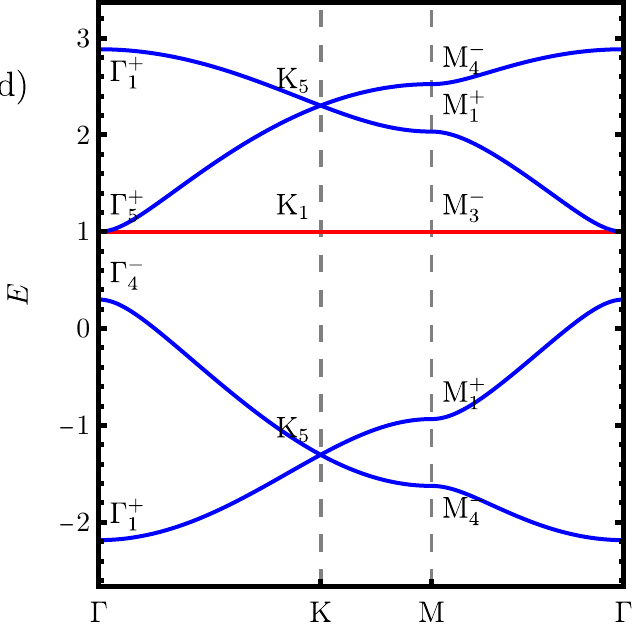}}}\setlength{\nsubht}{\ht\nsubbox}\centering\subfloat{\label{fig:kagome:a}\includegraphics[height=\nsubht]{1c.pdf}}\subfloat{\label{fig:kagome:b}\includegraphics[height=\nsubht]{1a.pdf}}\subfloat{\label{fig:kagome:c}\includegraphics[height=\nsubht]{1d.pdf}}\subfloat{\label{fig:kagome:d}\includegraphics[height=\nsubht]{1b.pdf}}
\caption{Examples of flat band constructions in generalized BCLs. Panels (a) and (c) illustrate the structure of two BCLs defined on the hexagonal lattice. In both cases, we  place $s$ orbitals at the $3f$ position within sublattice $L$ (black dots), while $\tilde{L}$ contains $s$ orbitals at the $1a$ and $2c$ position (red dots), in panels (a) and (c), respectively. The red and blue lines respectively denote intra-sublattice hopping with amplitude $t_2$ within $\tilde{L}$ and inter-sublattice hopping with amplitude $t_1$. There is a degenerate on-site energy term $\epsilon$ on sublattice $L$. The band structures corresponding to the BCLs from panels (a) and (c) are respectively shown in panels (b) and (d), for $t_1=-1$ and $t_2 =-0.1$, together with the corresponding (co)irreps~\citeBCS.}
\label{fig:kagome}
\end{figure*}

\emph{BTPs between the flat and dispersive bands.}
First, we consider the case when $\mathcal{BR}_{\tilde{L}}$ is a subset of $\mathcal{BR}_{L}$ at a given momentum point $\vec{K}$, such that \cref{eqn:br_subtraction} produces a \textit{bona fide} (co)representation at $\vec{K}$. This implies that any BTP between the flat and dispersive bands at $\vec{K}$ can be locally gapped. The proof follows by contradiction: assume the hopping matrix is not full-rank at $\vec{K}$ (\ie $r_{\vec{K}}<N_{\tilde{L}}$), and thus the flat bands of $T_{\vec{k}}$ are not gapped at $\vec{K}$. As such, there will be a (co)irrep of $\mathcal{G}_{\vec{K}}$, $\Xi \in \mathcal{BR}_{\tilde{L}}$, whose carrier space is given by the states $\psi_{\vec{K},\beta}$, for which $\tilde{T}_{\vec{K}} \psi_{\vec{K},\beta} = 0$. Because $\mathcal{BR}_{\tilde{L}} \subset \mathcal{BR}_{L}$, the (co)irrep $\Xi$ will also have a carrier space within the eigenstates of $T_{\vec{K}}$: $\phi_{\vec{K},\beta}$, with $T_{\vec{K}} \phi_{\vec{K},\beta} = 0$. As $\psi_{\vec{K},\beta}$ and $\phi_{\vec{K},\beta}$ form carrier spaces for the same (co)irrep $\Xi$, the inter-sublattice hopping matrix can be perturbed without breaking any crystalline symmetries 
\begin{equation}
	\label{eqn:smatrix_perturbation}
	S_{\vec{K}} \rightarrow S_{\vec{K}} + \mu \sum_{\beta} \phi_{\vec{K},\beta} \psi_{\vec{K},\beta}^\dagger,
\end{equation}
resulting in $\phi_{\vec{K},\beta}$ and $\psi_{\vec{K},\beta}$ becoming eigenstates with eigenvalue $\abs{\mu}^2$ of $T_{\vec{K}}$ and $\tilde{T}_{\vec{K}}$, respectively, and gapping the BCL flat bands at $\vec{K}$. This implies that there can be no locally-stable BTPs between the flat and dispersive bands of a BCL protected by crystalline symmetries, if the (co)irreps of $\mathcal{BR}_{\tilde{L}}$ form a subset of those of $\mathcal{BR}_{L}$. Additionally, as a linear combination of BRs, the BCL flat band(s) satisfy the compatibility relations~\cite{BRA17,ELC20a} and thus carry an \emph{bona fide} (co)representation at all momenta in the Brillouin zone, being generically gapped. Here, we define a locally-stable BTP as a BTP that can neither be gapped completely, nor have its location in momentum space changed by any symmetry-preserving perturbation to the $S_{\vec{k}}$ matrix. 

On the other hand, if $\mathcal{BR}_{\tilde{L}}$ is not a subset of $\mathcal{BR}_{L}$ at some momentum point $\vec{K}$, then the flat bands \emph{must} be degenerate with dispersive bands at $\vec{K}$. To see this, assume that at $\vec{K}$, the flat band is assigned the formal (co)irrep difference $\Xi \boxminus \Theta$, which cannot be simplified to a direct sum of (co)irreps. This means that the multiplicity of $\Xi$ ($\Theta$) in $\mathcal{BR}_{L}$ ($\mathcal{BR}_{\tilde{L}}$) is higher by one than in $\mathcal{BR}_{\tilde{L}}$ ($\mathcal{BR}_{L}$). Because of the one-to-one mapping of positive-energy eigenstates of the two effective Hamiltonians, it follows that the carrier space of $\Xi$ ($\Theta$) having dimension $d_{\Xi}$ ($d_{\Theta}$) is the kernel of $T_{\vec{K}}$ ($\tilde{T}_{\vec{K}}$), where $d_{\Xi} - d_{\Theta} = N_{L} - N_{\tilde{L}}$ (since $T_{\vec{K}}$ and $\tilde{T}_{\vec{K}}$ have the same rank). The kernel of $T_{\vec{K}}$ having dimension $d_{\Xi}$ includes the $N_{L}-N_{\tilde{L}}$ flat band eigenstates. Hence the flat bands in $T_{\vec{K}}$ (and in the corresponding generalized BCL) will touch exactly $d_{\Theta}$ dispersive bands at $\vec{K}$. There is no symmetry-preserving perturbation that can be added to the inter-sublattice hopping matrix $S_{\vec{K}}$ to gap this BTP. We thus conclude that it is locally-stable and protected by the crystalline symmetries of $\mathcal{G}$. At $\vec{K}$, the BTP is faithfully described only by the representation $\Xi$, but the complete formal difference $\Xi \boxminus \Theta$ is required to fully characterize the flat bands. This can be seen by considering two BTPs given by $\Xi \boxminus \Theta_1$ and $\Xi \boxminus \Theta_2$: in order to produce gapped flat bands in the two BCLs, different orbitals will need to be added in the $L$ lattice (\ie orbitals that induce the $\Theta_1$ and $\Theta_2$ (co)irreps, respectively). Moreover, in the vicinity of $\vec{K}$, $\Theta_1$ and $\Theta_2$ will subduce to potentially different (co)irreps of the corresponding little-group, effectively resulting in different (co)irreps being assigned to the flat bands. 

To illustrate the BR subtraction from \cref{eqn:br_subtraction}, in \cref{fig:kagome}, we present two generalized BCL examples on the two-dimensional hexagonal lattice in the $P6/mmm1'$ group (SSG 191.234 in the notation of Bilbao Crystallographic Server)~\citeBCS. In both examples, the $L$ sublattice contains $s$ orbitals at the $3f$ Wyckoff position. In \cref{fig:kagome:a}, we place $s$ orbitals at the $1a$ position in $\tilde{L}$, resulting in two degenerate flat bands with $\mathcal{B}_{\mathrm{FB}} = \left(\mathrm{A}_g \right)_{3f} \uparrow \mathcal{G} \boxminus \left(\mathrm{A}_{1_g} \right)_{3f} \uparrow \mathcal{G} = \left(\Gamma_{5}^+ \right) + \left( \mathrm{K}_5 \right) + \left( \mathrm{M}_{3}^{-} \oplus \mathrm{M}_{4}^{-}\right)$, as shown in \cref{fig:kagome:b}. Because all (co)irreps of $\mathcal{BR}_{\tilde{L}}$ are included in $\mathcal{BR}_{L}$, the flat bands are gapped. On the other hand, if $\tilde{L}$ contains $s$ orbitals at the $2c$ position (as shown in \cref{fig:kagome:c}), $\mathcal{B}_{\mathrm{FB}} = \left(\mathrm{A}_g \right)_{3f} \uparrow \mathcal{G} \boxminus \left(\mathrm{A}'_{1} \right)_{2c} \uparrow \mathcal{G} = \left(\Gamma_{5}^+ \boxminus \Gamma_{4}^- \right) + \left( \mathrm{K}_1 \right) + \left( \mathrm{M}_{3}^{-} \right)$ contains a formal (co)irrep difference at the $\Gamma$ point, signaling the presence of a BTP, as seen in \cref{fig:kagome:d}. 

We conclude that $\mathcal{B}_{\mathrm{FB}}$ contains all the information about the locally-stable BTP protected by crystalline symmetries, some of which were only partially understood in terms of \textit{ad-hoc} counting rules of real-space eigenstates with finite support~\cite{BER08,HWA21}. At a given momentum point $\vec{K}$, there are locally-stable BTPs if the BR subtraction results in a formal (co)irrep difference. On the other hand, if the subtraction rule in \cref{eqn:br_subtraction} generates direct sums of (co)irreps, the flat bands can always be locally gapped at $\vec{K}$. Note however that this does not preclude the existence of globally-stable BTP between the flat and dispersive bands (as shown in the example constructed in \cref{app:sec:example:C2T}). A globally-stable band touching point cannot be gapped completely by any symmetry-preserving perturbation to $S_{\vec{k}}$, but its position in the Brillouin zone \emph{can} be changed.

\emph{Classification.}~\Cref{eqn:br_subtraction} shows that the (co)irreps of the BCL flat band and corresponding BTPs in a generalized BCL Hamiltonian depend \emph{exclusively} on the orbital content of the two sublattices. The flat band (co)irreps ($\mathcal{B}_{\mathrm{FB}}$) are linear combination of EBRs, with positive or negative integer coefficients, depending on whether the corresponding orbitals belong to $L$ or $\tilde{L}$, respectively. This allows us to levy the power of MTQC which has catalogued all the EBRs in all 1651{} SSGs~\cite{BRA17,ELC20a}. The simple relation from \cref{eqn:br_subtraction} also hints that while the set of all possible perfectly flat bands for given SSG $\mathcal{G}$ ($\mathcal{F}_{\mathcal{G}}$) is \emph{infinite}, it is also \emph{finitely generated}.

To obtain $\mathcal{F}_{\mathcal{G}}$ explicitly, we introduce a $(d_{\cal G}+1)$-dimensional symmetry data vector $\overline{B}$, where $d_{\cal G}$ is the total number of (co)irreps of $\mathcal{G}$ for all high-symmetry momenta (see \cref{app:sec:fb_clas:sym_dat_vec}). The first component of $\overline{B}$ is a strictly positive entry specifying the number of flat bands indexed by $\mathcal{B}_{\mathrm{FB}}$, while the next $d_{\cal G}$ components specify the multiplicities of the flat band (co)irreps. Because $\mathcal{B}_{\mathcal{FB}}$ can contain formal differences of (co)irreps (corresponding to gapless flat bands), we allow the last $d_{\cal G}$ components of $\overline{B}$ to be negative. In \cref{app:sec:fb_clas:all}, we show that
\begin{equation}
	\mathcal{F}_{\mathcal{G}} = \left\{ \overline{B} \in \mathbb{Z}^{d_{\mathcal{G}}+1} \mid \overline{B} = \sum_{i=1}^{r} p_i e_i, p_i \in \mathbb{Z}, p_1 > 0  \right\},
\end{equation}
where $r$ is the number of linearly independent EBRs in the SSG $\mathcal{G}$ (\ie the rank of the $EBR$ matrix, as defined in \cref{app:sec:fb_clas:sym_dat_vec}), and $e_i$ ($1 \leq i \leq r$) is a set of integer $(d_{\cal G}+1)$-dimensional vectors which we term the flat band bases. In \cref{app:sec:fb_clas:table}, we tabulate the flat band bases for all 1651{} SSG, with or without spin-orbit coupling. This result is comprehensive over all symmetry groups and will be an invaluable tool in the experimental search for topological flat bands~\cite{REG21}.

With the space of all possible BCL flat bands in hand, obtaining the set of gapped flat bands $\mathcal{F}^{\mathrm{G}}_{\mathcal{G}}$ becomes a simple matter: to preclude band touching points, we restrict ourselves to the elements of $\mathcal{F}_{\mathcal{G}}$ with only nonnegative entries, such that the flat bands carry a \emph{bona fide} (co)representation at all momenta in the Brillouin zone. Using techniques of polyhedral computation first introduced to band theory by Ref.~\cite{SON20}, we show explicitly that $\mathcal{F}^{\mathrm{G}}_{\mathcal{G}}$ is also finitely generated and derive an algorithm for computing the corresponding bases in \cref{app:sec:fb_clas:gapped}. Moreover, as differences of BRs, we show in \cref{app:sec:fb_clas:fragile} that gapped BCL flat bands can also realize \emph{any} topologically fragile bands~\cite{BRA17,CAN18,PO18c,AHN19,SON20,SON20c,CHI20}, making them an ideal playground for strongly correlated phases of matter. In \cref{app:sec:fb_clas:relation_example} we present a simple examples illustrating the relation between gapped, gapless and topologically trivial bands.

\emph{Discussion.}~
We have presented a general technique for designing crystalline systems with perfectly flat bands. Unlike the various flat band models presented in literature, which it encapsulates, our method can be applied to systems with any orbital content, with or without spin-orbit coupling, and within any SSG. In particular, being less restrictive than the well-known line-graph construction, our procedure offers great hope for obtaining materials with flat bands near the Fermi energy, which realize exotic phases of matter. Ref.~\cite{REG21} highlights five prototypical compounds hosting flat bands (among many others) which can be explained with our formalism.

Within the framework of MTQC, the BCL flat bands can be understood as formal differences of BRs. This connection allow us to obtain criteria for identifying locally-stable BTPs between the flat and dispersive bands. In addition, we have shown that gapped BCL flat bands can realize any topologically fragile phase. Moreover, using the recently tabulated EBRs for all 1651{} SSG, we have constructed all possible symmetry data vectors that can be realized in flat bands, showing that the corresponding set in infinite, but finitely generated, and tabulating the corresponding bases.

\emph{Acknowledgments}~We thank Yuanfeng Xu, Ming-Rui Li, and Da-Shuai Ma for fruitful discussions and collaboration on related projects.
This work is part of a project that has received funding from the European Research Council (ERC) under the European Union's Horizon 2020 research and innovation programme (grant agreement no. 101020833). B.A.B. and N.R. were also supported by the U.S. Department of Energy (Grant No. DE-SC0016239), and were partially supported by the National Science Foundation (EAGER Grant No. DMR 1643312), a Simons Investigator grant (No. 404513), the Office of Naval Research (ONR Grant No. N00014-20-1-2303), the Packard Foundation, the Schmidt Fund for Innovative Research, the BSF Israel US foundation (Grant No. 2018226), the Gordon and Betty Moore Foundation through Grant No. GBMF8685 towards the Princeton theory program, and a Guggenheim Fellowship from the John Simon Guggenheim Memorial Foundation. B.A.B. and N.R. were supported by the NSF-MRSEC (Grant No. DMR-2011750). B.A.B. and N.R. gratefully acknowledge financial support from the Schmidt DataX Fund at Princeton University made possible through a major gift from the Schmidt Futures Foundation. L.E. was supported by the Government of the Basque Country (Project IT1301-19) and the Spanish Ministry of Science and Innovation (PID2019-106644GB-I00). B.A.B. received additional support from the Max Planck Society. Further support was provided by the NSF-MRSEC No. DMR-1420541, BSF Israel US foundation No. 2018226, and the Princeton Global Network Funds.

\let\oldaddcontentsline\addcontentsline
\renewcommand{\addcontentsline}[3]{}
\bibliographystyle{apsrev4-2}
\bibliography{FlatBandPaperBib}
\let\addcontentsline\oldaddcontentsline

\renewcommand{\thetable}{S\arabic{table}}
\renewcommand{\thefigure}{S\arabic{figure}}
\renewcommand{\theequation}{S\arabic{section}.\arabic{equation}}
\onecolumngrid
\pagebreak
\thispagestyle{empty}
\newpage
\begin{center}
	\textbf{\large Supplemental Material: \titlePaper}\\[.2cm]
\end{center}

\appendix
\renewcommand{\thesection}{\Roman{section}}
\tableofcontents
\let\oldaddcontentsline\addcontentsline
\interfootnotelinepenalty=10000
\NiceMatrixOptions{code-for-last-col=\scriptstyle,code-for-first-row=\scriptstyle}
\section{Bipartite Crystalline Lattices}\label{app:sec:bipartite_crystalline_lattices}
In this appendix, we provide a detailed discussion of the Bipartite Crystalline Lattice (BCL) construction. After defining our notation, we explain the presence of flat bands in a BCL Hamiltonian with chiral symmetry. We then introduce effective sublattice Hamiltonians and explain the relation between our construction and the line-graph and split-graph formalisms. Next, we discuss how the chiral symmetry can be broken in a BCL without perturbing the flat bands and derive the most general BCL Hamiltonian. Finally, we outline the implementation of crystalline symmetries in a BCL, provide an alternative proof of the band representation (BR) subtraction formula, and discuss its implications for the symmetry-protected locally-stable band touching points.

\subsection{Notation}\label{app:sec:bipartite_crystalline_lattices:notation}

Our method for constructing perfectly flat bands relies upon a family of quadratic, translation-invariant Hamiltonians defined on a BCL. A BCL is a translation-invariant collection of fermionic degrees of freedom which is partitioned into two sublattices, $L$ and $\tilde{L}$. We assume that the two sublattices obey the symmetries of the original lattice $L \oplus \tilde{L}$, with sublattice $L$ ($\tilde{L}$) containing $N_{L}$ ($N_{\tilde{L}}$) degrees of freedom within each unit cell. Without loss of generality, we will consider $N_{L} \geq N_{\tilde{L}}$. Denoting by $\hat{a}^\dagger_{\vec{R},i}$ ($1 \leq i \leq N_L$) and $\hat{b}^\dagger_{\vec{R},i}$ ($1 \leq i \leq N_{\tilde{L}}$) the creation operators corresponding to the $i$-th degree of freedom from unit cell $\vec{R}$ in sublattices $L$ and $\tilde{L}$, respectively, the most general BCL Hamiltonian can be written as
\begin{align}
	\mathcal{H} =& \sum_{\vec{R}_1, \vec{R}_2} \left[ \sum_{1 \leq i,j \leq N_L} \left[A_{\vec{R}_1 - \vec{R}_2}\right]_{ij} \hat{a}^\dagger_{\vec{R}_1,i}\hat{a}_{\vec{R}_2,j} + \sum_{1 \leq i,j \leq N_{\tilde{L}}} \left[B_{\vec{R}_1 - \vec{R}_2}\right]_{ij} \hat{b}^\dagger_{\vec{R}_1,i}\hat{b}_{\vec{R}_2,j} \right.\nonumber \\
	 +& \left.\sum_{\substack{1 \leq i \leq N_L \\ 1 \leq j \leq N_{\tilde{L}} }} \left( \left[S_{\vec{R}_1 - \vec{R}_2}\right]_{ij} \hat{a}^\dagger_{\vec{R}_1,i}\hat{b}_{\vec{R}_2,j} + \left[S^*_{\vec{R}_1 - \vec{R}_2}\right]_{ij} \hat{b}^\dagger_{\vec{R}_2,j}\hat{a}_{\vec{R}_1,i} \right) \right]. \label{app:eqn:gen_bcl_ham_rs}
\end{align}
In \cref{app:eqn:gen_bcl_ham_rs}, $A_{\vec{R}}$ ($B_{\vec{R}}$) is the intra-sublattice $N_{L} \times N_{L}$ ($N_{\tilde{L}} \times N_{\tilde{L}}$) hopping matrix corresponding to the fermions in sublattice $L$ ($\tilde{L}$) and $S_{\vec{R}}$ represents the $N_{L} \times N_{\tilde{L}}$ inter-sublattice hopping matrix. The index $i$ of the $\hat{a}^\dagger_{\vec{R},i}$ or $\hat{b}^\dagger_{\vec{R},i}$ operator incorporates the Wyckoff position, orbital and spin indices of the corresponding fermion. We note that given a partitioning of the lattice, one can rearrange any quadratic Hamiltonian with translation invariance into \cref{app:eqn:gen_bcl_ham_rs}. In \cref{app:sec:breaking_chiral}, we will derive sufficient conditions on the hopping matrices $A_{\vec{R}}$, $B_{\vec{R}}$, and $S_{\vec{R}}$ for $\mathcal{H}$ to have perfectly flat bands. For now, however, we will keep the discussion general in order to formalize our notation.

We introduce the Fourier-transformed operators 
\begin{equation}
	\label{app:eqn:momentum_ops}
	\begin{split}
		\hat{a}^\dagger_{\vec{k},i} &= \frac{1}{\sqrt{N}}\sum_{\vec{R}}\hat{a}^\dagger_{\vec{R},i} e^{-i \vec{k} \cdot \left( \vec{R} + \vec{r}^{\hat{a}}_i \right)}, \\
		\hat{b}^\dagger_{\vec{k},i} &= \frac{1}{\sqrt{N}}\sum_{\vec{R}}\hat{b}^\dagger_{\vec{R},i} e^{-i \vec{k} \cdot \left( \vec{R} + \vec{r}^{\hat{b}}_i \right)},
	\end{split}		
\end{equation}  
where $\vec{r}^{\hat{a}}_i$ ($\vec{r}^{\hat{b}}_i$) denotes the displacement of the $\hat{a}^\dagger_{\vec{R},i}$ ($\hat{b}^\dagger_{\vec{R},i}$) fermion relative to the unit cell origin, $\vec{k}$ represents the crystalline momentum, and $N$ represents the number of unit cells. Defining the Fourier-transformed hopping matrices
\begin{align}
	\left[ A_{\vec{k}} \right]_{ij} &= \sum_{\vec{R}} \left[ A_{\vec{R}} \right]_{ij} e^{i \vec{k} \cdot \left(\vec{R} + \vec{r}^{\hat{a}}_i - \vec{r}^{\hat{a}}_j \right)}, \label{app:eqn:momentum_A}  \\
	\left[ B_{\vec{k}} \right]_{ij} &= \sum_{\vec{R}} \left[ B_{\vec{R}} \right]_{ij} e^{i \vec{k} \cdot \left(\vec{R} + \vec{r}^{\hat{b}}_i - \vec{r}^{\hat{b}}_j \right)}, \label{app:eqn:momentum_B} \\
	\left[ S_{\vec{k}} \right]_{ij} &= \sum_{\vec{R}} \left[ S_{\vec{R}} \right]_{ij} e^{i \vec{k} \cdot \left(\vec{R} + \vec{r}^{\hat{a}}_i - \vec{r}^{\hat{b}}_j \right)}, \label{app:eqn:momentum_S}
\end{align} 
as well as the $\left( N_{L} + N_{\tilde{L}} \right)$-dimensional momentum space spinor
\begin{equation}
	\label{app:eqn:general_bas_spinor}
	\hat{\Psi}_{\vec{k}}^{T} = \left( \hat{a}_{\vec{k},1}, \hat{a}_{\vec{k},2}, \dots, \hat{a}_{\vec{k},N_{L}}, \hat{b}_{\vec{k},1}, \hat{b}_{\vec{k},2}, \dots, \hat{b}_{\vec{k},N_{\tilde{L}}} \right),
\end{equation}
the BCL Hamiltonian can be written compactly as 
\begin{equation}
	\label{app:eqn:gen_blc_ham_momentum}
	\mathcal{H} = \sum_{\vec{k}} \hat{\Psi}^\dagger_{\vec{k}} H_{\vec{k}} \hat{\Psi}_{\vec{k}},
\end{equation}
with the $\left( N_{L} + N_{\tilde{L}} \right) \times \left( N_{L} + N_{\tilde{L}} \right)$ first-quantized Hamiltonian
\begin{equation}
	\label{app:eqn:bip_ham_gen}
	H_{\vec{k}} = \begin{pNiceArray}{CC}[last-col,first-row]
		N_{L} & N_{\tilde{L}} \\
		A_{\vec{k}} & S_\vec{k} & N_{L}  \\
		S_\vec{k}^\dagger & B_{\vec{k}} & N_{\tilde{L}} \\
	\end{pNiceArray}.
\end{equation}
In \cref{app:eqn:bip_ham_gen}, the dimensions of the blocks in $H_{\vec{k}}$ have been written outside the matrix. Note that the diagonal blocks $A_{\vec{k}}$ and $B_{\vec{k}}$ are Hermitian.
\subsection{Flat bands in a chiral BCL}\label{app:sec:bipartite_crystalline_lattices:chiral} 
 
Before discussing the most general BCL Hamiltonian \cref{eqn:bip_ham_gen} for which perfectly flat bands emerge, it is useful to consider a simpler model featuring chiral symmetry in order to build an intuition on the origin of flat bands. We introduce a local chiral \emph{unitary} operator $C$ acting differently on the two sublattices,
\begin{equation}
    \label{app:eqn:definition_chiral_unitary}
	C \hat{a}^\dagger_{\vec{R},i} C^{-1} =  \hat{a}^\dagger_{\vec{R},i}, \qquad
	C \hat{b}^\dagger_{\vec{R},i} C^{-1} = - \hat{b}^\dagger_{\vec{R},i}, \qquad
	C i C = i.
\end{equation}
and require that $\left\lbrace C, \mathcal{H} \right\rbrace = 0$. The chiral symmetry can also be equivalently implemented by an \emph{anti-unitary} operator. We will discuss this possibility in \cref{app:eqn:definition_chiral_anti-unitary}. However, because only unitary operators admit eigenvalues, we will rely on the definition in \cref{app:eqn:definition_chiral_unitary} throughout this paper. The action of the chiral operator $C$ on the $\hat{\Psi}^\dagger_{\vec{k}}$ spinor reads
\begin{equation}
	C \hat{\Psi}^\dagger_{\vec{k}} C^{-1} = D \left( C \right) \hat{\Psi}^\dagger_{\vec{k}},
\end{equation}
where the representation matrix $D \left( C \right)$ is given by
\begin{equation}
	\label{app:eqn:rep_matrix_chiral}
	D \left( C \right) = \begin{pNiceArray}{CC}[last-col,first-row]
		N_{L} & N_{\tilde{L}} \\
		\mathbb{1} & \mathbb{0} & N_{L}  \\
		\mathbb{0} & -\mathbb{1} & N_{\tilde{L}} \\
	\end{pNiceArray}.
\end{equation}
The presence of chiral symmetry requires that $D \left( C \right) H_{\vec{k}}  D \left( C \right) = - H_{\vec{k}}$, and thus forbids intra-sublattice coupling. As a consequence, the first-quantized momentum-space Hamiltonian from \cref{app:eqn:bip_ham_gen} has the simple form
\begin{equation}
	\label{app:eqn:bip_ham_chiral}
	H_{\vec{k}} = \begin{pNiceArray}{CC}[last-col,first-row]
		N_{L} & N_{\tilde{L}} \\
		\mathbb{0} & S_\vec{k} & N_{L}  \\
		S_\vec{k}^\dagger & \mathbb{0} & N_{\tilde{L}} \\
	\end{pNiceArray}
\end{equation}
and corresponds to a BCL featuring hopping only between fermions belonging to different sublattices, which we term a chiral BCL. If we further impose that $N_{L} > N_{\tilde{L}}$, then $H_{\vec{k}}$ will necessarily have at least $N_{L} - N_{\tilde{L}}$ zero modes at each momentum point, implying that $\mathcal{H}$ will feature at least $N_{L} - N_{\tilde{L}}$ flat bands, as will be explained below. Such a construction is equivalent to the bipartite lattice with unequal number of lattice sites in each sublattice discussed in Ref.~\cite{LIE89}.

One way to understand the presence of flat bands in the chiral BCL Hamiltonian from \cref{app:eqn:bip_ham_chiral} is to consider the singular value decomposition (SVD) of the $N_L \times N_{\tilde L}$ matrix $S_{\vec{k}}$. Letting $r_{\vec{k}} \leq N_{\tilde{L}} < N_{L} $ be the rank of $S_{\vec{k}}$, we can write the SVD of $S_{\vec{k}}$ as 
\begin{equation}
	\label{app:eqn:s_svd}
	S_{\vec{k}} = W_{\vec{k}} \Sigma_{\vec{k}} V^{\dagger}_{\vec{k}} = \sum_{\alpha=1}^{r_{\vec{k}}} \epsilon_{\vec{k},\alpha} \phi_{\vec{k},\alpha} \psi^{\dagger}_{\vec{k},\alpha},
\end{equation}
where $W_{\vec{k}}$ ($V_{\vec{k}}$) is a $N_{L}\times N_{L}$ ($N_{\tilde{L}} \times N_{\tilde{L}}$) unitary matrix whose columns form the eigenstates of $S_{\vec{k}}S^\dagger_{\vec{k}}$ ($S^\dagger_{\vec{k}} S_{\vec{k}}$) and $\Sigma_{\vec{k}}$ is a $N_{L} \times N_{\tilde{L}}$ diagonal rectangular matrix containing $r_{\vec{k}}$ strictly positive entries on the main diagonal: the singular values of $S_{\vec{k}}$, denoted by $\epsilon_{\vec{k},\alpha}$, for $1 \leq \alpha \leq r_{\vec{k}}$. The $\alpha$-th column of $W_{\vec{k}}$ ($V_{\vec{k}}$) is denoted by $\phi_{\vec{k},\alpha}$ ($\psi_{\vec{k},\alpha}$) and is known as the $\alpha$-th left (right) singular eigenvector of $S_{\vec{k}}$. In what follows, we will assume that the matrix $S_{\vec{k}}$ is generically full rank (\ie with $r_{\vec{k}} = N_{\tilde{L}}$) throughout the Brillouin zone (with the exception of a finite number band touching points), such that there are exactly $N_{L}-N_{\tilde{L}}$ flat bands. 

To diagonalize $H_{\vec{k}}$, we use \cref{app:eqn:s_svd} and perform a unitary transformation of $H_{\vec{k}}$ 
\begin{equation}
	\label{app:eqn:svd_applied_H}
	H_{\vec{k}} = \begin{pmatrix}
		W_{\vec{k}} & \mathbb{0} \\
		\mathbb{0} & V_{\vec{k}} \\
	\end{pmatrix}
	\begin{pmatrix}
		\mathbb{0} & \Sigma_{\vec{k}} \\
		\Sigma^{T}_{\vec{k}} & \mathbb{0} \\
	\end{pmatrix}
	 \begin{pNiceArray}{CC}[last-col,first-row]
		N_{L} & N_{\tilde{L}} \\
		W^{\dagger}_{\vec{k}} & \mathbb{0} & N_{L}  \\
		\mathbb{0} & V^{\dagger}_{\vec{k}} & N_{\tilde{L}} \\
	\end{pNiceArray}.
\end{equation}
Because $H_{\vec{k}}$ is similar to a matrix containing $N_{L} - r_{\vec{k}}$ zero rows (columns), where $r_{\vec{k}} \leq N_{\tilde{L}}$, it follows that $H_{\vec{k}}$ must have at least $N_{L} - N_{\tilde{L}}$ zero modes for any $\vec{k}$. Consequently, $\mathcal{H}$ will have $N_{L}-N_{\tilde{L}}$ flat bands pinned at zero energy.

The spectrum of $H_{\vec{k}}$ can also be derived from \cref{app:eqn:svd_applied_H}. We denote the first-quantized eigenvectors and eigenvalues of $H_{\vec{k}}$ by $\Psi^{n}_{\vec{k},\alpha}$ and $\mathcal{E}^{n}_{\vec{k},\alpha}$, respectively, where 
\begin{equation}
	\label{app:eqn:eig_dec_H}
	H_{\vec{k}} \Psi^{n}_{\vec{k},\alpha} = \mathcal{E}^{n}_{\vec{k}, \alpha} \Psi^{n}_{\vec{k},\alpha},
\end{equation}
with $n=\pm$ and $\alpha$ indexing the eigenstates. For the nonzero eigenstates of $H_{\vec{k}}$ (\ie $\mathcal{E}^{n}_{\vec{k},\alpha} \neq 0$), the index $n$ is equal to the sign of $\mathcal{E}^{n}_{\vec{k},\alpha}$, while $1 \leq \alpha \leq r_{\vec{k}}$. For the zero eigenstates of $H_{\vec{k}}$, $\alpha > r_{\vec{k}}$, and we will show below that $n$ labels whether the corresponding eigenvectors have support on the $L$ ($n=+$) or $\tilde{L}$ ($n=-$) sublattice. We find that $H_{\vec{k}}$ has three types of eigenstates:
\begin{enumerate}
	\item For each $1 \leq \alpha \leq r_{\vec{k}}$, $H_{\vec{k}}$ has one pair of eigenstates
	\begin{equation}
		\label{app:eqn:eig_dispersive}
		\Psi_{\vec{k},\alpha}^{\pm} = \frac{1}{\sqrt{2}} \begin{pNiceArray}{C}[last-col]
			\pm \phi_{\vec{k},\alpha} & N_{L} \\
			\psi_{\vec{k},\alpha} & N_{\tilde{L}}  \\
		\end{pNiceArray}, \qquad \text{with} \qquad \mathcal{E}^{\pm}_{\vec{k},\alpha} = \pm \epsilon_{\vec{k},\alpha},
	\end{equation} 
	which are related by chiral symmetry (\ie $D \left( C \right) \Psi_{\vec{k},\alpha}^{\pm} = - \Psi_{\vec{k},\alpha}^{\mp} $). As they have non-zero energies and the flat bands of $H_{\vec{k}}$ are pinned at zero, the eigenstates in \cref{app:eqn:eig_dispersive} must correspond to the dispersive bands of $H_{\vec{k}}$.

	\item For $r_{\vec{k}} < \alpha \leq N_L$, $H_{\vec{k}}$ has $N_{L} - r_{\vec{k}}$ zero modes with support on sublattice $L$, corresponding to the kernel of $S_{\vec{k}}^{\dagger}$
	\begin{equation}
		\label{app:eqn:eig_zeroPlus}
		\Psi_{\vec{k},\alpha}^{+} = \begin{pNiceArray}{C}[last-col]
			\phi_{\vec{k},\alpha} & N_{L} \\
			\mathbb{0} & N_{\tilde{L}}  \\
		\end{pNiceArray}, \qquad \text{with} \qquad \mathcal{E}^{+}_{\vec{k},\alpha} = 0.
	\end{equation}
	The zero modes in \cref{app:eqn:eig_zeroPlus} are spanned by the $N_{L}-N_{\tilde{L}}$ flat band wave functions of $H_{\vec{k}}$ and, if $r_{\vec{k}} <N_{\tilde{L}} < N_{L}$ (\ie the $S_{\vec{k}}$ matrix is not full rank), by $N_{\tilde{L}}-r_{\vec{k}}$ additional zero modes which correspond to band touching points between the dispersive and flat bands.

	\item If $r_{\vec{k}} < N_{\tilde{L}}$, then $H_{\vec{k}}$ will also feature $N_{\tilde{L}}-r_{\vec{k}}$ zero modes (one for each $r_{\vec{k}} < \alpha \leq N_{\tilde{L}}$, corresponding to the kernel of $S_{\vec{k}}$) in addition to the ones in \cref{app:eqn:eig_zeroPlus},
	\begin{equation}
		\label{app:eqn:eig_zeroMinus}
		\Psi_{\vec{k},\alpha}^{-} = \begin{pNiceArray}{C}[last-col]
			\mathbb{0} & N_{L} \\
			\psi_{\vec{k},\alpha} & N_{\tilde{L}}  \\
		\end{pNiceArray}, \qquad \text{with} \qquad \mathcal{E}^{-}_{\vec{k},\alpha} = 0.
	\end{equation}	
	For general $\vec{k}$ points, $r_{\vec{k}} = N_{\tilde{L}}$ and so the zero modes in \cref{app:eqn:eig_zeroMinus} also constitute band touching points between the flat and dispersive bands, but have support on sublattice $\tilde{L}$. The total number of zero modes of $H_{\vec{k}}$ is thus $N_{L} + N_{\tilde{L}} - 2 r_{\vec{k}}$. At the $\vec{k}$ points where $S_{\vec{k}}$ is full-rank (\ie $r_{\vec{k}} = N_{\tilde{L}}$), $H_{\vec{k}}$ has exactly $N_{L} - N_{\tilde{L}}$ zero modes corresponding to the flat bands.  
\end{enumerate}  
In \cref{app:sec:bipartite_crystalline_lattices:effective,app:sec:breaking_chiral}, we will show that the chiral symmetry can be broken in a way that does not perturb the flat band eigenstates and corresponding band touching points from \cref{app:eqn:eig_zeroPlus} by the addition of intra-sublattice hopping inside the $\tilde{L}$ sublattice. On the other hand, this perturbation does gap the band touching points described by \cref{app:eqn:eig_zeroMinus}. Anticipating that discussion, we will refer to the band touching points from \cref{app:eqn:eig_zeroPlus} as ``stable'', and to the ones from \cref{app:eqn:eig_zeroMinus} as ``accidental'': the latter are only enforced when there is no intra-lattice hopping inside $\tilde{L}$.

Finally, we note that the chiral symmetry is sometimes implemented as an anti-unitary operator $\tilde{C}$ that switches creation and annihilation operators (and also acts differently on the two sublattices)~\cite{LUD15},
\begin{equation}
    \label{app:eqn:definition_chiral_anti-unitary}
	\tilde{C} \hat{a}^\dagger_{\vec{R},i} \tilde{C}^{-1} =  \hat{a}_{\vec{R},i}, \qquad
	\tilde{C} \hat{b}^\dagger_{\vec{R},i} \tilde{C}^{-1} = - \hat{b}_{\vec{R},i}, \qquad
	\tilde{C} i \tilde{C} = -i.
\end{equation}
The presence of chiral symmetry implies a commutation (rather than an anti-commutation) relation $\left[ \tilde{C}, \mathcal{H} \right] = 0$. The action of $\tilde{C}$ on the $\hat{\Psi}^\dagger_{\vec{k}}$ spinor is given by 
\begin{equation}
	\tilde{C} \hat{\Psi}^\dagger_{\vec{k}} \tilde{C}^{-1} = D ( \tilde{C} ) \hat{\Psi}_{\vec{k}},
\end{equation}
where the representation matrix of $\tilde{C}$ is the same as that of $C$ given in \cref{app:eqn:rep_matrix_chiral}, $D ( \tilde{C}) = D ( C )$. The presence of chiral symmetry $\tilde{C}$ imposes the same constraint as $C$ on the first-quantized Hamiltonian, requiring that $D ( \tilde{C} ) H_{\vec{k}}  D ( \tilde{C} ) = - H_{\vec{k}}$, and thus forbidding intra-sublattice coupling. However, because anti-unitary operators do not admit eigenvalues, we will implement the chiral symmetry operator as defined in \cref{app:eqn:definition_chiral_unitary}.

\subsection{Effective Hamiltonians. Relation with the line- and split-graph constructions}\label{app:sec:bipartite_crystalline_lattices:effective}

In \cref{app:sec:bipartite_crystalline_lattices:chiral}, we explained the origin of flat bands in a chiral BCL using the SVD of the inter-sublattice hopping matrix $S_{\vec{k}}$. Here, we introduce a different approach for understanding the BCL flat bands in terms of effective sublattice Hamiltonians. This alternative picture of BCL flat bands will not only enable us to explain the relation between our construction and the line-graph and split-graph formalisms~\cite{LIE89,MIE91,MIE91a,MIE92,MIE92a,MIE93,BER08,KOL19,KOL20,CHI20,MA20}, but also show how our construction encompasses and generalizes the latter. 

\subsubsection{Sublattice Hamiltonians}\label{app:sec:bipartite_crystalline_lattices:effective:def}

To derive the effective sublattice Hamiltonians, we start with the chiral BCL Hamiltonian from \cref{app:eqn:bip_ham_chiral} to which we add on-site chemical potential terms $\mu_{L}$ and $\mu_{\tilde{L}}$ for the atoms in sublattice $L$ and $\tilde{L}$, respectively. The corresponding Hamiltonian matrix reads
\begin{equation}
	\label{app:eqn:bip_ham_chiral_mu}
	H_{\vec{k}} = \begin{pNiceArray}{CC}[last-col,first-row]
		N_{L} & N_{\tilde{L}} \\
		\mu_{L} \mathbb{1} & S_\vec{k} & N_{L}  \\
		S_\vec{k}^\dagger & \mu_{\tilde{L}} \mathbb{1} & N_{\tilde{L}} \\
	\end{pNiceArray}.
\end{equation}
By analogy with \cref{app:eqn:svd_applied_H}, we perform a similarity transformation on $H_{\vec{k}}$
\begin{equation}
	H_{\vec{k}} = \begin{pmatrix}
		W_{\vec{k}} & \mathbb{0} \\
		\mathbb{0} & V_{\vec{k}} \\
	\end{pmatrix}
	\begin{pmatrix}
		\mu_L \mathbb{1} & \Sigma_{\vec{k}} \\
		\Sigma^{T}_{\vec{k}} & \mu_{\tilde{L}} \mathbb{1} \\
	\end{pmatrix}
	 \begin{pNiceArray}{CC}[last-col,first-row]
		N_{L} & N_{\tilde{L}} \\
		W^{\dagger}_{\vec{k}} & \mathbb{0} & N_{L}  \\
		\mathbb{0} & V^{\dagger}_{\vec{k}} & N_{\tilde{L}} \\
	\end{pNiceArray},
\end{equation}
which allows us to derive its eigenspectrum. Using the same notation as in \cref{app:eqn:eig_dec_H}, we find that the eigenstates of \cref{app:eqn:bip_ham_chiral_mu} are also of three types and can be written in terms of the left and right singular eigenvectors of $S_{\vec{k}}$:
\begin{enumerate}
	\item For each $1 \leq \alpha \leq r_{\vec{k}}$, $H_{\vec{k}}$ features two eigenstates
	\begin{equation}
	\label{app:eqn:eig_dispersive_mu} 
	\Psi_{\vec{k},\alpha}^{\pm} = \frac{1}{\sqrt{\epsilon_{\vec{k},\alpha}^2 + \left( u \pm \sqrt{u^2 +\epsilon_{\vec{k},\alpha}^2} \right)^2 }} \begin{pNiceArray}{C}[last-col]
		\left( u \pm \sqrt{u^2 + \epsilon_{\vec{k},\alpha}^2} \right) \phi_{\vec{k},\alpha} & N_{L} \\
		\epsilon_{\vec{k},\alpha} \psi_{\vec{k},\alpha} & N_{\tilde{L}}  \\
	\end{pNiceArray}, \qquad \text{with} \qquad \mathcal{E}^{\pm}_{\vec{k},\alpha} = \frac{\mu_{L} + \mu_{\tilde{L}}}{2} \pm \sqrt{u^2 + \epsilon_{\vec{k},\alpha}^2},
\end{equation}
where $u = \frac{\mu_{L}-\mu_{\tilde{L}}}{2}$. The eigenstates described in \cref{app:eqn:eig_dispersive_mu} correspond to the dispersive bands of $H_{\vec{k}}$, with \cref{app:eqn:eig_dispersive} being recovered by setting $\mu_{L} = \mu_{\tilde{L}} = 0$.

	\item For $r_{\vec{k}} < \alpha \leq N_L$, $H_{\vec{k}}$ has $N_{L} - r_{\vec{k}}$ eigenstates with support on the $L$ sublattice, with the same $\vec{k}$-independent eigenvalue 
\begin{equation}
	\label{app:eqn:eig_zeroPlus_mu}
	\Psi_{\vec{k},\alpha}^{+} = \begin{pNiceArray}{C}[last-col]
		\phi_{\vec{k},\alpha} & N_{L} \\
		\mathbb{0} & N_{\tilde{L}}  \\
	\end{pNiceArray}, \qquad \text{with} \qquad \mathcal{E}^{+}_{\vec{k},\alpha} = \mu_{L}.
\end{equation}
\Cref{app:eqn:eig_zeroPlus_mu} implies that the flat band modes and corresponding band touching points of $H_{\vec{k}}$ at specific $\vec{k}$ from \cref{app:eqn:eig_zeroPlus}, are unperturbed by the introduction of the chemical potential terms, up to a shift in energy (unlike the dispersive bands, whose wave functions change). 

	\item If $r_{\vec{k}} < N_{\tilde{L}}$, then $H_{\vec{k}}$ will inherit the $N_{\tilde{L}}-r_{\vec{k}}$ eigenstates from \cref{app:eqn:eig_zeroMinus}
\begin{equation}
	\label{app:eqn:eig_zeroMinus_mu}
	\Psi_{\vec{k},\alpha}^{-} = \begin{pNiceArray}{C}[last-col]
		\mathbb{0} & N_{L} \\
		\psi_{\vec{k},\alpha} & N_{\tilde{L}}  \\
	\end{pNiceArray}, \qquad \text{with} \qquad \mathcal{E}^{-}_{\vec{k},\alpha} = \mu_{\tilde{L}}.
\end{equation}
Unlike in the case with chiral symmetry (where $\mu_L=\mu_{\tilde{L}}=0$), the eigenstates in \cref{app:eqn:eig_zeroMinus_mu} are gapped from the flat band modes in \cref{app:eqn:eig_zeroPlus_mu} and therefore do not correspond to band touching points between the flat and dispersive bands.  
\end{enumerate} 

We stress that the difference between the band touching points described by the eigenstates from \cref{app:eqn:eig_zeroPlus_mu,app:eqn:eig_zeroMinus} arises because $N_{L} > N_{\tilde{L}}$: the flat bands are pinned at the chemical potential of the larger sublattice, $\mu_{L}$. If the situation were reversed and $N_{\tilde{L}} > N_L$, then the flat bands would have an energy $\mu_{\tilde{L}}$. Moreover, the band touching points from \cref{app:eqn:eig_zeroPlus_mu} would become gapped, while those described by \cref{app:eqn:eig_zeroMinus_mu} would remain gapless.

Having detailed the spectrum of $H_{\vec{k}}$ with on-site intra-sublattice terms, we now consider the limit $\abs{ \mu_{L} - \mu_{\tilde{L}} } \gg \norm{S_{\vec{k}}}$ for all $\vec{k}$, where $\norm{S_{\vec{k}}}$ represents the spectral norm (\ie the largest singular value) of $S_{\vec{k}}$. Physically, this is equivalent to introducing a large energy difference between the fermionic degrees of freedom from the two sublattices. As a consequence, inter-sublattice hopping processes become energetically suppressed and the low-energy spectrum of $H_{\vec{k}}$ can be computed using second-order degenerate perturbation theory~\cite{WIN03}:
\begin{itemize}
	\item If $\mu_{L}< \mu_{\tilde{L}}$ (in addition to $\abs{ \mu_{L} - \mu_{\tilde{L}} } \gg \norm{S_{\vec{k}}}$), then the lower $N_L$ bands are separated from the upper $N_{\tilde{L}}$ bands by a large gap of $\sim \abs{ \mu_{L} - \mu_{\tilde{L}} }$. Assuming that the lower $N_{L}$ bands are partially filled, a low-energy theory of the BCL Hamiltonian $\mathcal{H}$ is given by the effective Hamiltonian $\mathcal{T}_{\mathrm{eff}}$ which has support only on sublattice $L$ and is given by
	\begin{equation}
		\label{app:eqn:eff_Ham_L}
		\mathcal{T}_{\mathrm{eff}} = \sum_{\substack{\vec{k} \\1 \leq i,j \leq N_{L}}} \left(  \frac{1}{\mu_{L} - \mu_{\tilde{L}}}  \left[T_{\vec{k}} \right]_{ij} + \mu_L \delta_{ij} \right) \hat{a}^\dagger_{\vec{k},i} \hat{a}_{\vec{k},j} 
		= \sum_{\substack{\vec{R}_1,\vec{R}_2 \\1 \leq i,j \leq N_{L}}} \left(  \frac{1}{\mu_{L} - \mu_{\tilde{L}}}  \left[T_{\vec{R}_1 - \vec{R}_2} \right]_{ij} + \mu_L \delta_{ij} \delta_{\vec{R}_1, \vec{R}_2} \right) \hat{a}^\dagger_{\vec{R}_1,i} \hat{a}_{\vec{R}_2,j}
	\end{equation}
	In \cref{app:eqn:eff_Ham_L}, the effective Hamiltonian $T_{\vec{k}}$ ($T_{\vec{R}}$) matrix can be expressed in terms of $S_{\vec{k}}$ ($S_{\vec{R}}$) as
	\begin{equation}
		\label{app:eqn:T_deff}
		T_{\vec{k}} = S_{\vec{k}} S^{\dagger}_{\vec{k}}, \qquad T_{\vec{R}} = \sum_{\vec{R}'} S_{\vec{R}+\vec{R}'} S^{\dagger}_{\vec{R}'}.
	\end{equation}
	The degrees of freedom of the $T_{\vec{k}}$ Hamiltonian are the fermions in the $L$ sublattice. Hoppings to the energetically penalized $\tilde{L}$ lattice are virtual and lead to second-order terms in the effective theory.
	\item Alternatively, if $\mu_{L} > \mu_{\tilde{L}}$ (in addition to $\abs{ \mu_{L} - \mu_{\tilde{L}} } \gg \norm{S_{\vec{k}}}$), then the lower $N_{\tilde{L}}$ bands are separated from the upper $N_{L}$ bands by a large gap of $\sim \abs{ \mu_{L} - \mu_{\tilde{L}} }$. Assuming that we are partially filling the lower $N_{\tilde{L}}$ bands, then  $\mathcal{H}$ becomes governed by an effective Hamiltonian having support only on sublattice $\tilde{L}$, which reads
	\begin{equation}
		\label{app:eqn:eff_Ham_Ltilde}
		\tilde{\mathcal{T}}_{\mathrm{eff}} = \sum_{\substack{\vec{k} \\1 \leq i,j \leq N_{L}}} \left(  \frac{1}{\mu_{\tilde{L}} - \mu_{L}}  \left[\tilde{T}_{\vec{k}} \right]_{ij} + \mu_{\tilde{L}} \delta_{ij} \right) \hat{b}^\dagger_{\vec{k},i} \hat{b}_{\vec{k},j} 
		= \sum_{\substack{\vec{R}_1,\vec{R}_2 \\1 \leq i,j \leq N_{\tilde{L}}}} \left(  \frac{1}{\mu_{\tilde{L}} - \mu_{L}}  \left[\tilde{T}_{\vec{R}_1 - \vec{R}_2} \right]_{ij} + \mu_{\tilde{L}} \delta_{ij} \delta_{\vec{R}_1, \vec{R}_2} \right) \hat{b}^\dagger_{\vec{R}_1,i} \hat{b}_{\vec{R}_2,j}
	\end{equation}	
	with the corresponding Hamiltonian matrix being given by
	\begin{equation}
		\label{app:eqn:Ttilde_deff}
		\tilde{T}_{\vec{k}} = S^{\dagger}_{\vec{k}} S_{\vec{k}}, \qquad \tilde{T}_{\vec{R}} = \sum_{\vec{R}'} S^{\dagger}_{\vec{R}'} S_{\vec{R} + \vec{R}'}.
	\end{equation}
\end{itemize}

The two effective Hamiltonian matrices $T_{\vec{k}}$ and $\tilde{T}_{\vec{k}}$ can also be directly related to $H_{\vec{k}}$ by observing that 
\begin{equation}
	\left( H_{\vec{k}} \right)_{\mu_L=\mu_{\tilde{L}}=0}^2 = \begin{pNiceArray}{CC}[last-col,first-row]
		N_{L} & N_{\tilde{L}} \\
		T_{\vec{k}} & \mathbb{0} & N_{L}  \\
		\mathbb{0} & \tilde{T}_{\vec{k}} & N_{\tilde{L}} \\
	\end{pNiceArray}.
\end{equation}

The link between the spectra of $H_{\vec{k}}$ and the two effective Hamiltonians $T_{\vec{k}}$ and $\tilde{T}_{\vec{k}}$ can be made more rigorous by employing the SVD of the $S_{\vec{k}}$ matrix from \cref{app:eqn:s_svd}. The eigendecomposition of the effective Hamiltonians reads
\begin{align}
	T_{\vec{k}} &= \sum_{\alpha = 1}^{r_{\vec{k}}} \epsilon_{\vec{k},\alpha}^{2} \phi_{\vec{k},\alpha} \phi^{\dagger}_{\vec{k},\alpha}, \label{app:eqn:spectrum_T}\\
	\tilde{T}_{\vec{k}} &= \sum_{\alpha = 1}^{r_{\vec{k}}} \epsilon_{\vec{k},\alpha}^{2} \psi_{\vec{k},\alpha} \psi^{\dagger}_{\vec{k},\alpha}. \label{app:eqn:spectrum_Ttilde}
\end{align} 
The eigenvectors of $T_{\vec{k}}$ ($\tilde{T}_{\vec{k}}$) are the left (right) singular eigenvectors of $S_{\vec{k}}$, namely $\phi_{\vec{k},\alpha}$ for $1 \leq \alpha \leq N_L$ ($\psi_{\vec{k},\alpha}$ for $1 \leq \alpha \leq N_{\tilde{L}}$). Additionally, $T_{\vec{k}}$ and $\tilde{T}_{\vec{k}}$ have identical positive energy eigenvalues, with the $S_{\vec{k}}$ matrix acting as a map between the corresponding eigenvectors. More precisely, if $\psi_{\vec{k},\alpha}$ is a normalized eigenvector of $\tilde{T}_{\vec{k}}$ with nonzero eigenvalue $\epsilon_{\vec{k},\alpha}^2$ (\ie $\tilde{T}_{\vec{k}} \psi_{\vec{k},\alpha} = \epsilon_{\vec{k},\alpha}^2 \psi_{\vec{k},\alpha}$), then $\phi_{\vec{k,\alpha}} = \frac{1}{\epsilon_{\vec{k}}} S_{\vec{k}} \psi_{\vec{k},\alpha}$ is a normalized eigenvector of $T_{\vec{k}}$ with the same eigenvalue (\ie $T_{\vec{k}} \phi_{\vec{k},\alpha} = \epsilon_{\vec{k},\alpha}^2 \phi_{\vec{k},\alpha}$). The converse is also true, with $S_{\vec{k}}^{\dagger}$ acting as a map between the eigenvectors of $T_{\vec{k}}$ and $\tilde{T}_{\vec{k}}$.

For $N_{L}>N_{\tilde{L}}$ (which is the case we are considering throughout this paper), the eigenstates of $H_{\vec{k}}$ from \cref{app:eqn:eig_zeroPlus_mu}, including the flat bands and corresponding band touching points are zero modes of $T_{\vec{k}}$, implying that the flat bands of $H_{\vec{k}}$ and $T_{\vec{k}}$ are identical in terms of wave functions. On the other hand, the zero modes of $\tilde{T}_{\vec{k}}$ occur only at specific values of crystal momentum (\ie where $S_{\vec{k}}$ is not full-rank) and are identical to the eigenstates of $H_{\vec{k}}$ from \cref{app:eqn:eig_zeroMinus_mu}. Although degenerate with the flat bands of $H_{\vec{k}}$ in the chiral case, we will show in \cref{app:sec:breaking_chiral}, that they can be generically gapped and, as such, are not relevant for the topology of the BCL flat bands. This situation is reverse if $N_{L}<N_{\tilde{L}}$.

\subsubsection{Line- and split-graph constructions}\label{app:sec:bipartite_crystalline_lattices:effective:lin_split}

\begin{figure}[!t]
\captionsetup[subfloat]{farskip=0pt}\sbox\nsubbox{
		\resizebox{\textwidth}{!}
		{\includegraphics[height=6cm]{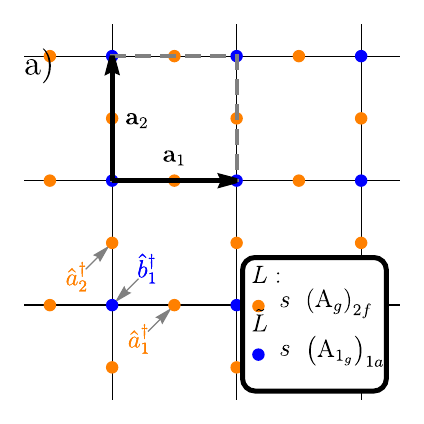}\includegraphics[height=6cm]{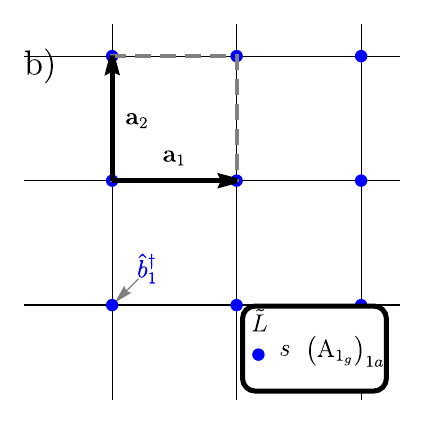}\includegraphics[height=6cm]{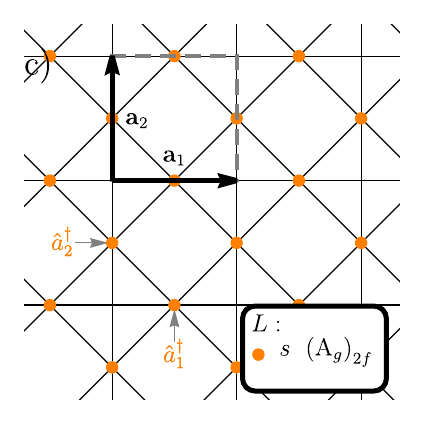}}}\setlength{\nsubht}{\ht\nsubbox}\centering\subfloat{\label{app:fig:split_root_line_lat:a}\includegraphics[height=\nsubht]{2f.pdf}}\subfloat{\label{app:fig:split_root_line_lat:b}\includegraphics[height=\nsubht]{2d.pdf}}\subfloat{\label{app:fig:split_root_line_lat:c}\includegraphics[height=\nsubht]{2e.pdf}}

\captionsetup[subfloat]{farskip=0pt}\sbox\nsubbox{
		\resizebox{\textwidth}{!}
		{\includegraphics[height=6cm]{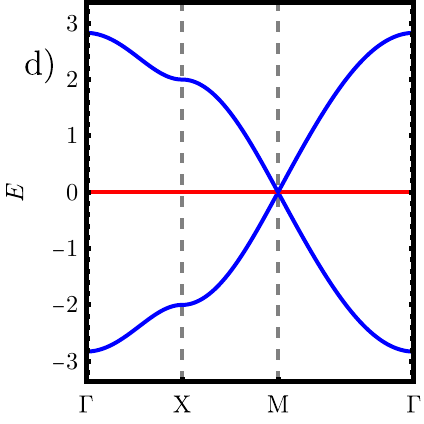}\includegraphics[height=6cm]{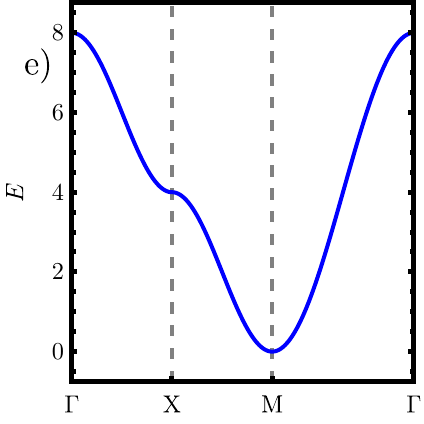}\includegraphics[height=6cm]{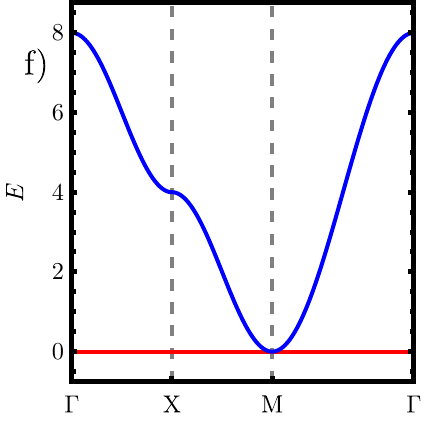}}}\setlength{\nsubht}{\ht\nsubbox}\centering\subfloat{\label{app:fig:split_root_line_band:a}\includegraphics[height=\nsubht]{2c.pdf}}\subfloat{\label{app:fig:split_root_line_band:b}\includegraphics[height=\nsubht]{2a.pdf}}\subfloat{\label{app:fig:split_root_line_band:c}\includegraphics[height=\nsubht]{2b.pdf}}
\caption{Split- and line-graph Hamiltonians of the two-dimensional square lattice. Panels a)-c) illustrate the lattice structure of the split-, root- and line-graph Hamiltonians defined for the two-dimensional square lattice according to \cref{app:eqn:ham_split_graph_rs,app:eqn:ham_root_graph_rs,app:eqn:ham_line_graph_rs}, respectively. The unit cell, spanned by the lattice vectors $\vec{a}_{1}$ and $\vec{a}_{2}$, is denoted by a dashed gray square, while the orbitals within each unit cell are represented by filled dots. The orbital corresponding to each fermionic operator $\hat{a}^\dagger_{1}$, $\hat{a}^\dagger_{2}$, or $\hat{b}^\dagger_{1}$ (where we suppress the unit cell index $\vec{R}$) is indicated by a gray arrow. The $\tilde{L}$ sublattice is formed of spinless $s$ orbitals at the $1a$ Wyckoff position (show in blue), which also form the vertices $\tilde{V}$ of the lattice graph $\tilde{G}\left( \tilde{V},\tilde{E} \right)$. The edges $\tilde{E}$ are defined by the lines connecting nearest neighbor $s$ orbitals in $\tilde{L}$. To each edge in $\tilde{E}$, we associate an $s$ orbitals (shown in orange), such that the $L$ sublattice is formed of $s$ orbitals at the $2f$ position. A black line connecting two orbitals corresponds to quadratic hopping of amplitude $t$ [in panel a)] or $t^2$ [in panels b) and c)]. The band structures corresponding to the split-, root-, and line-graph Hamiltonians are presented respectively in panels d)-f). Here, we consider $t=1$. As expected, the split- and line-graph Hamiltonians feature $N_{L} - N_{\tilde{L}} = 1$ perfectly flat band (shown in red). At the same time, the dispersive bands of the root- and line-graph Hamiltonians (shown in blue) are identical in energy (having energy $\epsilon_{\vec{k},\alpha}^2$), as expected from \cref{app:eqn:spectrum_T,app:eqn:spectrum_Ttilde}. The dispersive bands of the split-graph Hamiltonian come in pairs related by chiral symmetry and have energies $\pm \epsilon_{\vec{k},\alpha}$, as implied by \cref{app:eqn:eig_dispersive}.}
\label{app:fig:split_root_line}
\end{figure}

The effective Hamiltonians introduced in \cref{app:sec:bipartite_crystalline_lattices:effective:def} explain the link between BCLs and the line- and split-graph constructions and show how the latter are not only encapsulated, but also generalized by our formalism. Here, we briefly outline this relation, and point the reader to Refs.~\cite{MIE91,MIE91a,MIE92,MIE92a,MIE93,KOL20,CHI20,MA20} for a more in-depth discussions of line- and split-graphs. In short, the line- and split-graph construction correspond to a BCL construction that uses only $s$ orbitals, nearest-neighbor hoppings, and no spin-orbit coupling.

We consider $\tilde{L}$ to be formed of spinless $s$ orbitals~\footnote{Usually, the line- and split-graph construction are defined for systems without spin-orbit coupling, implying that the spin degree of freedom merely accounts for a two-fold degeneracy.} placed at the vertices of a (connected) lattice graph $ \tilde{G} \left(\tilde{V}, \tilde{E} \right)$, with $\tilde{V}$ and $\tilde{E}$ denoting the set of vertices and edges of the graph $\tilde{G}$. We take $\left(\vec{R},i \right) \in \tilde{V}$ to be a vertex of $\tilde{G}$ located in unit cell $\vec{R}$ and with $i$ indexing the corresponding $s$ orbital within the unit cell ($1 \leq i \leq N_{\tilde{L}}$). Similarly, we label the edges of $\tilde{G}$ by $\left[ \vec{R}, i \right] \in \tilde{E}$, with $\vec{R}$ labeling the unit cell and $i$ indexing the edge within the unit cell. Now define the $L$ sublattice to be formed of $s$ orbitals located at the midpoints of the edges of $\tilde{G}$. This allows for a natural definition of the inter-sublattice hopping matrix $S_{\vec{R}}$ as the rescaled (translation-invariant) incidence matrix of the graph $\tilde{G}$
\begin{equation}
	\label{app:eqn:s_matrix_incidence}
	\left[S_{\vec{R}_{2} - \vec{R}_1}\right]_{ij} = \begin{cases}
		t, & $ if the edge $\left[\vec{R}_1,i \right]$ is incident on the vertex $\left(\vec{R}_2,j \right) \\
		0, & $ otherwise $
	\end{cases},
\end{equation}
where $t$ represents a real hopping parameter, $1 \leq i \leq N_L$, and $1 \leq j \leq N_{\tilde{L}}$. Taking $A_{\vec{R}} = B_{\vec{R}} = \mathbb{0}$ in \cref{app:eqn:gen_bcl_ham_rs}, we obtain the following quadratic Hamiltonian
\begin{equation}
	\label{app:eqn:ham_split_graph_rs}
	\mathcal{H}_{\mathrm{split}} = \sum_{\vec{R}_1, \vec{R}_2} \left[ \sum_{\substack{1 \leq i \leq N_L \\ 1 \leq j \leq N_{\tilde{L}} }} \left( \left[S_{\vec{R}_1 - \vec{R}_2}\right]_{ij} \hat{a}^\dagger_{\vec{R}_1,i}\hat{b}_{\vec{R}_2,j} + \left[S_{\vec{R}_1 - \vec{R}_2}\right]_{ij} \hat{b}^\dagger_{\vec{R}_2,j}\hat{a}_{\vec{R}_1,i} \right) \right], 
\end{equation}
which is known as the \emph{split-graph Hamiltonian} corresponding to the lattice graph $\tilde{G}$~\cite{MA20}. As detailed in \cref{app:sec:bipartite_crystalline_lattices:chiral}, $\mathcal{H}_{\mathrm{split}}$, which is just a BCL Hamiltonian with chiral symmetry, will feature $N_{L} - N_{\tilde{L}}$ flat bands pinned at zero energy (provided $N_L > N_{\tilde{L}}$).

Additionally, we can also define the \emph{root-} and \emph{line-graph Hamiltonians} on the $\tilde{L}$ and $L$ sublattices in terms of the $S_{\vec{R}}$ matrix, in accordance to \cref{app:eqn:Ttilde_deff,app:eqn:T_deff}, respectively. Employing \cref{app:eqn:s_matrix_incidence}, we find
\begin{equation}
	\left[ T_{\vec{R}_2 - \vec{R}_1} \right]_{ij} = \begin{cases}
		2 t^2, & $ if $ \vec{R}_1 = \vec{R}_2$ and $i=j \\
		t^2, & $ if $ \left[\vec{R}_1,i \right] $ and $ \left[\vec{R}_2,j \right] $ are adjacent in $ \tilde G \\
		0, & $ otherwise $
	\end{cases},
\end{equation}
and
\begin{equation}
	\left[ \tilde{T}_{\vec{R}_2 - \vec{R}_1} \right]_{ij} = \begin{cases}
		n_i t^2, & $ if $ \vec{R}_1 = \vec{R}_2$ and $i=j \\
		t^2, & $ if $ \left(\vec{R}_1,i \right) $ and $ \left(\vec{R}_2,j \right) $ are connected in$ \tilde G \\
		0, & $ otherwise $
	\end{cases}.
\end{equation}
where $n_i$ denotes the degree of a vertex $\left(\vec{R},i \right)$ within the graph $\tilde{G}$ (the degree of a vertex is the number of edges adjacent to it). Two vertices (edges) are connected (adjacent) if they share an edge (vertex) in $\tilde G$. Letting $G$ be the line-graph of $\tilde{G}$, we see that the off-diagonal elements of $T_{\vec{R}}$ and $\tilde{T}_{\vec{R}}$ are equal to the off-diagonal adjacency matrix elements of $G$ and $\tilde{G}$, respectively, up to a factor of $t^2$. The diagonal terms of $T_{\vec{R}}$ and $\tilde{T}_{\vec{R}}$ represent chemical potential terms. The corresponding quadratic Hamiltonians, 
\begin{align}
	\tilde{\mathcal{T}}_{\mathrm{root}} &= \sum_{\substack{\vec{R}_1, \vec{R}_2 \\ 1 \leq i,j \leq N_L}} \left[\tilde{T}_{\vec{R}_1 - \vec{R}_2}\right]_{ij} \hat{b}^\dagger_{\vec{R}_1,i}\hat{b}_{\vec{R}_2,j}, \label{app:eqn:ham_root_graph_rs} \\ 
	\mathcal{T}_{\mathrm{line}} &= \sum_{\substack{\vec{R}_1, \vec{R}_2 \\ 1 \leq i,j \leq N_{\tilde{L}}}} \left[T_{\vec{R}_1 - \vec{R}_2}\right]_{ij} \hat{a}^\dagger_{\vec{R}_1,i}\hat{a}_{\vec{R}_2,j}, \label{app:eqn:ham_line_graph_rs}
\end{align}
are known, respectively, as the root- and line-graph Hamiltonians. Usually, a line-graph is constructed starting from a certain root-graph without reference to the split-graph or the inter-sublattice hopping matrix $S_{\vec{R}}$ from \cref{app:eqn:s_matrix_incidence}. As shown in \cref{app:sec:bipartite_crystalline_lattices:effective:def} more generally for the effective Hamiltonians $T_{\vec{k}}$ and $\tilde{T}_{\vec{k}}$, the root- and line-graph Hamiltonians have identical positive spectra, but additionally the line-graph Hamiltonian features $N_{L} - N_{\tilde{L}}$ perfectly flat bands pinned at zero energy. We exemplify the the split- and line-graph constructions in the case of a two-dimensional square lattice in \cref{app:fig:split_root_line}.

The split- and line-graph constructions are entirely included in our formalism. However, while the former are essentially limited to binary (and real) incidence matrices, as well as spinless $s$ orbitals, in a BCL, the orbital content of the two sublattices, as well as the inter-sublattice hoppings can be chosen arbitrarily. Additionally, while a simple gapless criterion can be defined for line-graphs depending on whether the root-graph is bipartite or not~\cite{MIE91}, our construction allows us to identify locally stable gapless points enforced by crystalline symmetries, as well as their Brillouin zone momentum $\vec{k}$. Our formalism also allows a clear diagnosis of the fragile topology of the flat bands (see \cref{app:sec:fb_clas:fragile,app:sec:fb_clas:relation_example}).  Preliminary signs that flat bands supported (more often than not) topological states were given in Ref.~\cite{CHI20,MA20}. Our construction provides a clear reason for this: a BR difference that gives flat bands also can create fragile topological bands! Moreover, as shown in \cref{app:sec:breaking_chiral}, the chiral symmetry of a split-graph (which is not a symmetry of real materials) can be broken without perturbing the flat bands, thus increasing the likelihood for identifying real material realizations for our construction~\cite{REG21}. 

\subsection{Breaking the chiral symmetry in a BCL}\label{app:sec:breaking_chiral}
Returning to the most general BCL Hamiltonian as defined in \cref{app:eqn:gen_bcl_ham_rs}, we here present a sufficient condition on the intra-sublattice hopping matrices $A_{\vec{R}}$ and $B_{\vec{R}}$ for the emergence of perfectly flat bands. More precisely, we will now show that if $A_{\vec{k}}$ has an eigenvalue $a$ with degeneracy $n_a$ at every momentum point (where $N_{\tilde{L}}< n_a \leq N_L$), then the BCL Hamiltonian will have at least $n_a - N_{\tilde{L}}$ perfectly flat bands at energy $a$.

To prove this statement, we start by diagonalizing the $A_{\vec{k}}$ and $B_{\vec{k}}$ hopping matrices: since $A_{\vec{k}}$ has a $\vec{k}$-independent eigenvalue $a$ with degeneracy $n_a$, there exist unitary matrices $Q_{A,\vec{k}}$ and $Q_{B,\vec{k}}$ such that
\begin{equation}
		A_{\vec{k}}= Q_{A,\vec{k}}^\dagger 
			\begin{pNiceArray}{CC}[last-col,first-row]
				n_a & N_{L} - n_a \\
				a \mathbb{1} & \mathbb{0} & n_a \\
				\mathbb{0} & D_{A,\vec{k}} & N_{L} - n_a \\
			\end{pNiceArray}		
		Q_{A,\vec{k}} \quad \text{and} \quad
		B_{\vec{k}}= Q_{B,\vec{k}}^\dagger D_{B,\vec{k}} Q_{B,\vec{k}},
\end{equation}
with $D_{A,\vec{k}}$ and $D_{B,\vec{k}}$ being diagonal square matrices. At each $\vec{k}$-point, we define a new Hamiltonian matrix $H'_{\vec{k}}$, which is related to the $H_{\vec{k}}$ by a similarity transformation
\begin{equation}
	H'_{\vec{k}} =
	\begin{pmatrix}
		Q_{A,\vec{k}} & \mathbb{0} \\
		\mathbb{0} & Q_{B,\vec{k}}
	\end{pmatrix}
	H_{\vec{k}}
	\begin{pmatrix}
		Q^\dagger_{A,\vec{k}} & \mathbb{0} \\
		\mathbb{0} & Q^\dagger_{B,\vec{k}}
	\end{pmatrix}
	= \begin{pNiceArray}{CC}[last-col,first-row]
		N_{L} & N_{\tilde{L}} \\
		Q_{A,\vec{k}} A_{\vec{k}} Q^\dagger_{A,\vec{k}} & Q_{A,\vec{k}} S_{\vec{k}} Q_{B,\vec{k}}^\dagger & N_{L}  \\
		Q_{B,\vec{k}} S^{\dagger}_{\vec{k}} Q_{A,\vec{k}}^\dagger & Q_{B,\vec{k}} B_{\vec{k}} Q_{B,\vec{k}}^\dagger & N_{\tilde{L}} \\
	\end{pNiceArray},
\end{equation}
and hence shares exactly the same spectrum. By block-decomposing the matrix $Q_{A,\vec{k}} S_{\vec{k}} Q_{B,\vec{k}}^\dagger$
\begin{equation}
	Q_{A,\vec{k}} S_{\vec{k}} Q_{B,\vec{k}}^\dagger = 
	\begin{pNiceArray}{C}[last-col,first-row]
		N_{\tilde{L}} \\
		S'_{1,\vec{k}} & n_a \\
		S'_{2,\vec{k}} & N_{L}-n_a \\
	\end{pNiceArray}
\end{equation}
we can write $H'_{\vec{k}}$ as
\begin{equation}
	H'_{\vec{k}} = \begin{pNiceArray}{CCC}[last-col,first-row]
		n_a & N_{L}-n_a & N_{\tilde{L}} \\
		a \mathbb{1} & \mathbb{0} & S'_{1,\vec{k}} & n_a\\
		\mathbb{0} & D_{A,\vec{k}} & S'_{2,\vec{k}} & N_{L}-n_a \\
		S_{1,\vec{k}}^{\prime\dagger} & S_{2,\vec{k}}^{\prime\dagger} & D_{B,\vec{k}} & N_{\tilde{L}} 
	\end{pNiceArray}.		
\end{equation}
Because $S'_{1,\vec{k}}$ is an $n_a \times N_{\tilde{L}}$ matrix (and $n_a - N_{\tilde{L}} > 0$), its rank must be equal or smaller than $N_{\tilde{L}}$, and so there must exist at least $n_a-N_{\tilde{L}}$ linearly independent solutions $\phi_{1,\vec{k},\alpha}$ ($1 \leq \alpha \leq n_a-N_{\tilde{L}}$) to the equation $S^{\dagger\prime}_{1,\vec{k}} \phi_{1,\vec{k},\alpha} = 0$. This implies that 
\begin{equation}
	H'_{\vec{k}} \begin{pmatrix}
		\phi_{1,\vec{k},\alpha} \\		
		\mathbb{0} \\
		\mathbb{0}
	\end{pmatrix} = a \begin{pNiceArray}{C}[last-col]
		\phi_{1,\vec{k},\alpha} & n_a\\
		\mathbb{0} & N_{L}-n_a \\
		\mathbb{0} & N_{\tilde{L}}
	\end{pNiceArray}
\end{equation}
for every value of $\vec{k}$, so that $H'_{\vec{k}}$ and correspondingly $H_{\vec{k}}$ have $n_a-N_{\tilde{L}}$ exactly flat bands at energy $a$. 

We see that a sufficient condition for the existence of perfectly flat bands in a generalized BCL as defined in \cref{app:eqn:bip_ham_gen} is for the $A_{\vec{k}}$ matrix to have $n_a$ degenerate $\vec{k}$-independent eigenvalues (with $n_a > N_{\tilde{L}}$). At the same time, remarkably, there is no requirement on the intra-sublattice hoppings within the smaller sublattice $\tilde{L}$. 

The simplest case in which $A_{\vec{k}}$ features degenerate $\vec{k}$-independent eigenvalues is to have $A_{\vec{k}} = a \mathbb{1}$, implying that $n_a = L$. The corresponding BCL Hamiltonian from \cref{app:eqn:bip_ham_gen} reads
\begin{equation}
	\label{app:eqn:bip_ham_gen_with_id}
	H_{\vec{k}} = \begin{pNiceArray}{CC}[last-col,first-row]
		N_{L} & N_{\tilde{L}} \\
    	a \mathbb{1} & S_\vec{k} & N_{L}  \\
		S_\vec{k}^\dagger & B_{\vec{k}} & N_{\tilde{L}} \\
	\end{pNiceArray},
\end{equation}
and has $n_a - N_{\tilde{L}} = N_{L} - N_{\tilde{L}}$ flat bands. Compared to the chiral case discussed in \cref{app:sec:bipartite_crystalline_lattices:chiral}, the presence of intra-sublattice hopping inside the $\tilde{L}$ sublattice does not perturb the flat band and potential band touching point eigenstates from \cref{app:eqn:eig_zeroPlus}, whose energy is merely shifted from zero to $a$.  

In what follows, we will identify and discuss three other (physical) possibilities in which $A_{\vec{k}}$ features degenerate $\vec{k}$-independent eigenvalues of the appropriate multiplicity for $H_{\vec{k}}$ as defined in \cref{app:eqn:bip_ham_gen} to have perfectly flat bands. We will then show that each of them reduces to the BCL Hamiltonian from \cref{app:eqn:bip_ham_gen_with_id}, by arguing that an equivalent generalized BCL Hamiltonian of the form in \cref{app:eqn:bip_ham_gen_with_id} featuring identical flat band and potential band touching point wave functions can always be constructed. 

\begin{enumerate}
	\item\label{app:enum:BCL_atomic} One simple case is to have the orbitals in $L$ only contain onsite (but not necessarily identical) potential terms. The BCL matrix can then be written in block form as 
	\begin{equation}
		\label{app:eqn:gen_bcl_atomic}
		H_{\vec{k}} = \begin{pNiceArray}{CCCCC}[last-col,first-row]
		N_{L_1}	& N_{L_2} & \cdots & N_{L_m} & N_{\tilde{L}} \\
		\epsilon_1 \mathbb{1} & \mathbb{0} & \cdots & \mathbb{0} & S_{1,\vec{k}} & N_{L_1} \\
		\mathbb{0} & \epsilon_2 \mathbb{1} & \cdots & \mathbb{0} & S_{2,\vec{k}} & N_{L_2} \\
		\vdots & \vdots & \ddots & \vdots & \vdots & \vdots \\
		\mathbb{0} & \mathbb{0} & \cdots &  \epsilon_m \mathbb{1} & S_{m,\vec{k}} & N_{L_m} \\
		S^\dagger_{1,\vec{k}} & S^\dagger_{2,\vec{k}} & \cdots & S^\dagger_{m,\vec{k}} & B_{\vec{k}} & N_{\tilde{L}} \\
	\end{pNiceArray},
	\end{equation}
	where $L=\bigoplus_{i=1}^{m} L_{i}$ is a partitioning of the sublattice $L$ according to the onsite energies $\epsilon_i$. For each sublattice $L_i$ with $N_{L_i}>N_{\tilde{L}}$, there will generically be $N_{L_i}-N_{\tilde{L}}$ flat bands pinned at energy $\epsilon_i$. The corresponding flat band eigenmodes, as well as the eigenmodes of the potential band touching points depend exclusively on the kernel of the matrix $S^{\dagger}_{i,\vec{k}}$, as argued in \cref{app:eqn:eig_zeroPlus}, and can be understood in terms of a simplified generalized BCL Hamiltonian defined only on the $L_i \oplus \tilde{L}$ lattice, namely
	\begin{equation}
		\label{app:eqn:gen_bcl_identity}
	    	H^i_{\vec{k}} = \begin{pNiceArray}{CC}[last-col,first-row]
	    	N_{L_i} & N_{\tilde{L}} \\
	    	\epsilon_i \mathbb{1} & S_{i,\vec{k}} & N_{L_i} \\
	    	S_{i,\vec{k}}^\dagger & B_{\vec{k}} & N_{\tilde{L}} \\
	    \end{pNiceArray},
	\end{equation} 
	which is of the same form as \cref{app:eqn:bip_ham_gen_with_id}.
	
	\item\label{app:enum:BCL_molecular} Another way in which the Hamiltonian $A_{\vec{k}}$ can have $\vec{k}$-independent degenerate eigenvalues is for the intra-sublattice hopping terms in $L$ to lead to the formation of uncoupled degenerate \emph{molecular} orbitals. Such an example was constructed recently by Ref.~\cite{QI20}. We define a connectivity graph for the orbitals in $L$, such that each orbital corresponds to a node and two nodes are connected if and only if there is hopping between them (as described by $A_{\vec{k}}$). This possibility arises if the connectivity graph has an extensive number of disconnected components. Molecular orbitals with finite support can be formed by hybridizing the original atomic orbitals within each disconnected component. From a symmetry standpoint, there is no difference between these molecular orbitals with finite support and atomic orbitals. Thus this possibility is entirely contained in the \cref{app:enum:BCL_atomic}, and therefore also reduces to \cref{app:eqn:bip_ham_gen_with_id}. We illustrate this case with an example in \cref{app:fig:Lieb_molecular} of \cref{app:sec:example:noChiral}.
	\item\label{app:enum:nested_doll} Finally, $A_\vec{k}$ can \emph{itself} be a Hamiltonian with $n_a$ perfectly flat bands pinned at energy $a$ (that are neither atomic nor molecular orbitals, but arise from wave function interference). Crucially, however, because $N_{\tilde{L}} \geq 1$, the condition $n_a>N_{L}$ can only be achieved if $n_a \geq 2$. In the case of systems with spinful time-reversal symmetry, $N_{\tilde{L}} \geq 2$ and so $n_a \geq 3$ is required for the existence of flat bands. To our knowledge, the only crystalline models reported in literature featuring degenerate flat bands are either (a) described by a BCL Hamiltonian $H_{\vec{k}}$, or (b) described by an effective Hamiltonian $T_{\vec{k}}$ obtained from a BCL Hamiltonian. We substantiate this claim in \cref{app:sec:example:AZ_fb}, where we consider the possibility of flat band protection in the ten Altland–Zirnbauer classes with inversion symmetry~\cite{ZIR96, ALT97a, KIT09a, SCH08, QI08, RYU10, CHI13, MOR13, SHI14, CHI16}. We find that only BCL Hamiltonians can host flat bands with a degeneracy higher than one (or two, for systems with spinful time-reversal symmetry)~\footnote{More precisely, we find that inversion ($\mathcal{I}$) combined with particle-hole ($\mathcal{P}$) symmetry can also protect perfectly flat bands. However, while BCLs can host any number of (possibly degenerate) flat bands, $\mathcal{IP}$ can only protect a single (Kramers degenerate) flat band in spinless (spinful with time-reversal symmetry) systems.}. As such, we can restrict ourselves to the two cases (a) and (b), which will be discussed below:
	
	\begin{enumerate}
	    \item\label{app:enum:nested_doll_sMat} If $A_{\vec{k}}$ is a BCL Hamiltonian, then we can also reduce $H_{\vec{k}}$ to the form of \cref{app:eqn:bip_ham_gen_with_id} through a redefinition of the two sublattices. To see this, assume that $A_{\vec{k}}$ is a BCL Hamiltonian defined on two sublattices $M$ and $\tilde{M}$ (with $M \oplus \tilde{M} = L$)
        \begin{equation}
	        A_{\vec{k}} = \begin{pNiceArray}{CC}[last-col,first-row]
		        N_{M} & N_{\tilde{M}} \\
		        a \mathbb{1} & \overline{S}_\vec{k} & N_{M}  \\
		        \overline{S}^{\dagger}_\vec{k} & \overline{B}_{\vec{k}} & N_{\tilde{M}} \\
		        \end{pNiceArray},
        \end{equation}
        where $N_{M}-N_{\tilde{M}}=n_a$, $N_{\tilde{M}}+N_{M}=N_{L}$, and $\overline{B}_{\vec{k}}$ being a Hermitian matrix denoting the intra-sublattice hopping inside sublattice $\tilde{M}$. Writing the original $S_{\vec{k}}$ matrix from \cref{app:eqn:bip_ham_gen} in block form
        \begin{equation}
        	S_{\vec{k}} = \begin{pNiceArray}{C}[last-col,first-row]
	        	N_{\tilde{L}} \\
	        	S_{1,\vec{k}} & N_{M} \\
		        S_{2,\vec{k}} & N_{\tilde{M}} \\
        	\end{pNiceArray}
        \end{equation}
        one can express the original BCL Hamiltonian from \cref{app:eqn:bip_ham_gen} as 
        \begin{equation}
	        H_{\vec{k}} = \begin{pNiceArray}{CCC}[last-col,first-row]
		        N_{M} & N_{\tilde{M}} & N_{\tilde{L}}\\
	        	a \mathbb{1} & \overline{S}_{\vec{k}} & S_{1,\vec{k}} & N_{M} \\
        		\overline{S}^{\dagger}_{\vec{k}} & \overline{B}_{\vec{k}} & S_{2,\vec{k}} & N_{\tilde{M}} \\
		        S^{\dagger}_{1,\vec{k}} & S^{\dagger}_{2,\vec{k}} & B_{\vec{k}} & N_{\tilde{L}}
        	\end{pNiceArray}.		
        \end{equation}

After defining $L' \equiv M$, $\tilde{L}' \equiv \tilde{M} \oplus \tilde{L}$, as well as 
\begin{equation}
	S'_{\vec{k}} \equiv \begin{pNiceArray}{CC}[last-col,first-row]
		N_{\tilde{M}} & N_{\tilde{L}} \\
		\overline{S}_\vec{k} & S_{1,\vec{k}} & N_{M}
	\end{pNiceArray} \quad \text{and} \quad
	B'_{\vec{k}} \equiv \begin{pNiceArray}{CC}[last-col,first-row]
		N_{\tilde{M}} & N_{\tilde{L}}\\
		\overline{B}_{\vec{k}} & S_{2,\vec{k}} & N_{\tilde{M}} \\
		S^{\dagger}_{2,\vec{k}} & B_{\vec{k}} & N_{\tilde{L}} 
	\end{pNiceArray}
\end{equation}
the Hamiltonian $H_{\vec{k}}$ can be rewritten as
\begin{equation}
	\label{app:eqn:gen_bcl_identity_general}
	H_{\vec{k}} = \begin{pNiceArray}{CC}[last-col,first-row]
		N_{L} & N_{\tilde{L}} \\
		a \mathbb{1} & S'_{\vec{k}} & N_{L}  \\
		S^{\prime \dagger}_{\vec{k}} & B'_{\vec{k}} & N_{\tilde{L}} \\
	\end{pNiceArray},
\end{equation} 
thus reducing this case to \cref{app:eqn:bip_ham_gen_with_id}.
		\item\label{app:enum:nested_doll_tMat} The other possibility is for $A_{\vec{k}}$ to be an effective BCL Hamiltonian as defined in \cref{app:sec:bipartite_crystalline_lattices:effective:def} (up to rescaling and chemical potential terms)
		\begin{equation}
		    \label{app:eqn:gen_bcl_with_eff_A}
			A_{\vec{k}} = a \mathbb{1} + \overline{T}_{\vec{k}}.
		\end{equation}
		In \cref{app:eqn:gen_bcl_with_eff_A}, $\overline{T}_{\vec{k}} = \overline{S}_{\vec{k}} \overline{S}_{\vec{k}}^{\dagger}$ is an effective Hamiltonian corresponding to a certain $M \oplus \tilde{M}$ BCL, where $M = L$ and $N_{M} - N_{\tilde{M}} = n_a$. As required, $A_{\vec{k}}$ has a $\vec{k}$-independent degenerate eigenvalue $a$ with multiplicity $n_{a}$. The eigenstates of $H_{\vec{k}}$ [as defined in \cref{app:eqn:bip_ham_gen}] corresponding to its flat bands and potential band touching points between its flat and dispersive bands have energy $a$ and a wave function of the form 
		\begin{equation}
		    \label{app:eqn:gen_bcl_with_eff_orig_fb}
			\Psi_{\vec{k}} = \begin{pNiceArray}{C}[last-col]
				\phi_{\vec{k}} & N_{L}  \\
				\mathbb{0} & N_{\tilde{L}} \\
			\end{pNiceArray},
		\end{equation}
		where $\phi_{\vec{k}}$ obeys 
		\begin{align}
			A_{\vec{k}} \phi_{\vec{k}} &= a \phi_{\vec{k}} \label{app:eqn:gen_bcl_with_eff_A_eig_phi_A} \\ 
			S^{\dagger}_{\vec{k}} \phi_{\vec{k}} &= \mathbb{0}. \label{app:eqn:gen_bcl_with_eff_A_ker_S}
		\end{align}
		\Cref{app:eqn:gen_bcl_with_eff_A,app:eqn:gen_bcl_with_eff_A_eig_phi_A}, together with the condition that $\overline{T}_{\vec{k}}$ is a positive semidefinite matrix imply that $\phi_{\vec{k}}$ is in the kernel of $\overline{S}_{\vec{k}}^{\dagger}$, \ie
		\begin{equation}
		    \label{app:eqn:gen_bcl_with_eff_A_ker_Soverline}
			\overline{S}_{\vec{k}}^{\dagger} \phi_{\vec{k}} = \mathbb{0}.
		\end{equation}
		Armed with this information, one can construct a BCL Hamiltonian $\overline{H}_{\vec{k}}$ on the $L' \oplus \tilde{L}'$ BCL, where $L' \equiv L$ and $\tilde{L}' = \tilde{M} \oplus \tilde{L} $, whose block structure is given by
		\begin{equation}
			\overline{H}_{\vec{k}} = \begin{pNiceArray}{CCC}[last-col,first-row]
				N_{L} & N_{\tilde{M}} & N_{\tilde{L}} \\
				a \mathbb{1} & \overline{S}_{\vec{k}} & S_{\vec{k}} & N_{L}\\
				\overline{S}^{\dagger}_{\vec{k}} & \mathbb{0} & \mathbb{0} & N_{\tilde{M}} \\
				S^{\dagger}_{\vec{k}} & \mathbb{0} & B_{\vec{k}} & N_{\tilde{L}}
			\end{pNiceArray}.
		\end{equation}
		For any $\phi_{\vec{k}}$ obeying \cref{app:eqn:gen_bcl_with_eff_A_eig_phi_A,app:eqn:gen_bcl_with_eff_A_ker_S}, we can define
		\begin{equation}
		   \label{app:eqn:gen_bcl_with_eff_eng_fb}
			\overline{\Psi}_{\vec{k}} = \begin{pNiceArray}{C}[last-col]
				\phi_{\vec{k}} & N_{L}  \\
				\mathbb{0} & N_{\tilde{M}} \\
				\mathbb{0} & N_{\tilde{L}} \\
			\end{pNiceArray},
		\end{equation}
		which, as a consequence of \cref{app:eqn:gen_bcl_with_eff_A_ker_S,app:eqn:gen_bcl_with_eff_A_ker_Soverline}, satisfies $\overline{H}_{\vec{k}} \overline{\Psi}_{\vec{k}} = a\overline{\Psi}_{\vec{k}}$. Otherwise stated, for any flat band and corresponding band touching point eigenstate of $H_{\vec{k}}$ with energy $a$ and wave function given in \cref{app:eqn:gen_bcl_with_eff_orig_fb,}, $\overline{H}_{\vec{k}}$ will have an eigenstate at energy $a$ with an identical wave function, as seen in \cref{app:eqn:gen_bcl_with_eff_eng_fb}. Since $\overline{H}_{\vec{k}}$ is manifestly of the form in \cref{app:eqn:bip_ham_gen_with_id}, we conclude that this case also reduces to \cref{app:eqn:bip_ham_gen_with_id}. 
	\end{enumerate}
\end{enumerate}

Because all cases considered above reduce to the \cref{app:eqn:bip_ham_gen_with_id}, we can restrict ourselves to generalized BCL Hamiltonians which have onsite terms in the larger sublattice $A_{\vec{k}} = a \mathbb{1}$ and arbitrary hopping in the smaller sublattice. Compared to the chiral BCL Hamiltonian from \cref{app:eqn:bip_ham_chiral}, in a generalized BCL, the presence of inter-sublattice hopping in the smaller sublattice $\tilde{L}$ gaps the accidental band touching points from \cref{app:eqn:eig_zeroMinus}. On the other hand, the flat band eigenstates as well as the eigenstates of the corresponding stable band touching points from \cref{app:eqn:eig_zeroPlus} remain unperturbed in a generalized BCL, up to a shift in energy. Because they are only dependent on the kernel of $S^\dagger_{\vec{k}}$, for the purpose of our classification of flat bands, we can consider only the chiral BCL Hamiltonians from \cref{app:eqn:bip_ham_chiral}, without loss of generality.  

\subsection{Symmetries in a BCL}\label{app:sec:unitary_sym}

In the main text, we have proved the flat band subtraction rule
\begin{equation}
	\label{app:eqn:br_subtraction}
	\mathcal{B}_{\mathrm{FB}} = \mathcal{BR}_{L} \boxminus \mathcal{BR}_{\tilde{L}}.
\end{equation} 
invoking the fact that the non-zero energy eigenstates of the two effective Hamiltonians $T_{\vec{k}}$ and $\tilde{T}_{\vec{k}}$ defined in \cref{app:sec:bipartite_crystalline_lattices:effective} not only share the same eigenvalues, but also form carrier spaces of identical (co)irreps of the group $\mathcal{G}$. Additionally, $T_{\vec{k}}$ contains the flat band modes, as well as the corresponding (stable) band touching points between dispersive bands and flat bands of $H_{\vec{k}}$ from \cref{app:eqn:eig_zeroPlus}, whereas $\tilde{T}_{\vec{k}}$ does not. We concluded that the (co)irreps of the perfectly flat bands and their band touching points were given by a formal difference between the BRs induced from the orbitals of $L$ and those induced from the orbitals of $\tilde{L}$, as shown in \cref{app:eqn:br_subtraction}. In this appendix, we detail the implementation of crystalline symmetries in a BCL and formulate an alternative proof which does not rely on the introduction of effective sublattice Hamiltonians.

We start from the chiral BCL Hamiltonian given by $\mathcal{H}=\sum_{\vec{k}} \hat{\Psi}^\dagger_{\vec{k}} H_{\vec{k}} \hat{\Psi}_{\vec{k}}$, where 
\begin{equation}
	\label{app:eqn:bip_ham_chiral_sym}
	H_{\vec{k}} = \begin{pNiceArray}{CC}[last-col,first-row]
		N_{L} & N_{\tilde{L}} \\
		\mathbb{0} & S_\vec{k} & N_{L}  \\
		S_\vec{k}^\dagger & \mathbb{0} & N_{\tilde{L}} \\
	\end{pNiceArray}
\end{equation}
and $\hat{\Psi}_{\vec{k}}^T = \left( \hat{a}_{\vec{k},1}, \dots, \hat{a}_{\vec{k},N_L}, \hat{b}_{\vec{k},1}, \dots, \hat{b}_{\vec{k},N_{\tilde{L}}} \right)$, as defined in \cref{app:eqn:bip_ham_chiral}. As discussed in \cref{app:sec:bipartite_crystalline_lattices:chiral,app:sec:breaking_chiral}, the wave functions of the flat band and corresponding stable band touching points are identical in chiral and generalized BCLs. We assume that $\mathcal{H}$ is invariant under a certain Shubnikov Space Group (SSG) $\mathcal{G}$. Moreover, since the partitioning of the BCL into sublattices $L$ and $\tilde{L}$ respects the symmetries of $\mathcal{G}$, it follows that the transformations of $\mathcal{G}$ do not mix degrees of freedom belonging to different sublattices. At a given high-symmetry momentum point $\vec{K}$, for any (unitary or anti-unitary) symmetry operation in the little group of $\vec{K}$, $g \in \mathcal{G}_{\vec{K}}$, there exist unitary matrices $U(g)$ and $\tilde{U}(g)$, respectively implementing the transformation $g$ in the $L$ and $\tilde{L}$ sublattices
\begin{equation}
	\label{app:eqn:definition:U}
	\begin{split}
		\hat{g} \hat{a}^\dagger_{i,\vec{K}} \hat{g}^{-1} &= \sum_{j=1}^{N_{L}} U_{ji}(g)  \hat{a}^\dagger_{j,\vec{K}}, \\
		\hat{g} \hat{b}^\dagger_{i,\vec{K}} \hat{g}^{-1} &= \sum_{j=1}^{N_{\tilde{L}}} \tilde{U}_{ji}(g)  \hat{b}^\dagger_{j,\vec{K}}.
	\end{split}
\end{equation}
In general, the same symmetry transformation $g$ can have different representation matrices depending on the momentum $\vec{K}$, meaning that $U(g)$ and $\tilde{U}(g)$ implicitly depend on $\vec{K}$. As $g$ is a symmetry of the BCL Hamiltonian (meaning that $\left[g, \mathcal{H} \right]=0$), it follows that 
\begin{equation}
	\label{app:eqn:invarianceBCL_Symmetry}
	\mathcal{U}(g) H^{(*)}_{\vec{K}} \mathcal{U}^\dagger(g) = H_{\vec{K}},
\end{equation}
where we have defined the unitary symmetry representation matrix for the BCL to be
\begin{equation}
	\label{app:eqn:definition:bigU}
	\mathcal{U}(g) = \begin{pNiceArray}{CC}[last-col,first-row]
		N_{L} & N_{\tilde{L}} \\
		U(g) & \mathbb{0} & N_{L}  \\
		\mathbb{0} & \tilde{U}(g) & N_{\tilde{L}} \\
	\end{pNiceArray}.
\end{equation}
In \cref{app:eqn:invarianceBCL_Symmetry} and in the rest of the appendix, the asterisk ${}^{(*)}$ indicates that an additional complex conjugation is to be performed if $g$ is an anti-unitary symmetry transformation (\ie a unitary transformation combined with time reversal $\mathcal{T}$).

\begin{table*}[!t]
\begin{tabular}{l | c | c | c | c | c} 
	Hamiltonian & Energy & Wave Function & Index & (Co)Representation & Flat Band  \\ 
\hline \hline 
	\multirow{2}{*}{$H_{\vec{K}} = \begin{pmatrix}
		\mathbb{0} & S_\vec{K} \\
		S_\vec{K}^\dagger & \mathbb{0} \\
	\end{pmatrix}$} & 
	\multirow{2}{*}{$0$} & 
	$\Psi_{\vec{K},\alpha}^{+} = \begin{pmatrix}
			\phi_{\vec{K},\alpha} \\
			\mathbb{0} \\
		\end{pmatrix}$ & $ r_{\vec{K}} < \alpha \leq N_{L}$ & $\mathcal{R}_{\vec{K}}^{+,0}$ & \cmark \\
	\cline{3-6}
	& & 
	$\Psi_{\vec{K},\alpha}^{-} = \begin{pmatrix}
			\mathbb{0} \\
			\psi_{\vec{K},\alpha} \\
		\end{pmatrix}$ & $r_{\vec{K}} < \alpha \leq N_{\tilde{L}}$ & $\mathcal{R}_{\vec{K}}^{-,0}$ & \xmark \\
	\hline
	\multirow{2}{*}{$H_{\vec{K}} = \begin{pmatrix}
		\mu_{L}\mathbb{1} & S_\vec{K} \\
		S_\vec{K}^\dagger & \mu_{\tilde{L}}\mathbb{1} \\
	\end{pmatrix}$} & 
	$\mu_L$ & 
	$\Psi_{\vec{K},\alpha}^{+} = \begin{pmatrix}
			\phi_{\vec{K},\alpha} \\
			\mathbb{0} \\
		\end{pmatrix}$ & $r_{\vec{K}} < \alpha \leq N_{L}$ & $\mathcal{R}_{\vec{K}}^{+,0}$ & \cmark \\
	\cline{2-6}
	& $\mu_{\tilde{L}}$ & 
	$\Psi_{\vec{K},\alpha}^{-} = \begin{pmatrix}
			\mathbb{0} \\
			\psi_{\vec{K},\alpha} \\
		\end{pmatrix} $ & $r_{\vec{K}} < \alpha \leq N_{\tilde{L}}$ & $\mathcal{R}_{\vec{K}}^{-,0}$ & \xmark \\
	\hline	
	$H_{\vec{K}} = \begin{pmatrix}
		a\mathbb{1} & S_\vec{K} \\
		S_\vec{K}^\dagger & B_{\vec{K}} \\
	\end{pmatrix}$ & 
	$a$ & 
	$\Psi_{\vec{K},\alpha}^{+} = \begin{pmatrix}
			\phi_{\vec{K},\alpha} \\
			\mathbb{0} \\
		\end{pmatrix}$ & $r_{\vec{K}} < \alpha \leq N_{L}$ & $\mathcal{R}_{\vec{K}}^{+,0}$ & \cmark \\
	\hline	
	$T_{\vec{K}}$ & 
	$ 0 $&
	$\phi_{\vec{K},\alpha}$ & $r_{\vec{K}} < \alpha \leq N_{L}$ & $\mathcal{R}_{\vec{K}}^{+,0}$ & \cmark \\
	\hline		
	$\tilde{T}_{\vec{K}}$ & 
	$ 0 $ &
	$\psi_{\vec{K},\alpha}$ & $r_{\vec{K}} < \alpha \leq N_{\tilde{L}}$ & $\mathcal{R}_{\vec{K}}^{-,0}$ & \xmark \\  
\end{tabular} 
\caption{Interpreting formal (co)representation differences. We consider the $L \oplus \tilde{L}$ (with $N_{L} > N_{\tilde{L}}$) BCL Hamiltonians $H_{\vec{k}}$ either in the chiral case (first row), or with sublattice chemical potential terms (second row), or with intra-sublattice hopping in $\tilde{L}$ (third row), as well as the two effective Hamiltonians $T_{\vec{k}}$ and $\tilde{T}_{\vec{k}}$. We assume that the BCL flat bands are assigned a formal (co)representation difference at $\vec{K}$ given by $\mathcal{R}_{\vec{K}}^{+,0} \boxminus \mathcal{R}_{\vec{K}}^{-,0}$, which indicates the presence of a band touching point. In the chiral case, the flat band and corresponding band touching point wave functions carry the (co)representation $\mathcal{R}_{\vec{K}}^{+,0} \oplus \mathcal{R}_{\vec{K}}^{-,0}$. In a generalized BCL, the flat bands and corresponding band touching points carry a (co)representation $\mathcal{R}_{\vec{K}}^{+,0}$, which is shared by the flat bands and band touching points of $T_{\vec{k}}$. Additionally, the effective Hamiltonian $\tilde{T}_{\vec{k}}$ has zero modes which carry the (co)representation $\mathcal{R}_{\vec{K}}^{-,0}$.}
\label{app:tab:relation_between_flatbands}
\end{table*}

Our goal will be to show that at $\vec{K}$, the eigenspace of $H_{\vec{K}}$ corresponding to states with strictly positive energies transforms according to the same (co)representation as the eigenspace with strictly negative ones. For a given singular value of $S_{\vec{K}}$, $\mathcal{E} \geq 0$, and $n=\pm$, we follow the notation defined in \cref{app:eqn:eig_dec_H} and denote the energy eigenstates of $H_{\vec{K}}$ by $\Psi^n_{\vec{K},\alpha}$, where $\alpha$ indexes all the (possibly degenerate) states having energy $\mathcal{E}^{n}_{\vec{K},\alpha} = n \mathcal{E}$. Because $g$ is a symmetry of $\mathcal{H}$, for every eigenstate $\Psi^{n}_{\vec{K},\alpha}$ of $H_{\vec{K}}$ with energy $n \mathcal{E}$, $\mathcal{U}(g)\Psi^{n(*)}_{\vec{K},\alpha}$ will also be an eigenstate of $H_{\vec{K}}$ with the same energy $n \mathcal{E}$. This implies that 
 \begin{equation}
	\label{app:eqn:action_g_BCL_eigs}
	\mathcal{U}(g) \Psi^{n(*)}_{\vec{K},\alpha} = \sum_{\beta} \left[\mathcal{R}^{n,\mathcal{E}}_{\vec{K}}(g)\right]_{\alpha \beta} \Psi^{n}_{\vec{K},\beta},
\end{equation}
where the matrix
\begin{equation}
	\label{app:eqn:sewing_definition}
	\left[\mathcal{R}^{n,\mathcal{E}}_{\vec{K}}(g)\right]_{\alpha \beta} = \Psi^{n \dagger }_{\vec{K},\beta}	\mathcal{U}(g) \Psi^{n(*)}_{\vec{K},\alpha}
\end{equation}
is known as the sewing matrix~\footnote{Usually, the sewing matrices are denoted by $\left[B_{\vec{K}}(g)\right]_{\alpha \beta}$. Throughout this work, we will employ the notation $\left[\mathcal{R}_{\vec{K}}(g)\right]_{\alpha \beta}$ for sewing matrices, to avoid confusing them with the intrasublattice hopping matrix $B_{\vec{k}}$ defined in \cref{app:sec:bipartite_crystalline_lattices:notation}, or with the (augmented) symmetry data vector $B$ ($\overline{B}$), which will be introduced in \cref{app:sec:fb_clas}.} of the symmetry transformation $g$ for the eigenstates of $H_{\vec{K}}$ at energy $n\mathcal{E}$. Note that if $\mathcal{E}>0$, the symmetry transformation cannot change the index $n$, as that would map positive and negative energy states into one another. Additionally, $\left[\mathcal{R}^{n,\mathcal{E}}_{\vec{K}}(g)\right]_{\alpha \beta}$ only has support for $1 \leq \alpha,\beta \leq r_{\vec{K}}$ for which $\epsilon_{\vec{K},\alpha} = \epsilon_{\vec{K},\beta} = \mathcal{E}$. For ${\mathcal E} = 0$, the index $n$ no longer corresponds to positive or negative energies, as all the states $\Psi^{n}_{\vec{K},\alpha}$ with $\mathcal{E}^{n}_{\vec{k},\alpha} = n \mathcal{E}$ have zero energy. Instead, as defined in \cref{app:eqn:eig_zeroPlus,app:eqn:eig_zeroMinus}, $n$ specifies the sublattice on which the state $\Psi^{n}_{\vec{K},\alpha}$ is supported: $n=+$ ($n=-$), if the state $\Psi^{n}_{\vec{K},\alpha}$ is supported on sublattice $L$ ($\tilde{L}$). As the symmetry $g$ does not mix degrees of freedom belonging to different sublattices, $\mathcal{U} (g)$ still cannot change the index $n$ even when $\mathcal{E}=0$.	 Additionally, the matrix $\left[\mathcal{R}_{\vec{K}}^{+,0} (g) \right]_{\alpha \beta}$ is only supported for $r_{\vec{k}} < \alpha,\beta \leq N_L$, while the matrix $\left[ \mathcal{R}_{\vec{K}}^{-,0} (g) \right]_{\alpha \beta} $ only has support for $r_{\vec{k}} < \alpha, \beta \leq N_{\tilde{L}}$.

Finally, we note that when restricted on the indices on which they are supported, the sewing matrices in \cref{app:eqn:sewing_definition} are unitary
\begin{align}
	\sum_{\alpha} \left[ \mathcal{R}_{\vec{K}}^{n,\mathcal{E}} (g) \right]_{\alpha \beta}  \left[\mathcal{R}^{n,\mathcal{E}*}_{\vec{K}} (g) \right]_{\alpha \gamma} 
	&=  \sum_{\alpha} \Psi^{n \dagger }_{\vec{K},\beta}	\mathcal{U}(g) \Psi^{n(*)}_{\vec{K},\alpha}  \Psi^{n T }_{\vec{K},\gamma}	\mathcal{U}^{*}(g) \left[ \Psi^{n(*)}_{\vec{K},\alpha} \right]^* \nonumber \\
	&=  \sum_{\alpha} \Psi^{n \dagger }_{\vec{K},\beta}	\mathcal{U}(g) \Psi^{n(*)}_{\vec{K},\alpha} \left[ \Psi^{n(*)}_{\vec{K},\alpha} \right]^{\dagger} \mathcal{U}^{\dagger}(g) \Psi^{n}_{\vec{K},\gamma} \nonumber \\
	&=  \Psi^{n \dagger }_{\vec{K},\beta}	\mathcal{U}(g) \mathcal{U}^{\dagger}(g) \Psi^{n}_{\vec{K},\gamma} = \delta_{\beta \gamma}. \label{app:eqn:sewing_unitary}	
\end{align}

\Cref{app:eqn:action_g_BCL_eigs} implies that each degenerate eigenspace of $H_{\vec{K}}$ forms a carrier space for a (co)representation of the little group $\mathcal{G}_{\vec{K}}$. In an abuse of notation, we will denote the (co)representation carried by the states in a degenerate eigenspace of $H_{\vec{K}}$ by the set of sewing matrices corresponding to the eigenstates. For example, the (co)representation carried by the BCL eigenstates $\Psi^{n}_{\vec{K},\alpha}$ having energy $n \mathcal{E}$ will be denoted by $\mathcal{R}^{n,\mathcal{E}}_{\vec{K}}$. An important property (which will be derived in \cref{app:sec:gapless}) is that, generically, $\mathcal{R}^{+,0}_{\vec{K}}$ and $\mathcal{R}^{-,0}_{\vec{K}}$ contain no common (co)irreps of $\mathcal{G}_{\vec{K}}$, \ie
\begin{equation}
	\label{app:eqn:no_common_irreps}
	\mathcal{R}^{+,0}_{\vec{K}} \cap \mathcal{R}^{-,0}_{\vec{K}} = \emptyset.
\end{equation}

If the matrix $S_{\vec{K}}$ is full-rank then the flat bands of $\mathcal{H}$ are gapped, and carry the (co)representation $\mathcal{R}_{\vec{K}}^{+,0}$ at $\vec{K}$. On the other hand, if the matrix $S_{\vec{K}}$ is not full-rank, there are band touching points between the flat and dispersive bands of $\mathcal{H}$. The flat band and corresponding band touching point wave functions will carry the (co)representation $\mathcal{R}_{\vec{K}}^{+,0} \oplus \mathcal{R}_{\vec{K}}^{-,0}$. However, as outlined in \cref{app:tab:relation_between_flatbands}, upon breaking the chiral symmetry of $\mathcal{H}$ (which is not a symmetry of crystalline materials) and obtaining a generalized BCL Hamiltonian, the flat bands and corresponding band touching points will carry just the (co)representation $\mathcal{R}_{\vec{K}}^{+,0}$ at $\vec{K}$. As argued in the main text, however, we will assign the formal (co)representation difference $\mathcal{R}_{\vec{K}}^{+,0} \boxminus \mathcal{R}_{\vec{K}}^{-,0}$ to the BCL flat bands at $\vec{K}$, as this provides a more complete description of the band touching point: while $\mathcal{R}_{\vec{K}}^{+,0}$ faithfully describes the flat band and corresponding band touching points at $\vec{K}$, the full formal difference is required in the vicinity of $\vec{K}$, where the (co)representations get subduced (and the flat band might gap from the dispersive bands). Moreover, the (co)representation $\mathcal{R}_{\vec{K}}^{-,0}$ gives information about the degrees of freedom that should be added in sublattice $L$ in order to produce gapped flat bands at $\vec{K}$ [\ie the orbitals that induce the (co)representation $\mathcal{R}_{\vec{K}}^{-,0}$ at $\vec{K}$].

To obtain an expression for the formal (co)representation difference $\mathcal{R}_{\vec{K}}^{+,0} \boxminus \mathcal{R}_{\vec{K}}^{-,0}$ (where $\mathcal{R}_{\vec{K}}^{-,0} = \emptyset$ if $S_{\vec{K}}$ is full-rank), we note that \cref{app:eqn:eig_dispersive,app:eqn:definition:bigU,app:eqn:action_g_BCL_eigs} imply that for any singular value of $S_{\vec{K}}$, $\mathcal{E} > 0$, we must have
\begin{equation}
	\label{app:eqn:same_sewing}
	\begin{pmatrix}
		U(g) & \mathbb{0} \\
		\mathbb{0} & \tilde{U}(g)
	\end{pmatrix}
	\begin{pmatrix}
		\pm \phi_{\vec{K},\alpha} \\
		\psi_{\vec{K},\alpha} \\
	\end{pmatrix}^{(*)}
	=\sum_{\beta} 
	\left[\mathcal{R}^{\pm,\mathcal{E}}_{\vec{K}} (g) \right]_{\alpha \beta}
	\begin{pNiceArray}{C}[last-col]
		\pm \phi_{\vec{K},\beta} & N_{L} \\
		\psi_{\vec{K},\beta} & N_{\tilde{L}}  \\
	\end{pNiceArray},
\end{equation}
for $1 \leq \alpha \leq r_{\vec{K}}$, with $\epsilon_{\vec{K},\alpha} = \mathcal{E}$. In \cref{app:eqn:same_sewing}, ${}^{(*)}$ denotes an additional complex conjugation in the case of anti-unitary operators. As a consequence of the block diagonal structure of $\mathcal{U}(g)$, we conclude that $\mathcal{R}^{+,\mathcal{E}}_{\vec{K}} (g) = \mathcal{R}^{-,\mathcal{E}}_{\vec{K}} (g) \equiv \mathcal{R}^{\mathcal{E}}_{\vec{K}} (g)$. In other words, the nonzero eigenstates of $H_{\vec{K}}$ corresponding to the same singular value of $S_{\vec{K}}$, but having opposite energies, carry identical (co)representations of $\mathcal{G}_{\vec{K}}$. \Cref{app:eqn:same_sewing} also implies the conclusion that was derived in the main text, namely that the eigenvectors of $T_{\vec{K}}$ and $\tilde{T}_{\vec{K}}$ defined in \cref{app:eqn:spectrum_T,app:eqn:spectrum_Ttilde} with the same nonzero eigenvalue $\mathcal{E}^2$ carry identical (co)representations of $\mathcal{G}_{\vec{K}}$ 
\begin{align}
	U(g) \phi_{\vec{K},\alpha}^{(*)} &=\sum_{\beta} \left[\mathcal{R}_{\vec{K}}^{\mathcal{E}} (g) \right]_{\alpha \beta} \phi_{\vec{K},\beta}, \label{app:eqn:sewingL} \\
	\tilde{U} (g) \psi_{\vec{K},\alpha}^{(*)} &=\sum_{\beta} \left[\mathcal{R}_{\vec{K}}^{\mathcal{E}} (g) \right]_{\alpha \beta}  \psi_{\vec{K},\beta} \label{app:eqn:sewingLtilde} .
\end{align}
For future reference, we summarize the relation between the dispersive band wave functions, energy eigenstates and (co)representations of Hamiltonians $H_{\vec{K}}$ in the chiral limit, $H_{\vec{K}}$ with chemical potential terms, $T_{\vec{K}}$, and $\tilde{T}_{\vec{K}}$ in \cref{app:tab:relation_between_eigenstates}.

\begin{table*}[!t]
\begin{tabular}{l | c | r} 
	Hamiltonian & Energy & Wave Function  \\ 
\hline \hline 
	\multirow{2}{*}{$H_{\vec{K}} = \begin{pmatrix}
		\mathbb{0} & S_\vec{K} \\
		S_\vec{K}^\dagger & \mathbb{0} \\
	\end{pmatrix}$} & 
	$\mathcal{E}$ & 
	$\Psi_{\vec{K},\alpha}^{+} = \begin{pmatrix}
			\phi_{\vec{K},\alpha} \\
			\psi_{\vec{K},\alpha} \\
		\end{pmatrix}$  \\
	\cline{2-3}
	& $-\mathcal{E}$ & 
	$\Psi_{\vec{K},\alpha}^{-} = \begin{pmatrix}
			-\phi_{\vec{K},\alpha} \\
			\psi_{\vec{K},\alpha} \\
		\end{pmatrix}$  \\
	\hline
	\multirow{2}{*}{$H_{\vec{K}} = \begin{pmatrix}
		\mu_{L}\mathbb{1} & S_\vec{K} \\
		S_\vec{K}^\dagger & \mu_{\tilde{L}}\mathbb{1} \\
	\end{pmatrix}$} & 
	$\frac{\mu_{L} + \mu_{\tilde{L}}}{2} + \sqrt{u^2 + \mathcal{E}^2}$ & 
	$\Psi_{\vec{K},\alpha}^{+} \propto \begin{pmatrix}
			\left( u + \sqrt{u^2 + \mathcal{E}^2} \right) \phi_{\vec{K},\alpha} \\
			\mathcal{E} \psi_{\vec{K},\alpha} \\
		\end{pmatrix}$  \\	
	\cline{2-3}
	& $\frac{\mu_{L} + \mu_{\tilde{L}}}{2} - \sqrt{u^2 + \mathcal{E}^2}$ & 
	$\Psi_{\vec{K},\alpha}^{-} \propto \begin{pmatrix}
			\left( u - \sqrt{u^2 + \mathcal{E}^2} \right) \phi_{\vec{K},\alpha} \\
			\mathcal{E} \psi_{\vec{K},\alpha} \\
		\end{pmatrix}$  \\	
	\hline		
	$T_{\vec{K}}$ & 
	$\mathcal{E}^2$ &
	$\phi_{\vec{K},\alpha}$ \\
	\hline		
	$\tilde{T}_{\vec{K}}$ & 
	$\mathcal{E}^2$ &
	$\psi_{\vec{K},\alpha}$ \\  
\end{tabular} 
\caption{The dispersive spectra of $H_{\vec{k}}$ in the chiral case (first row), or with sublattice chemical potential terms (second row), as well as of the two effective sublattice Hamiltonians $T_{\vec{k}}$ and $\tilde{T}_{\vec{k}}$ (last two rows) at $\vec{k}= \vec{K}$. For a given singular value $\mathcal{E}>0$ of $S_{\vec{K}}$, we list the energy and wave functions of the corresponding degenerate eigenspace of each Hamiltonian matrix (here $u = \frac{\mu_{L}-\mu_{\tilde{L}}}{2}$). The set of wave functions is restricted to those indices $1 \leq \alpha \leq r_{\vec{K}}$ for which $\epsilon_{\vec{K},\alpha}=\mathcal{E}$. All wave functions within a given set carry the same (co)representation of $\mathcal{G}_{\vec{K}}$, $\mathcal{R}_{\vec{K}}^{\mathcal{E}}$, as defined in the text surrounding \cref{app:eqn:same_sewing}.}
\label{app:tab:relation_between_eigenstates}
\end{table*}

By construction, the representations induced from all the orbitals of the BCL are simply given by 
\begin{equation}
	\label{app:eqn:br_total}
	\bigoplus_{\left( \rho \right)_{w} \in \left(L \oplus \tilde{L} \right)} \left[ \left( \rho \right)_{w} \uparrow \mathcal{G} \right] \downarrow \mathcal{G}_{\vec{K}} = \left(\bigoplus_{\mathcal{E}>0} 2 \mathcal{R}_{\vec{K}}^{\mathcal{E}} \right) \oplus \mathcal{R}_{\vec{K}}^{+,0} \oplus \mathcal{R}_{\vec{K}}^{-,0},
\end{equation}
where $\left[ \left( \rho \right)_{w} \uparrow \mathcal{G} \right] \downarrow \mathcal{G}_{\vec{K}}$ denotes the (co)irrep subduced in $\mathcal{G}_{\vec{K}}$ from the BR $\left( \rho \right)_{w} \uparrow \mathcal{G}$ which is induced from the (co)irrep $\rho$ of the site symmetry group corresponding to the Wyckoff position $w$.

To obtain another identity of the form of \cref{app:eqn:br_total}, we introduce the wave functions $\Psi^{\prime T}_{\vec{K},\alpha} = \left( \mathbb{0}, \psi_{\vec{K},\alpha}^T \right)$, for $1 \leq \alpha \leq N_{\tilde{L}}$. $\Psi^{\prime T}_{\vec{K},\alpha}$ are \emph{not} eigenstates of $H_{\vec{K}}$ for $1 \leq \alpha \leq r_{\vec{K}}$, but rather a basis spanning the states with momentum $\vec{K}$ supported on the $\tilde{L}$ sublattice. For every singular value of $S_{\vec{K}}$, $\mathcal{E} > 0$, \cref{app:eqn:sewingLtilde,app:eqn:definition:bigU} imply that
\begin{equation}
	\label{app:eqn:sewingFake_nonzero} 
	\mathcal{U}(g) \Psi^{\prime (*)}_{\vec{K},\alpha} = \left[\mathcal{R}^{\mathcal{E}}_{\vec{K}} (g) \right]_{\alpha \beta} \Psi^{\prime}_{\vec{K},\beta},
\end{equation} 
for $1 \leq \alpha \leq r_{\vec{K}}$, with $\epsilon_{\vec{K},\alpha} = \mathcal{E}$. Correspondingly, for $r_{\vec{K}} < \alpha \leq N_{\tilde{L}}$, \cref{app:eqn:action_g_BCL_eigs} implies that 
\begin{equation}
	\label{app:eqn:sewingFake_zero} 
	\mathcal{U}(g) \Psi^{\prime (*)}_{\vec{K},\alpha} = \left[ \mathcal{R}^{-,0}_{\vec{K}} (g) \right]_{\alpha \beta} \Psi^{\prime}_{\vec{K},\beta}.
\end{equation} 
\Cref{app:eqn:sewingFake_nonzero,app:eqn:sewingFake_zero} combined with the fact that the vectors $\Psi^{\prime}_{\vec{K},\alpha}$ form a complete basis of all states with momentum $\vec{K}$ having weight only on sublattice $\tilde{L}$ enable us to write
\begin{equation}
	\label{app:eqn:br_small}
	\bigoplus_{\left( \rho \right)_{w} \in \tilde{L}} \left[ \left( \rho \right)_{w} \uparrow \mathcal{G} \right] \downarrow \mathcal{G}_{\vec{K}} = \left(\bigoplus_{\mathcal{E}>0} \mathcal{R}^{\mathcal{E}}_{\vec{K}} \right) \oplus \mathcal{R}_{\vec{K}}^{-,0}.
\end{equation}
Combining \cref{app:eqn:br_total,app:eqn:br_small}, we also find that 
\begin{equation}
	\label{app:eqn:br_big}
	\bigoplus_{\left( \rho \right)_{w} \in L} \left[ \left( \rho \right)_{w} \uparrow \mathcal{G} \right] \downarrow \mathcal{G}_{\vec{K}} = \left(\bigoplus_{\mathcal{E}>0} \mathcal{R}^{\mathcal{E}}_{\vec{K}} \right) \oplus \mathcal{R}_{\vec{K}}^{+,0}.
\end{equation}
Finally, the formal difference of (co)representations assigned to the flat bands and corresponding band touching points of $\mathcal{H}$ at $\vec{K}$, namely $\mathcal{R}^{+,0}_{\vec{K}} \boxminus \mathcal{R}^{-,0}_{\vec{K}}$, can be computed from \cref{app:eqn:br_big,app:eqn:br_small}
	\begin{equation}
		\mathcal{R}^{+,0}_{\vec{K}} \boxminus \mathcal{R}^{-,0}_{\vec{K}} = \bigoplus_{\left( \rho \right)_{w} \in L} \left[ \left( \rho \right)_{w} \uparrow \mathcal{G} \right] \downarrow \mathcal{G}_{\vec{K}}
		\boxminus
		\bigoplus_{\left( \rho \right)_{w} \in \tilde{L}} \left[ \left( \rho \right)_{w} \uparrow \mathcal{G} \right] \downarrow \mathcal{G}_{\vec{K}}.
	\end{equation}
	Repeating this argument at all high-symmetry momentum points proves the BR subtraction rule for the flat bands \cref{app:eqn:br_subtraction}.  

\subsection{Band touching points in a BCL}\label{app:sec:gapless}

In the main text, we outlined how the flat band subtraction rule from \cref{app:eqn:br_subtraction} can be used to diagnose locally-stable band touching points between the flat and dispersive BCL bands. In this appendix, we give a detailed proof of this statement by considering the the flat band (co)irreps at a given momentum $\vec{K}$ (with little group $\mathcal{G}_{\vec{K}}$). 

We start by deriving a useful intermediate result, which was already employed in \cref{app:sec:unitary_sym}, namely that $\mathcal{R}^{+,0}_{\vec{K}}$ and $\mathcal{R}^{-,0}_{\vec{K}}$, in general, contain no common (co)irreps of $\mathcal{G}_{\vec{K}}$. To see this, we assume the converse to be true, such that there exists a (co)representation of $\mathcal{G}_{\vec{K}}$, $\Xi \neq \emptyset$ with
\begin{equation}
	\label{app:eqn:no_common_irreps_contradiction}
	\Xi = \mathcal{R}^{+,0}_{\vec{K}} \cap \mathcal{R}^{-,0}_{\vec{K}}.
\end{equation}
As summarized in \cref{app:tab:relation_between_flatbands}, \cref{app:eqn:no_common_irreps_contradiction} implies that there are eigenstates within the kernels of both $T_{\vec{K}}$ and $\tilde{T}_{\vec{K}}$, which form carrier spaces for $\Xi$. Letting $d_{\Xi}$ denote the dimension of the (co)representation $\Xi$, a gauge choice for the SVD of $S_{\vec{K}}$ can always be found in which the carrier space for $\Xi$ in the eigenspectrum of $T_{\vec{K}}$ ($\tilde{T}_{\vec{K}}$) is given by $\phi_{\vec{K},\alpha}$ ($\psi_{\vec{K},\alpha}$) for $r_{\vec{K}} < \alpha \leq r_{\vec{K}} + d_{\Xi}$, with the action of any symmetry operation $g \in \mathcal{G}_{\vec{K}}$ reading as
\begin{align}
	U (g) \phi_{\vec{K},\alpha}^{(*)} &= \sum_{\beta = r_{\vec{k}} + 1 }^{r_{\vec{K}} + d_{\Xi}} \left[\mathcal{R}_{\vec{K}}^{\Xi} (g) \right]_{\alpha \beta} \phi_{\vec{K},\beta},\label{app:eqn:gapped_irrep_space_L}  \\
	\tilde{U} (g) \psi_{\vec{K},\alpha}^{(*)} &= \sum_{\beta = r_{\vec{k}} + 1 }^{r_{\vec{K}} + d_{\Xi}} \left[\mathcal{R}_{\vec{K}}^{\Xi} (g) \right]_{\alpha \beta} \psi_{\vec{K},\beta}. \label{app:eqn:gapped_irrep_space_Ltilde}
\end{align}
In \cref{app:eqn:gapped_irrep_space_L,app:eqn:gapped_irrep_space_Ltilde} the sewing matrix $\mathcal{R}_{\vec{K}}^{\Xi} (g)$ can be chosen to be identical for the two carrier spaces of $\Xi$, as a consequence of the gauge freedom in defining the SVD of the $S_{\vec{K}}$ matrix. This is because $\phi_{\vec{K},\alpha}$ for $r_{\vec{K}} < \alpha \leq N_{L}$ ($\psi_{\vec{K},\alpha}$ for $r_{\vec{K}} < \alpha \leq N_{\tilde{L}}$) are left (right) singular eigenvectors of $S_{\vec{K}}$ with zero singular value and thus can be arbitrarily rotated by a $\mathrm{U}(N_{L}-r_{\vec{K}})$ [$\mathrm{U}(N_{\tilde{L}}-r_{\vec{K}})$] transformation.

Now, we consider a perturbation $\Delta_{\vec{K}}$ defined by
	\begin{equation}
		\Delta_{\vec{K}} = \mu \sum_{\beta = r_{\vec{k}} + 1 }^{r_{\vec{K}} + d_{\Xi}}  \phi_{\vec{K},\beta} \psi^{\dagger}_{\vec{K},\beta},
	\end{equation}
	where $ \mu\neq 0$. Using \cref{app:eqn:gapped_irrep_space_Ltilde,app:eqn:gapped_irrep_space_L}, as well as the unitarity of the sewing matrix $R_{\vec{K}}^{\Xi}(g)$, we find that the perturbation $\Delta_{\vec{K}}$ is invariant under any group transformation $g \in \mathcal{G}_{\vec{K}}$
	\begin{equation}
		U(g) \Delta^{(*)}_{\vec{K}} \tilde{U}(g) = \Delta_{\vec{K}}.
	\end{equation}
	As such, one can always find an $N_{L} \times N_{\tilde{L}}$ inter-sublattice hopping matrix $\delta S_{\vec{R}}$ with finite-range hopping such that $\delta S_{\vec{K}} = \Delta_{\vec{K}}$. On the other hand,
	\begin{alignat}{4}
		& S_{\vec{K}} \psi_{\vec{K},\alpha} \neq 0, \quad && \Delta_{\vec{K}} \psi_{\vec{K},\alpha} = 0, \quad && \text{for} \quad && 1 \leq \alpha \leq r_{\vec{K}} \text{ and } r_{\vec{K}} + d_{\Xi} < \alpha \leq N_{\tilde{L}}, \nonumber \\
		& S_{\vec{K}} \psi_{\vec{K},\alpha} = 0, \quad && \Delta_{\vec{K}} \psi_{\vec{K},\alpha} \neq 0, \quad && \text{for} \quad &&r_{\vec{K}} < \alpha \leq r_{\vec{K}} + d_{\Xi},
	\end{alignat}
	and so one can define a perturbed inter-sublattice hopping matrix 
	\begin{equation}
		\label{app:eqn:perturb_to_gap}
		S'_{\vec{k}} = S_{\vec{k}} + \delta S_{\vec{k}},
	\end{equation}
	such that $S'_{\vec{K}} \psi_{\vec{K},\alpha} \neq 0$ for any $r_{\vec{K}} < \alpha \leq r_{\vec{K}} + d_{\Xi}$. In the perturbed BCL corresponding to $S'_{\vec{K}}$, the carrier spaces of $\Xi$ are no longer in the kernels of $T'_{\vec{K}} = S'_{\vec{k}} S^{\prime \dagger}_{\vec{K}}$ and $\tilde{T}'_{\vec{K}} = S^{\prime \dagger}_{\vec{K}} S'_{\vec{K}}$. Instead, the kernel of $T'_{\vec{K}}$ ($\tilde{T}'_{\vec{K}}$) will form a carrier space for the \emph{bona fide} (co)representation $\mathcal{R}^{\prime +,0}_{\vec{K}} \equiv \mathcal{R}^{+,0}_{\vec{K}} \ominus \Xi$ ($\mathcal{R}^{\prime -,0}_{\vec{K}} \equiv \mathcal{R}^{-,0}_{\vec{K}} \ominus \Xi$). In the perturbed BCL, 
	\begin{equation}
		\mathcal{R}^{\prime +,0}_{\vec{K}} \cap \mathcal{R}^{\prime -,0}_{\vec{K}} = \emptyset.
	\end{equation}
	
	Therefore, we have shown that \emph{any} common (co)irrep of $\mathcal{R}^{+,0}_{\vec{K}}$ and $\mathcal{R}^{-,0}_{\vec{K}}$ in a BCL can be eliminated through a symmetry-allowed perturbation. As such, one generically has $\mathcal{R}^{+,0}_{\vec{K}} \cap \mathcal{R}^{-,0}_{\vec{K}} = \emptyset$. 

Returning to the question of diagnosing band touching points using the subtraction rule from \cref{app:eqn:br_subtraction}, there are two possibilities depending on the relation between the BRs of the $L$ and $\tilde{L}$ sublattices subduced in the $\mathcal{G}_{\vec{K}}$ group (denoted by $\mathcal{BR}_{\tilde{L}} \downarrow \mathcal{G}_{\vec{K}}$ and $\mathcal{BR}_{L} \downarrow \mathcal{G}_{\vec{K}}$, respectively):
\begin{enumerate}
	\item 	$\mathcal{BR}_{\tilde{L}} \downarrow \mathcal{G}_{\vec{K}} \subset \mathcal{BR}_{L} \downarrow \mathcal{G}_{\vec{K}}$. If the BR subtraction from \cref{app:eqn:br_subtraction} produces a (co)representation at $\vec{K}$, then there can be no locally-stable band touching points between the flat and dispersive BCL bands. The proof follows by contradiction. We assume there to be a band touching point between the flat and dispersive bands at $\vec{K}$. This is equivalent to the rank of $S_{\vec{K}}$ being strictly smaller than $N_{\tilde{L}}$. As such, $\tilde{T}_{\vec{K}}$ has zero modes forming a carrier space for $\mathcal{R}^{-,0}_{\vec{K}} \neq \emptyset$. However, the condition $\mathcal{BR}_{\tilde{L}} \downarrow \mathcal{G}_{\vec{K}} \subset \mathcal{BR}_{L} \downarrow \mathcal{G}_{\vec{K}}$ coupled with \cref{app:eqn:br_small,app:eqn:br_big} implies that $R^{-,0}_{\vec{K}} \subset R^{+,0}_{\vec{K}}$, which leads us to conclude that $R^{-,0}_{\vec{K}} \cap R^{+,0}_{\vec{K}} = R^{-,0}_{\vec{K}} \neq \emptyset$, which is in contradiction with the condition that one generically has $\mathcal{R}^{+,0}_{\vec{K}} \cap \mathcal{R}^{-,0}_{\vec{K}} = \emptyset$.
		
	\item $\mathcal{BR}_{\tilde{L}} \downarrow \mathcal{G}_{\vec{K}} \not\subset \mathcal{BR}_{L} \downarrow \mathcal{G}_{\vec{k}}$. If the BR subtraction assigns a formal (co)representation difference at $\vec{K}$, then a band touching point between the flat and dispersive BCL bands is enforced. This can be seen directly from the fact that a formal difference $\Xi \boxminus \Theta$ assigned to the flat bands at $\vec{K}$ implies through the one-to-one mapping between the positive eigenstates of $T_{\vec{K}}$ and $\tilde{T}_{\vec{K}}$ outlined in \cref{app:tab:relation_between_eigenstates} that the carrier space of $\Theta$ must be the kernel of $\tilde{T}_{\vec{K}}$ and thus $S_{\vec{K}}$ is not full-rank. This band touching point cannot be gapped by any symmetry-preserving perturbation to the inter-sublattice hopping matrix. To see this, assume that there is such a perturbation that gaps the band touching point. The inter-sublattice hopping matrix would be come full-rank at $\vec{K}$, implying that $\tilde{T}_{\vec{K}}$ has only nonzero eigenstates. However, as a consequence of the one-to-one mapping between the positive eigenstates of $T_{\vec{K}}$ and $\tilde{T}_{\vec{K}}$ outlined in \cref{app:tab:relation_between_eigenstates}, this would imply that $\mathcal{BR}_{\tilde{L}} \downarrow \mathcal{G}_{\vec{K}} \subset \mathcal{BR}_{L} \downarrow \mathcal{G}_{\vec{k}}$, which leads to a contradiction.
\end{enumerate}

\section{Examples of Flat Band Constructions}\label{app:sec:example}

In this appendix, we present a series of tight-binding Hamiltonians in two dimensions that exemplify our BCL construction. First, we illustrate  \cref{app:enum:BCL_molecular} of \cref{app:sec:breaking_chiral} through a BCL model featuring flat bands and having intra-sublattice hopping in both $L$ and $\tilde{L}$. In \cref{app:sec:example:TBG}, we show that the ten-band tight-binding model for twisted bilayer graphene near the first magic angle constructed by Ref.~\cite{PO19} is a chiral BCL Hamiltonian and explicitly verify the BR subtraction prescription from \cref{app:eqn:br_subtraction}. We then relate the $p$-orbital model introduced by Ref.~\cite{WU07} (a model built of $p_x$ and $p_y$ orbitals on the honeycomb lattice) to our BCL construction and provide a rigorous justification for its flat bands and corresponding band touching points in \cref{app:sec:example:honeycombP}. An example of a BCL Hamiltonian featuring globally stable band touching points between the flat and dispersive bands, which cannot be diagnosed through (co)irreps \`a la \cref{app:eqn:br_subtraction}, is presented in \cref{app:sec:example:C2T}. We review one of the few examples from literature~\cite{GRE10} featuring flat bands that are neither a modified BCL, nor an effective Hamiltonian as defined in \cref{app:sec:bipartite_crystalline_lattices:effective:def}. In this example, the existence of flat bands is protected by the combined operation of particle-hole symmetry and inversion. After explaining the origin of the flat band in this model, we construct an equivalent BCL Hamiltonian within the same SSG that features the exact same flat band in terms of wave function. Finally, we explore the possibility of flat band protection in the ten Altland–Zirnbauer classes with inversion symmetry~\cite{ALT97a, KIT09a, SCH08, QI08, RYU10, CHI13, MOR13, SHI14, CHI16}

For the examples that we present here, we will follow the notation introduced in \cref{app:sec:bipartite_crystalline_lattices:notation}. 
In addition, when discussing the symmetries of the lattice Hamiltonians, we will employ the conventions of the Bilbao Crystallographic Server (BCS)~\citeBCS{}. Despite only discussing two-dimensional models, we will always quote a three-dimensional SSG, where we implicitly mod out translations along the third crystallographic axis (in the notation of BCS). Since all the examples featured here also belong to symmorphic SSG, with an appropriate choice of origin, the action of any point group transformation $g$ on the real-space BCL fermions reads as 
\begin{equation}
	g \hat{f}^\dagger_{\vec{R},i} g^{-1} = \sum_{j} D^{\hat{f}}_{ij}(g) \hat{f}^\dagger_{g \left(\vec{R} + \vec{r}^{\hat{f}}_i \right)-\vec{r}^{\hat{f}}_j,j},
\end{equation} 
for $\hat{f}^\dagger_{\vec{R},i}=\hat{a}^\dagger_{\vec{R},i},\hat{b}^\dagger_{\vec{R},i}$, where $g \vec{r} $ denotes the action of $g$ on the real-space vector $\vec{r}$. For example, if $g=\mathcal{T}$ (where $\mathcal{T}$ denotes time-reversal), $g \vec{r} = \vec{r}$. The representation matrix for the transformation $g$ is defined by 
\begin{equation}
	D^{\hat{f}}_{ij} ( g ) = \begin{cases}
		\bra{\Phi^{\hat{f}}_{j}} \left( g \ket{\Phi^{\hat{f}}_{i}} \right), & \quad \text{if } g \left(\vec{R} + \vec{r}^{\hat{f}}_i \right)-\vec{r}^{\hat{f}}_j \text{ is a lattice vector}\\
		0, &  \quad \text{otherwise}
	\end{cases},
\end{equation}
where $\ket{\Phi^{\hat{f}}_{i}}$ denotes the orbital wave function corresponding to a fermion $\hat{f}^\dagger_{\vec{R},i}$. Using the Fourier transform convention from \cref{app:eqn:momentum_ops}, the action of $g$ on the momentum-space BCL operators is given by 
\begin{equation}
    \label{app:eqn:symmetry_action_examples}
	g \hat{f}^\dagger_{\vec{k},i} g^{-1} = \begin{cases}
		\sum_{j} D^{\hat{f}}_{ij}(g) \hat{f}^\dagger_{g \vec{k},j}, & \quad \text{if } g \text{ is unitary} \\
		\sum_{j} D^{\hat{f}}_{ij}(g) \hat{f}^\dagger_{-g \vec{k},j}, & \quad \text{if } g \text{ is anti-unitary}
	\end{cases},
\end{equation}
where the action of $g\vec{k}$ is defined in the same way as the action of $g$ on a real-space vector (\eg $\mathcal{T} \vec{k} = \vec{k}$). 

Throughout this appendix, we will also denote point group operations as follows: For any axis $\hat{\vec{v}}$, $C_{nv}$ denotes $n$-fold rotations about $\hat{\vec{v}}$, while $m_v$ denotes mirror reflections perpendicular to the axis $\hat{\vec{v}}$. Additionally, $\mathcal{I}$ denotes three-dimensional spatial inversions and $\mathcal{T}$ represents time-reversal.

\subsection{Flat bands in a BCL without chiral symmetry}\label{app:sec:example:noChiral}

\begin{figure}[!t]
\captionsetup[subfloat]{farskip=0pt}\sbox\nsubbox{
		\resizebox{\textwidth}{!}
		{\includegraphics[height=6cm]{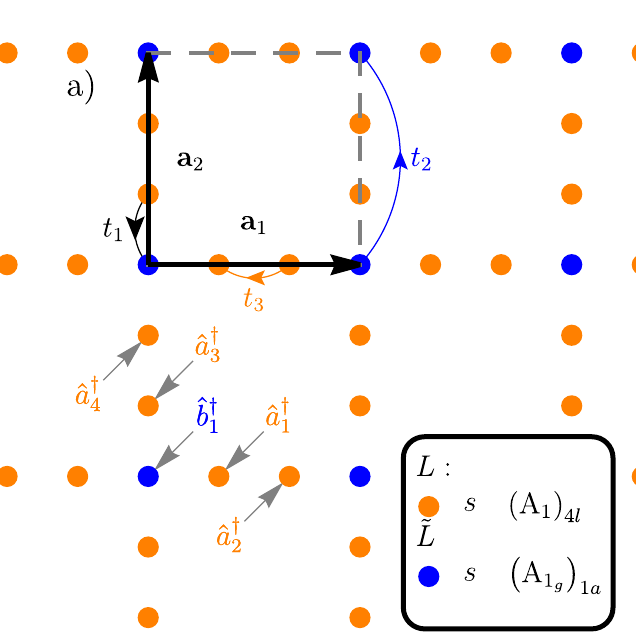}\includegraphics[height=6cm]{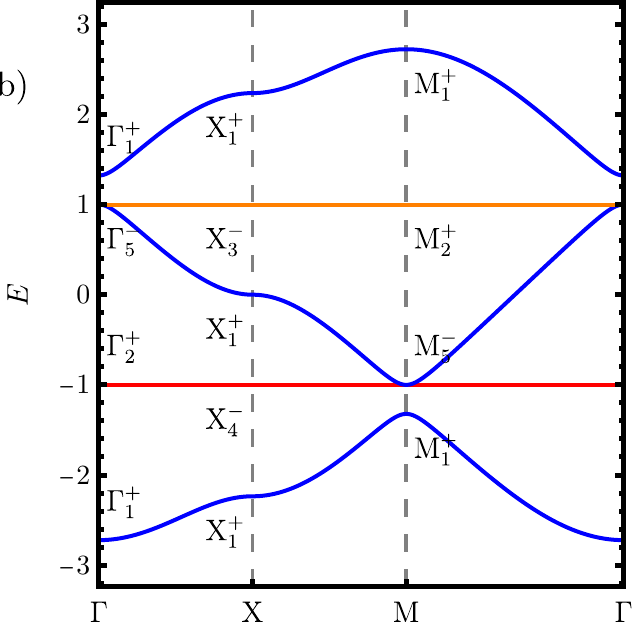}}}\setlength{\nsubht}{\ht\nsubbox}\centering\subfloat{\label{app:fig:Lieb_molecular:a}\includegraphics[height=\nsubht]{3b.pdf}}\subfloat{\label{app:fig:Lieb_molecular:b}\includegraphics[height=\nsubht]{3a.pdf}}
\caption{Flat bands in a BCL without chiral symmetry. In panel a), we illustrate the lattice structure of the model. The unit cell, spanned by the lattice vectors $\vec{a}_{1}$ and $\vec{a}_{2}$, is denoted by a dashed gray square. We represent the five spinless orbitals within each unit cell with filled dots. The orbital content of each sublattice is detailed in the inset at the bottom-right of the panel. For each orbital, we specify the wave function, as well as the irrep and Wyckoff position. The orbital corresponding to each fermionic operator $\hat{a}^\dagger_{i}$, with $1 \leq i \leq 4$, or $\hat{b}^\dagger_{1}$ (where we suppress the unit cell index $\vec{R}$) is shown by a gray arrow. Additionally, an arrow connecting two orbitals represents a hopping process between the orbitals at the start and end positions of the arrow. The hopping amplitude is given as a label for the arrow. By applying the symmetry operations of the $P4/mmm1'$ group (SSG 123.340 in the notation of BCS)~\citeBCS{}, the entire tight-binding Hamiltonian of the model can be recovered from the three hopping amplitudes shown. Panel b) shows the band structure of the model for $t_1=-1$, $t_2=-0.1$, and $t_3=-1$. At each high-symmetry momentum point, we also indicate the (co)irreps of the corresponding bands. The model features two gapless flat bands at energies $E=+t_3$ and $E=-t_3$, respectively plotted in red and orange.}
\label{app:fig:Lieb_molecular}
\end{figure}

In \cref{app:sec:breaking_chiral}, we showed that the chiral symmetry in a BCL can be broken without perturbing the flat bands by adding arbitrary hopping in the smaller sublattice, $\tilde{L}$. Additionally, we argued that if the intra-sublattice hopping matrix $A_{\vec{k}}$ contains a $\vec{k}$-independent eigenvalue $a$ with degeneracy $n_{a} > N_{\tilde{L}}$, then the BCL Hamiltonian is guaranteed to have $n_a-N_{\tilde{L}}$ flat bands. Flat bands can also arise if $A_{\vec{R}}$ leads to the formation of uncoupled degenerate molecular orbitals. In this appendix, we illustrate this case with a simple example defined on the two-dimensional square lattice, and show explicitly how the BR subtraction from \cref{app:eqn:br_subtraction} is recovered.

We consider a BCL in the $P4/mmm1'$ group (SSG 123.340 in the notation of BCS)~\citeBCS{}, where the $L$ sublattice contains spinless $s$ orbitals at the $4l$ Wyckoff position, while the $\tilde{L}$ sublattice has spineless $s$ orbitals at the $1a$ position, as shown in \cref{app:fig:Lieb_molecular:a}. The symmetries of $\mathcal{H}$ are generated by two-dimensional translations, $C_{4z}$, $C_{2y}$, and $\mathcal{I}$. We take the $\hat{a}^\dagger_{\vec{R},1}$ fermions to be located at $\vec{r}^{\hat{a}}_{1}=\alpha \vec{a}_1$ within each unit cell, where, without loss of generality, we restrict to $0 < \alpha < 1/2$. The first-quantized Hamiltonian of the BCL in the basis of \cref{app:eqn:general_bas_spinor} is given by 
\begin{equation}
	H_{\vec{k}}=\begin{pmatrix}
 0 & t_3 e^{i (2 \alpha -1) k_x} & 0 & 0 & t_1 e^{i \alpha  k_x} \\
 t_3 e^{i (1-2 \alpha ) k_x} & 0 & 0 & 0 & t_1 e^{-i \alpha  k_x} \\
 0 & 0 & 0 & t_3 e^{i (2 \alpha -1) k_y} & t_1 e^{i \alpha  k_y} \\
 0 & 0 & t_3 e^{i (1-2 \alpha ) k_y} & 0 & t_1 e^{-i \alpha  k_y} \\
 t_1 e^{-i \alpha  k_x} & t_1 e^{i \alpha  k_x} & t_1 e^{-i \alpha  k_y} & t_1 e^{i \alpha  k_y} & 2 t_2 \left(\cos \left(k_x\right)+\cos \left(k_y\right)\right) \\
\end{pmatrix} ,
\end{equation}
where $k_x$ and $k_y$ are the two Cartesian components of the $\vec{k}$ momentum vector and $t_i$ ($i=1,2,3$) represent real hopping parameters.

The band structure of $H_{\vec{k}}$ is shown in \cref{app:fig:Lieb_molecular:b}: the BCL Hamiltonian features two gapless flat bands located at different energies. To understand their origin, we first note that the intra-sublattice hopping in $L$ merely hybridizes the $s$ orbitals at the $4l$ Wyckoff position into $s$ and $p$ orbitals with energies $+t_3$ and $-t_3$, respectively, located at the $2f$ position. Therefore, we can define a set of molecular orbitals
\begin{equation}
	\label{app:eqn:noChiral_molecular_bas}
	\hat{a}^\dagger_{\vec{R},\pm,1} = \frac{1}{\sqrt{2}}\left(\hat{a}^\dagger_{\vec{R},1} \pm \hat{a}^\dagger_{\vec{R},2} \right) \qquad \text{and} \qquad
	\hat{a}^\dagger_{\vec{R},\pm,2} = \frac{1}{\sqrt{2}}\left(\hat{a}^\dagger_{\vec{R},3} \pm \hat{a}^\dagger_{\vec{R},4} \right), 
\end{equation} 
or alternatively, in momentum space,
\begin{equation}
    \renewcommand\arraystretch{2}
	\begin{pmatrix}
		\hat{a}^\dagger_{\vec{k},+,1} \\
		\hat{a}^\dagger_{\vec{k},+,2} \\
		\hat{a}^\dagger_{\vec{k},-,1} \\
		\hat{a}^\dagger_{\vec{k},-,2} \\
	\end{pmatrix}
	=	\begin{pmatrix}
 \frac{e^{i \left(\alpha -\frac{1}{2}\right) k_x}}{\sqrt{2}} & \frac{e^{i \left(\frac{1}{2}-\alpha \right) k_x}}{\sqrt{2}} & 0 & 0 \\
 0 & 0 & \frac{e^{i \left(\alpha -\frac{1}{2}\right) k_y}}{\sqrt{2}} & \frac{e^{i \left(\frac{1}{2}-\alpha \right) k_y}}{\sqrt{2}} \\
 -\frac{e^{i \left(\alpha -\frac{1}{2}\right) k_x}}{\sqrt{2}} & \frac{e^{i \left(\frac{1}{2}-\alpha \right) k_x}}{\sqrt{2}} & 0 & 0 \\
 0 & 0 & \frac{e^{i \left(\alpha -\frac{1}{2}\right) k_y}}{\sqrt{2}} & -\frac{e^{i \left(\frac{1}{2}-\alpha \right) k_y}}{\sqrt{2}} \\
\end{pmatrix} 	\begin{pmatrix}
		\hat{a}^\dagger_{\vec{k},1} \\
		\hat{a}^\dagger_{\vec{k},2} \\
		\hat{a}^\dagger_{\vec{k},3} \\
		\hat{a}^\dagger_{\vec{k},4} \\
	\end{pmatrix},
\end{equation}
and rewrite the BCL Hamiltonian as 
\begin{equation}
	\mathcal{H} = \sum_{\vec{k}} \hat{\Psi'}^\dagger_{\vec{k}} H'_{\vec{k}} \hat{\Psi'}_{\vec{k}},
\end{equation}
where we have introduced the five-dimensional spinor $\hat{\Psi'}_{\vec{k}} = \left( \hat{a}_{\vec{k},+,1},\hat{a}_{\vec{k},+,2},\hat{a}_{\vec{k},-,1},\hat{a}_{\vec{k},-,2},\hat{b}_{\vec{k},1} \right)^{T}$, as well as the Hamiltonian matrix
\begin{equation}
	\label{app:eqn:hamiltonian_noChiral_transformed}
	H'_{\vec{k}} = \begin{pNiceArray}{CC|CC|C}[first-row]
 \multicolumn{2}{C}{L_+} & \multicolumn{2}{C}{L_-} & \tilde{L} \\
 t_3 & 0 & 0 & 0 & \sqrt{2} t_1 \cos \left(\frac{k_x}{2}\right) \\
 0 & t_3 & 0 & 0 & \sqrt{2} t_1 \cos \left(\frac{k_y}{2}\right) \\
\hline
 0 & 0 & -t_3 & 0 & -i \sqrt{2} t_1 \sin \left(\frac{k_x}{2}\right) \\
 0 & 0 & 0 & -t_3 & i \sqrt{2} t_1 \sin \left(\frac{k_y}{2}\right) \\
\hline
 \sqrt{2} t_1 \cos \left(\frac{k_x}{2}\right) & \sqrt{2} t_1 \cos \left(\frac{k_y}{2}\right) & i \sqrt{2} t_1 \sin \left(\frac{k_x}{2}\right) & -i \sqrt{2} t_1 \sin \left(\frac{k_y}{2}\right) & 2 t_2 \left(\cos \left(k_x\right)+\cos \left(k_y\right)\right) \\
\end{pNiceArray}
 .
\end{equation}
	Expressed in the new basis defined in \cref{app:eqn:noChiral_molecular_bas}, the BCL Hamiltonian from \cref{app:eqn:hamiltonian_noChiral_transformed} is decoupled into the form of \cref{app:eqn:gen_bcl_atomic}: the larger sublattice ($N_{L} = 4$) can be decomposed as $L=L_+ \oplus L_-$, where $L_+$ and $L_-$ respectively contain $s$ and $p$ orbitals at the $2f$ position and only onsite ``chemical potential'' terms. For simplicity, we have labeled the blocks of $H'_{\vec{k}}$ in \cref{app:eqn:hamiltonian_noChiral_transformed} according to the sublattices that they couple. We therefore expect that $\mathcal{H}$ has a flat band located at energy $+t_3$ ($-t_3$) with an identical wave function to the flat band arising in a $L_{+} \oplus \tilde{L}$ ($L_{-} \oplus \tilde{L}$) BCL that contains only the $L_+$ and $\tilde{L}$ ($L_-$ and $\tilde{L}$) sublattices and the corresponding hoppings. This is possible because the flat bands of $H'_{\vec{k}}$ only depend on the kernels of the $2 \times 1$ inter-sublattice hopping matrices which connect the $L_{+}$ and the $\tilde{L}$ sublattices (for the flat band with energy $+t_3$), or the $L_{-}$ and the $\tilde{L}$ sublattices (for the flat band with energy $-t_3$). Analyzing $\mathcal{H}$ within the $P4/mmm1'$ group  (SSG 123.340 in the notation of BCS)~\citeBCS{}, we find
\begin{align}
	\mathcal{BR}_{L_+} &= \left(\mathrm{A}_{g}\right)_{2f} \uparrow \mathcal{G} = \left(\Gamma^+_1 \oplus \Gamma^+ _2\right) + \left(\mathrm{X}^+_1 \oplus \mathrm{X}^-_4 \right) + \left(\mathrm{M}^-_5 \right),\\
	\mathcal{BR}_{L_-} &= \left(\mathrm{B}_{2_u}\right)_{2f} \uparrow \mathcal{G} = \left(\Gamma^-_5 \right) + \left(\mathrm{X}^+_1 \oplus \mathrm{X}^-_3 \right) + \left(\mathrm{M}^+_1 \oplus \mathrm{M}^+_2 \right),\\
	\mathcal{BR}_{\tilde{L}} &= \left(\mathrm{A}_{1_g}\right)_{1a} \uparrow \mathcal{G} = \left(\Gamma^+_1\right) + \left(\mathrm{X}^+_1 \right) + \left(\mathrm{M}^+_1  \right),  
\end{align}
where $\left(\rho \right)_{w} \uparrow \mathcal{G}$ denotes the band representation induced from the (co)irrep $\rho$ of the site symmetry group of the Wyckoff position $w$ (see \cref{app:tab:char_p4mmm} of \cref{app:sec:example:char} for the character table). Using \cref{app:eqn:br_subtraction}, we can find the (co)irreps for the two flat bands located at $\pm t_3$, $\mathcal{B}_{\mathrm{FB}_{\pm}} = \mathcal{BR}_{L_{\pm}} \boxminus \mathcal{BR}_{\tilde{L}}$,
\begin{align}
	\mathcal{B}_{\mathrm{FB}_{+}} &= \left(\Gamma^+ _2\right) + \left(\mathrm{X}^-_4 \right) + \left(\mathrm{M}^-_5 \boxminus \mathrm{M}^+_1 \right),\\
	\mathcal{B}_{\mathrm{FB}_{-}} &= \left(\Gamma^-_5 \boxminus \Gamma^+_1 \right) + \left(\mathrm{X}^-_3 \right) + \left(\mathrm{M}^+_2 \right).
\end{align}

As reproduced in the band structure of \cref{app:fig:Lieb_molecular:b}, the presence of a formal irrep difference in $\mathcal{B}_{\mathrm{FB}_{+}}$ ($\mathcal{B}_{\mathrm{FB}_{-}}$) indicates a band touching point at the $\mathrm{M}$ ($\Gamma$) point for the flat band at energy $+t_3$ ($-t_3$). 
\subsection{Flat bands in a model of twisted bilayer graphene}\label{app:sec:example:TBG}

\begin{figure}[!t]
\captionsetup[subfloat]{farskip=0pt}\sbox\nsubbox{
		\resizebox{\textwidth}{!}
		{\includegraphics[height=6cm]{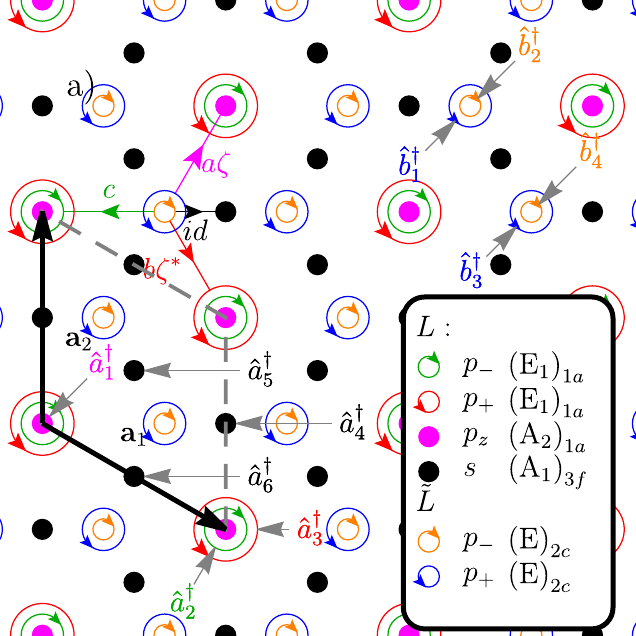}\includegraphics[height=6cm]{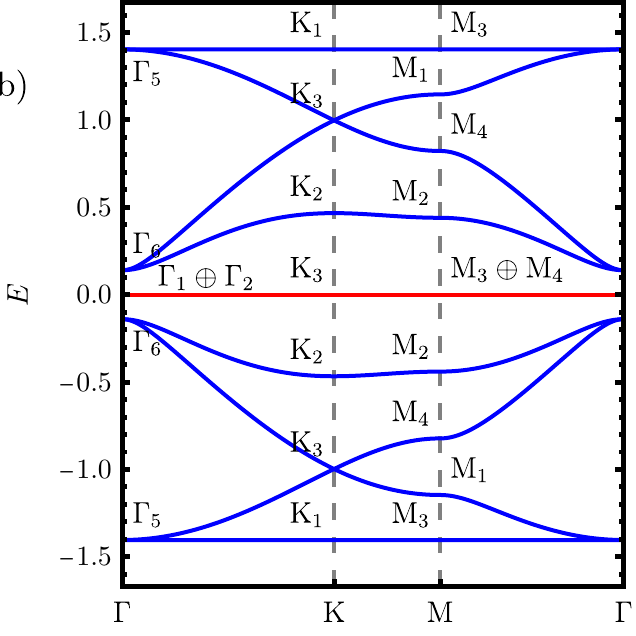}}}\setlength{\nsubht}{\ht\nsubbox}\centering\subfloat{\label{app:fig:TBG_tight_binding:a}\includegraphics[height=\nsubht]{4b.pdf}}\subfloat{\label{app:fig:TBG_tight_binding:b}\includegraphics[height=\nsubht]{4a.pdf}}
\caption{The ten-band PZSV model of single-valley TBG from Ref.~\cite{PO19}. In panel a), we illustrate the lattice structure of the model. The meaning of the symbols is the same as in \cref{app:fig:Lieb_molecular:a}. We use the convention of Ref.~\cite{PO19}, where $p_{\pm} = \frac{1}{\sqrt{2}} \left( p_x \pm i p_y \right)$. By applying the the symmetry operations of the $P6/221'$ group (SSG 177.150 in the notation of BCS)~\citeBCS{}, the entire tight-binding Hamiltonian of the PZSV model can be recovered from the four hopping amplitudes shown. Panel b) shows the band structure of the PZSV model, which features two gapped, perfectly flat bands at zero energy (in red) as well as eight dispersive bands (in blue). At each high-symmetry momentum point, we also indicate the (co)irreps of the corresponding bands. Note that, unlike the bands at zero energy, the top and bottom bands are not flat, but have a finite bandwidth, $\delta E \sim 0.001$. The presence of these almost flat bands can be related to the fact that $S_{\vec{k}}$ features a $\vec{k}$-independent singular value if $b=c=0$. For the parameters chosen in Ref.~\cite{PO19}, $b,c \ll a,d$, and so these bands remain quasi-flat.}
\label{app:fig:TBG_tight_binding}
\end{figure}

Near the first magic angle ($\theta \approx 1.05 \degree$), twisted bilayer graphene (TBG) exhibits two almost perfectly flat bands for each spin and valley degree of freedom~\cite{BIS11}. Within each valley, these so-called active TBG bands have been shown to display fragile~\footnote{Recently, Ref.~\cite{SON20b} proved that owing to its (approximate) particle-hole symmetry~\cite{SON19}, the single-valley continuum Bistritzer-MacDonald model of TBG~\cite{BIS11} is anomalous, and, as such, eludes any symmetry-preserving lattice regularization. Its topology was shown to be stable and characterized by a $\mathbb{Z}_2$ invariant. This is not inconsistent with Ref.~\cite{PO19}, as the models presented in the latter lack the the particle-hole symmetry of TBG. For the purpose of this appendix, we will completely ignore these details and instead focus solely on exemplifying the BR subtraction with one of the lattice models presented in Ref.~\cite{PO19}.} topology~\cite{PO18c,CAN18,SON20} protected by $C_{2z}\mathcal{T}$ and characterized by an integer-valued winding number~\cite{SON19,AHN19}. The resulting Wannier obstruction can however be lifted by the addition of certain trivial bands, as shown explicitly in Ref.~\cite{PO19}, which constructs a series of lattice tight-binding models capturing the low-energy physics of single-valley TBG (but breaking the particle-hole symmetry~\cite{SON19}). In this appendix, we show that the ten-band model derived in Ref.~\cite{PO19} is a special case of the BCL construction and explicitly prove the BR subtraction from \cref{app:eqn:br_subtraction}. 

We start by giving a short overview of the ten-band particle-hole-breaking Po-Zou-Senthil-Vishwanath (PZSV) lattice model of single-valley TBG~\cite{PO19}. As shown in \cref{app:fig:TBG_tight_binding:a}, the PZSV lattice model is essentially a chiral BCL defined on the hexagonal lattice where the two sublattices $L$ and $\tilde{L}$ contain $N_{L}=6$ and $N_{\tilde{L}}=4$ (spinless) orbitals per unit cell. More precisely, sublattice $L$ contains $p_x$, $p_y$, and $p_z$ orbitals at the $1a$ position, in addition to $s$ orbitals at the $3f$ position, while sublattice $\tilde{L}$ has $p_x$ and $p_y$ orbitals at the $2c$ position. The Hamiltonian of the model is given by the chiral BCL Hamiltonian from \cref{app:eqn:bip_ham_chiral_sym}, with the $6 \times 4$ $S_{\vec{k}}$ matrix in the basis of \cref{app:eqn:general_bas_spinor}\ reading
\begin{align}
	\left[S_{\vec{k}}\right]_{:,1}=&\begin{pmatrix}
 -a \zeta ^* e^{-\frac{1}{6} i \left(\sqrt{3} k_x+3 k_y\right)} \left(\omega ^* e^{\frac{1}{2} i \left(\sqrt{3} k_x+k_y\right)}+e^{i k_y}+\omega \right) \\
 b \zeta  e^{-\frac{1}{6} i \left(\sqrt{3} k_x+3 k_y\right)} \left(\omega ^*+\omega  e^{\frac{1}{2} i \left(\sqrt{3} k_x+k_y\right)}+e^{i k_y}\right) \\
 c e^{-\frac{1}{6} i \left(\sqrt{3} k_x+3 k_y\right)} \left(e^{\frac{1}{2} i \left(\sqrt{3} k_x+k_y\right)}+e^{i k_y}+1\right) \\
 -i d e^{-\frac{i k_x}{2 \sqrt{3}}} \\
 -i d \omega  e^{\frac{1}{12} i \left(\sqrt{3} k_x-3 k_y\right)} \\
 -i d \omega ^* e^{\frac{1}{12} i \left(\sqrt{3} k_x+3 k_y\right)} \\
\end{pmatrix},\nonumber \\ 
\left[S_{\vec{k}}\right]_{:,2}=&\begin{pmatrix}
 a \zeta ^* e^{-\frac{1}{6} i \left(\sqrt{3} k_x+3 k_y\right)} \left(\omega ^* e^{\frac{1}{2} i \left(\sqrt{3} k_x+k_y\right)}+\omega  e^{i k_y}+1\right) \\
 c e^{-\frac{1}{6} i \left(\sqrt{3} k_x+3 k_y\right)} \left(e^{\frac{1}{2} i \left(\sqrt{3} k_x+k_y\right)}+e^{i k_y}+1\right) \\
 b \zeta  e^{-\frac{1}{6} i \left(\sqrt{3} k_x+3 k_y\right)} \left(\omega ^* e^{i k_y}+\omega  e^{\frac{1}{2} i \left(\sqrt{3} k_x+k_y\right)}+1\right) \\
 -i d e^{-\frac{i k_x}{2 \sqrt{3}}} \\
 -i d \omega ^* e^{\frac{1}{12} i \left(\sqrt{3} k_x-3 k_y\right)} \\
 -i d \omega  e^{\frac{1}{12} i \left(\sqrt{3} k_x+3 k_y\right)} \\
\end{pmatrix},\nonumber \\ 
\left[S_{\vec{k}}\right]_{:,3}=&\begin{pmatrix}
 a \zeta  e^{-\frac{i k_x}{\sqrt{3}}} \left(\omega ^* e^{\frac{1}{2} i \left(\sqrt{3} k_x-k_y\right)}+e^{\frac{1}{2} i \left(\sqrt{3} k_x+k_y\right)}+\omega \right) \\
 b \zeta ^* e^{-\frac{1}{6} i \left(2 \sqrt{3} k_x+3 k_y\right)} \left(\omega ^* e^{\frac{i k_y}{2}}+e^{\frac{1}{2} i \sqrt{3} k_x} \left(\omega +e^{i k_y}\right)\right) \\
 c \left(e^{\frac{1}{6} i \left(\sqrt{3} k_x-3 k_y\right)}+e^{\frac{1}{6} i \left(\sqrt{3} k_x+3 k_y\right)}+e^{-\frac{i k_x}{\sqrt{3}}}\right) \\
 i d e^{\frac{i k_x}{2 \sqrt{3}}} \\
 i d \omega  e^{-\frac{1}{12} i \left(\sqrt{3} k_x-3 k_y\right)} \\
 i d \omega ^* e^{-\frac{1}{12} i \left(\sqrt{3} k_x+3 k_y\right)} \\
\end{pmatrix},\nonumber \\ 
\left[S_{\vec{k}}\right]_{:,4}=&\begin{pmatrix}
 -a \zeta  e^{-\frac{1}{6} i \left(2 \sqrt{3} k_x+3 k_y\right)} \left(\omega ^* e^{\frac{1}{2} i \left(\sqrt{3} k_x+2 k_y\right)}+e^{\frac{1}{2} i \sqrt{3} k_x}+\omega  e^{\frac{i k_y}{2}}\right) \\
 c \left(e^{\frac{1}{6} i \left(\sqrt{3} k_x-3 k_y\right)}+e^{\frac{1}{6} i \left(\sqrt{3} k_x+3 k_y\right)}+e^{-\frac{i k_x}{\sqrt{3}}}\right) \\
 b \zeta ^* e^{-\frac{1}{6} i \left(2 \sqrt{3} k_x+3 k_y\right)} \left(\omega ^* e^{\frac{i k_y}{2}}+e^{\frac{1}{2} i \sqrt{3} k_x} \left(1+\omega  e^{i k_y}\right)\right) \\
 i d e^{\frac{i k_x}{2 \sqrt{3}}} \\
 i d \omega ^* e^{-\frac{1}{12} i \left(\sqrt{3} k_x-3 k_y\right)} \\
 i d \omega  e^{-\frac{1}{12} i \left(\sqrt{3} k_x+3 k_y\right)} \\
\end{pmatrix} , \label{app:eqn:sMatrix_TBG}
\end{align}
where $\left[S_{\vec{k}} \right]_{:,i}$ denotes the $i$-th column of the $S_{\vec{k}}$ matrix. Additionally, we take $A_{\vec{k}} = B_{\vec{k}} = \mathbb{0}$. The orbitals corresponding to each $\hat{a}^\dagger_{\vec{R},i}$ ($1 \leq i \leq 6$) and $\hat{b}^\dagger_{\vec{R},i}$ ($1 \leq i \leq 4$) operators are also shown explicitly in \cref{app:fig:TBG_tight_binding:a}. The PZSV model features four hopping parameters which are given (in dimensionless units) by $a=0.110$, $b=0.033$, $c=0.033$, and $d=0.573$~\cite{PO19}. Using the same conventions as in Ref.~\cite{PO19}, we have also denoted the two complex roots of unity by 
\begin{equation}
	\label{app:eqn:unit_roots}
	\zeta = e^{2\pi i/6 } \quad \text{and} \quad \omega=\zeta^2.
\end{equation}

The resulting BCL Hamiltonian has $N_{L}-N_{\tilde{L}}=2$ perfectly flat bands, as shown in \cref{app:fig:TBG_tight_binding:b}. It also features two accidental symmetries, $C_{2z}$ and $\mathcal{T}$~\cite{PO19}, in addition to the symmetries of single valley TBG, which is only symmetric under $C_{2z}\mathcal{T}$, $C_{3z}$, and $C_{2x}$ (since both $C_{2z}$ and $\mathcal{T}$ exchange valleys)~\cite{SON19,SON20b}. Ref.~\cite{PO19} breaks these two accidental symmetries by the addition of a small perturbation, which also breaks the chiral symmetry of the BCL and renders the two perfectly flat bands dispersive (though their bandwidth remains small). 

Here, we focus on the BCL \emph{with} the accidental symmetries as it features the perfectly flat bands. As such, we will analyze the model within the the $P6/221'$ group (SSG 177.150 in the notation of BCS)~\citeBCS{}. Additionally, due to the negligible spin-orbit coupling of regular TBG (and lack of spin-orbit coupling in the PZSV model), we will consider spinless orbitals within each sublattice. The BRs corresponding to the two sublattices can be found directly through induction (by referencing the BCS~\citeBCS{}; see also \cref{app:tab:char_p622} of \cref{app:sec:example:char} for the character table of the SSG)
\begin{align}
	\mathcal{BR}_{L} &= \left(\mathrm{E}_1\right)_{1a} \uparrow \mathcal{G} \oplus \left(\mathrm{A}_2\right)_{1a} \uparrow \mathcal{G} \oplus \left(\mathrm{A}_1\right)_{3f} \uparrow \mathcal{G} = \nonumber \\
		&= \left(\Gamma_1 \oplus \Gamma_2 \oplus \Gamma_5 \oplus \Gamma _6\right) + \left(\mathrm{K}_1 \oplus \mathrm{K}_2 \oplus 2 \mathrm{K}_3 \right) + \left(\mathrm{M}_1 \oplus \mathrm{M}_2 \oplus 2 \mathrm{M}_3 \oplus 2\mathrm{M}_4 \right), \\
	\mathcal{BR}_{\tilde{L}} &= \left(\mathrm{E}\right)_{2c} \uparrow \mathcal{G} = \left(\Gamma_5 \oplus \Gamma _6\right) + \left(\mathrm{K}_1 \oplus \mathrm{K}_2 \oplus \mathrm{K}_3 \right) + \left(\mathrm{M}_1 \oplus \mathrm{M}_2 \oplus \mathrm{M}_3 \oplus \mathrm{M}_4 \right). 
\end{align}
Because $\mathcal{BR}_{\tilde{L}} \subset \mathcal{BR}_{L}$, we expect the two flat bands to be gapped from the dispersive bands at the high-symmetry momentum points and lines, and by employing \cref{app:eqn:br_subtraction}, we can directly determine their (co)irrep to be
\begin{equation}
	\mathcal{B}_{\mathrm{FB}} = \left(\Gamma_1 \oplus \Gamma_2\right) + \left( \mathrm{K}_3 \right) + \left(\mathrm{M}_3 \oplus \mathrm{M}_4 \right),
\end{equation} 
which is a fragile band. As expected, this is accurately reproduced in the full band structure shown in \cref{app:fig:TBG_tight_binding:b}. 

\subsection{Flat bands in a honeycomb lattice}\label{app:sec:example:honeycombP}

\begin{figure}[!t]
\captionsetup[subfloat]{farskip=0pt}\sbox\nsubbox{
		\resizebox{\textwidth}{!}
		{\includegraphics[height=6cm]{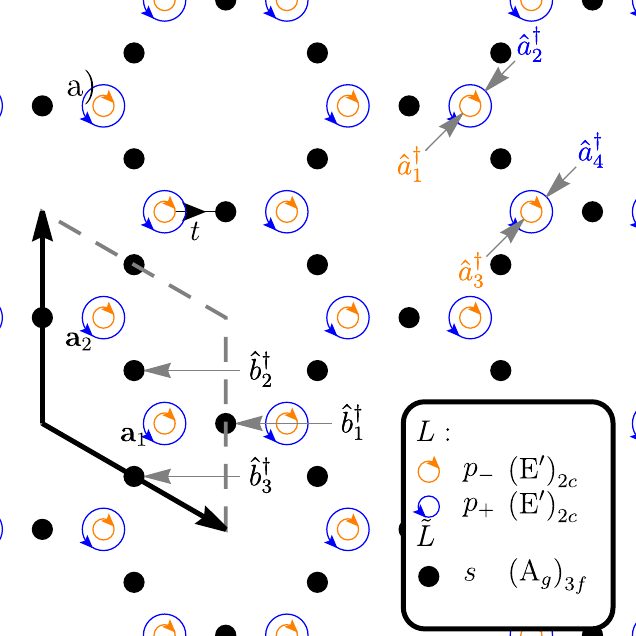}\includegraphics[height=6cm]{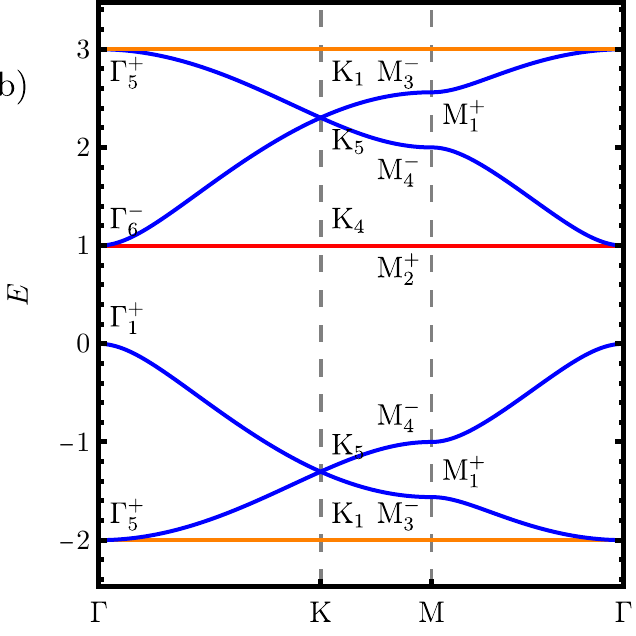}}}\setlength{\nsubht}{\ht\nsubbox}\centering\subfloat{\label{app:fig:honeycombP:a}\includegraphics[height=\nsubht]{5b.pdf}}\subfloat{\label{app:fig:honeycombP:b}\includegraphics[height=\nsubht]{5a.pdf}}
\caption{Flat bands in a BCL Hamiltonian for which the effective sublattice Hamiltonian was constructed in Ref.~\cite{WU07} in an \emph{ad-hoc} manner. We take the intra-sublattice hopping matrices to be $A_{\vec{k}}=\epsilon \mathbb{1}$, $B_{\vec{k}} = \mathbb{0}$, while the $S_{\vec{k}}$ matrix is given in \cref{app:eqn:honeycombP_S_k}. In panel a), we illustrate the lattice structure of the model. The meaning of the symbols is the same as in \cref{app:fig:Lieb_molecular:a}. By applying the symmetry operations of the $P6/mmm1'$ group (SSG 191.234 in the notation of BCS)~\citeBCS{}, the entire tight-binding Hamiltonian of the model can be recovered from the single hopping amplitude shown. Panel b) shows the band structure of the model, which features three gapless, perfectly flat bands at different energies. Here, we take $\epsilon=1$ and $t=-1$. At each high-symmetry momentum point, we also indicate the (co)irreps of the corresponding bands. The flat band shown in red is enforced by the rank of the $S_{\vec{k}}$ matrix, while the flat bands in orange stem from a $\vec{k}$-independent singular value of $S_{\vec{k}}$.}
\label{app:fig:honeycombP}
\end{figure}

Another model featuring completely flat bands was proposed in Ref.~\cite{WU07} as a means of realizing Wigner crystalline phases of matter upon the introduction of on-site Hubbard-like repulsion. The tight-binding Hamiltonian is built from spinless $p_x$ and $p_y$ orbitals on the two-dimensional honeycomb lattice and features two flat bands with distinct energies, whose origin was explained in terms of real space eigenstates with finite support~\cite{WU07,BER08}. In this appendix, we provide a rigorous justification for the two flat bands, showing by explicit construction that the model is essentially an effective BCL Hamiltonian $T_{\vec{k}}$ (up to rescaling and chemical potential terms), explaining the origin of its band touching points, and verifying our BR subtraction prescription.

We first discuss the original Hamiltonian of the model~\cite{WU07,BER08} and consider placing $p_x$ and $p_y$ orbitals at the sites of the $2c$ Wyckoff position of the hexagonal lattice (see \cref{app:fig:honeycombP:a}). Within each unit cell, there are two such lattice sites, which we label by $\mathrm{A}$ and $\mathrm{B}$, and which are located respectively at $\vec{r}_{\mathrm{A}} = \frac{1}{3} \vec{a}_1 + \frac{2}{3} \vec{a}_2$ and $\vec{r}_{\mathrm{B}} = \frac{2}{3} \vec{a}_1 + \frac{1}{3} \vec{a}_2$ (where $\vec{a}_1$ and $\vec{a}_{2}$ are the primitive lattice vectors). We denote by $\hat{p}^\dagger_{\vec{R}, \alpha, f}$ the creation operator corresponding to the fermion in unit cell $\vec{R}$, orbital $p_f$, and site $\alpha$, where $f=x,y$ and $\alpha = \mathrm{A},\mathrm{B}$. We also define three unit vectors
\begin{equation}
	\hat{\vec{e}}_{1} = \hat{\vec{e}}_{x} \qquad \text{and} \qquad
	\hat{\vec{e}}_{2,3} = -\frac{1}{2} \hat{\vec{e}}_{x} \pm \frac{\sqrt{3}}{2} \hat{\vec{e}}_{y},
\end{equation}
where $\hat{\vec{e}}_{x}$ and $\hat{\vec{e}}_{y}$ are the cartesian unit vectors. $\hat{\vec{e}}_i$ ($i=1,2,3$) correspond to the three directions that link $\mathrm{B}$ sites to neighboring $\mathrm{A}$ sites. Following Ref.~\cite{WU07}, we introduce an overcomplete set of fermionic operators
\begin{equation}
	\label{app:eqn:honeycombP_overcomplete}
	\hat{p}^\dagger_{\vec{R}, \alpha, i} = \left( \hat{p}^\dagger_{\vec{R}, \alpha, x} \hat{\vec{e}}_{x} + \hat{p}^\dagger_{\vec{R}, \alpha, y} \hat{\vec{e}}_{y} \right) \cdot \hat{\vec{e}}_{i},
\end{equation}
for $i=1,2,3$, representing the ``projections'' of the $p$ orbitals at each lattice site along the unit vectors $\hat{\vec{e}}_i$. For each bond between two nearest-neighboring honeycomb sites, we will couple the corresponding $p$ orbital projections \emph{along} the direction of the bond leading to the following quadratic Hamiltonian~\cite{WU07,BER08}
\begin{equation}
	\label{app:eqn:honeycombP_ham}
	\mathcal{H}_{0}=t_{\parallel} \sum_{\vec{R}} \sum_{i=1}^{3} \left[ \hat{p}^\dagger_{\vec{R} + \vec{r}_{\mathrm{B}} - \vec{r}_{\mathrm{A}} + \lambda \hat{\vec{e}}_{i},\mathrm{A},i} \hat{p}_{\vec{R},\mathrm{B},i } + \mathrm{h.c.} \right] - \mu \sum_{\vec{R}} \sum_{\substack{\alpha=A,B \\f=x,y}} \hat{p}^\dagger_{\vec{R},\alpha,f}\hat{p}_{\vec{R},\alpha,f},
\end{equation} 
where $\lambda = 1/\sqrt{3}$ is the nearest-neighbor distance between two honeycomb lattice sites, $\mu$ is the chemical potential, and $t_{\parallel}$ is the $\sigma$-bonding term describing the hopping between neighboring $p$ orbitals. The symmetry group of the Hamiltonian $\mathcal{H}_{0}$ is generated by two-dimensional translations, $C_{6z}$, $C_{2x}$, $\mathcal{I}$, and $\mathcal{T}$. Thus, we will analyze this example within the $P6/mmm1'$ group (SSG 191.234 in the notation of BCS)~\citeBCS{}. 

The spectrum of $\mathcal{H}_{0}$ was discussed in Refs.~\cite{WU07,BER08} and was shown to feature two gapless flat bands located at energies $\pm \frac{3}{2} t_{\parallel} - \mu$. Here, we will first show that $\mathcal{H}_{0}$ is essentially an effective BCL Hamiltonian $\mathcal{T}_{\mathrm{eff}}$ from \cref{app:eqn:eff_Ham_L} (up to rescaling and chemical potential terms). As such, we will not study the Hamiltonian from \cref{app:eqn:honeycombP_ham} directly, but instead construct the corresponding BCL Hamiltonian $\mathcal{H}$ and recover all its properties from there. 

To begin with, we note that the presence of the projected $\sigma$-bonding between neighboring $p$ orbitals is reminiscent of a virtual second-order hopping process, as will be explained below. Indeed, we can imagine starting from uncoupled $p_x$ and $p_y$ orbitals at the $2c$ Wyckoff position of the hexagonal lattice. We then place $s$ orbitals at the $3f$ Wyckoff position (\ie in the middle of the bond between any two nearest-neighbor honeycomb sites) and introduce hopping between nearest-neighbor $p$ and $s$ orbitals. For any pair of nearest-neighbor $s$ and $p$ orbitals, the only symmetry-allowed hopping is between the $s$ orbital and the $p$ orbital projection \emph{along} the $s$-$p$ bond. If the $s$ orbitals are ``integrated'' out, then one obtains an effective BCL Hamiltonian featuring only the projected $\sigma$-bonding between neighboring $p$ orbitals from \cref{app:eqn:honeycombP_ham}.

To put these ideas on rigorous grounds, we introduce the following fermionic operators
\begin{equation}
	\hat{p}^\dagger_{\vec{R},\alpha,\pm} = \frac{1}{\sqrt{2}} \left( \hat{p}^\dagger_{\vec{R},\alpha,x} \pm i \hat{p}^\dagger_{\vec{R},\alpha,y} \right),
\end{equation}
which allow us to rewrite the overcomplete set of orbital projection operators from \cref{app:eqn:honeycombP_overcomplete} as 
\begin{equation}
	\label{app:eqn:honeycombP_overcomplete_rotation}
	\hat{p}^\dagger_{\vec{R},\alpha,j} = \frac{1}{\sqrt{2}} \left[ \left(\omega^{*}\right)^{j-1} \hat{p}^\dagger_{\vec{R},\alpha,+} + \omega^{j-1} \hat{p}^\dagger_{\vec{R},\alpha,-} \right],
\end{equation}
where $\omega$ is the complex root of unity defined in \cref{app:eqn:unit_roots} and $j=1,2,3$. In constructing the BCL Hamiltonian, we employ the notation from \cref{app:sec:bipartite_crystalline_lattices:notation} and define the creation operators on the $L$ sublattice to be
\begin{alignat}{2}
	&\hat{a}^\dagger_{\vec{R},1} = \hat{p}^\dagger_{\vec{R},\mathrm{A},-}, \qquad && \hat{a}^\dagger_{\vec{R},2} = \hat{p}^\dagger_{\vec{R},\mathrm{A},+}, \nonumber \\
	&\hat{a}^\dagger_{\vec{R},3} = \hat{p}^\dagger_{\vec{R},\mathrm{B},-}, \qquad && \hat{a}^\dagger_{\vec{R},4} = \hat{p}^\dagger_{\vec{R},\mathrm{B},+}, \label{app:eqn:honeycombP_ops}
\end{alignat}
with the corresponding displacement vectors given by $\vec{r}^{\hat{a}}_{1}=\vec{r}^{\hat{a}}_{2} = \vec{r}_{\mathrm{A}}$ and $\vec{r}^{\hat{a}}_{3}=\vec{r}^{\hat{a}}_{4} = \vec{r}_{\mathrm{B}}$, as shown in \cref{app:fig:honeycombP:a}. Inside the $\tilde{L}$ sublattice, we place $s$ orbitals at the $3f$ Wyckoff position and denote by $\hat{b}^\dagger_{\vec{R},i}$ the creation operator corresponding to the $s$ orbital in unit cell $\vec{R}$ and position $\vec{r}^{\hat{b}}_{i} = \vec{r}_{\vec{B}} + \frac{\lambda}{2} \hat{\vec{e}}_i$ within the unit cell. We now define the BCL Hamiltonian of the model to be given by $\mathcal{H}$ from \cref{app:eqn:gen_bcl_ham_rs} with $A_{\vec{R}}= \epsilon \delta_{\vec{R},\vec{0}} \mathbb{1} $, $B_{\vec{R}} = \mathbb{0}$, and 
\begin{equation}
	\label{app:eqn:honeycombP_S}
	S_{\vec{R}} = t \begin{pmatrix}
		-\delta_{\vec{R}+\vec{r}^{\hat{a}}_{1} - \frac{\lambda}{2} \hat{\vec{e}}_1,\vec{r}^{\hat{b}}_{1}} &
		-\omega \delta_{\vec{R}+\vec{r}^{\hat{a}}_{1} - \frac{\lambda}{2} \hat{\vec{e}}_2,\vec{r}^{\hat{b}}_{2}} &
		-\omega^* \delta_{\vec{R}+\vec{r}^{\hat{a}}_{1} - \frac{\lambda}{2} \hat{\vec{e}}_3,\vec{r}^{\hat{b}}_{3}} \\
		-\delta_{\vec{R}+\vec{r}^{\hat{a}}_{2} - \frac{\lambda}{2} \hat{\vec{e}}_1,\vec{r}^{\hat{b}}_{1}}  &		 
		-\omega^* \delta_{\vec{R}+\vec{r}^{\hat{a}}_{2} - \frac{\lambda}{2} \hat{\vec{e}}_2,\vec{r}^{\hat{b}}_{2}} &
		-\omega \delta_{\vec{R}+\vec{r}^{\hat{a}}_{2} - \frac{\lambda}{2} \hat{\vec{e}}_3,\vec{r}^{\hat{b}}_{3}} \\
		\delta_{\vec{R}+\vec{r}^{\hat{a}}_{3} + \frac{\lambda}{2} \hat{\vec{e}}_1,\vec{r}^{\hat{b}}_{1}} & 
		\omega \delta_{\vec{R}+\vec{r}^{\hat{a}}_{3} + \frac{\lambda}{2} \hat{\vec{e}}_2,\vec{r}^{\hat{b}}_{2}} & 
		\omega^* \delta_{\vec{R}+\vec{r}^{\hat{a}}_{3} + \frac{\lambda}{2} \hat{\vec{e}}_3,\vec{r}^{\hat{b}}_{3}} \\
		\delta_{\vec{R}+\vec{r}^{\hat{a}}_{4} + \frac{\lambda}{2} \hat{\vec{e}}_1,\vec{r}^{\hat{b}}_{1}} &
		\omega^* \delta_{\vec{R}+\vec{r}^{\hat{a}}_{4} + \frac{\lambda}{2} \hat{\vec{e}}_2,\vec{r}^{\hat{b}}_{2}} &
		\omega \delta_{\vec{R}+\vec{r}^{\hat{a}}_{4} + \frac{\lambda}{2} \hat{\vec{e}}_3,\vec{r}^{\hat{b}}_{3}} \\
	\end{pmatrix},
\end{equation}
where $t$ is a real hopping parameter and $\epsilon$ is the on-site energy of the of the $p$ orbitals. In momentum space, the inter-sublattice hopping matrix reads
\begin{equation}
	S_{\vec{k}}=t\begin{pmatrix}
 -e^{\frac{i k_x}{2 \sqrt{3}}} & \omega  \left(-e^{-\frac{1}{12} i \left(\sqrt{3} k_x-3 k_y\right)}\right) & \omega ^* \left(-e^{-\frac{1}{12} i \left(\sqrt{3} k_x+3 k_y\right)}\right) \\
 -e^{\frac{i k_x}{2 \sqrt{3}}} & \omega ^* \left(-e^{-\frac{1}{12} i \left(\sqrt{3} k_x-3 k_y\right)}\right) & \omega  \left(-e^{-\frac{1}{12} i \left(\sqrt{3} k_x+3 k_y\right)}\right) \\
 e^{-\frac{i k_x}{2 \sqrt{3}}} & \omega  e^{\frac{1}{12} i \left(\sqrt{3} k_x-3 k_y\right)} & \omega ^* e^{\frac{1}{12} i \left(\sqrt{3} k_x+3 k_y\right)} \\
 e^{-\frac{i k_x}{2 \sqrt{3}}} & \omega ^* e^{\frac{1}{12} i \left(\sqrt{3} k_x-3 k_y\right)} & \omega  e^{\frac{1}{12} i \left(\sqrt{3} k_x+3 k_y\right)} \\
\end{pmatrix} . \label{app:eqn:honeycombP_S_k}
\end{equation}
in the basis of \cref{app:eqn:general_bas_spinor}. Using \cref{app:eqn:T_deff}, we can derive the effective Hamiltonian matrix 
\begin{equation}
	\label{app:eqn:honeycombP_T}
	T_{\vec{R}} = -t^{2} \sum_{j=1}^{3} \left[ \begin{pmatrix}
		 0 & 0 & 1 & \omega^{2 \left( j-1 \right)}  \\
		 0 & 0 & \left(\omega^*\right)^{2 \left( j-1 \right)}  & 1 \\
		 1 & \omega^{2 \left( j-1 \right)} & 0 & 0 \\
		  \left(\omega^* \right)^{2 \left( j-1 \right)} & 1 & 0 & 0\\
	\end{pmatrix} 
	\delta_{\vec{R}+\vec{r}_{\mathrm{A}}-\vec{r}_{\mathrm{B}}, \lambda \hat{\vec{e}}_j} + 
	\delta_{\vec{R},\vec{0}} \mathbb{1} 
	\right].
\end{equation}
On the other hand, by employing \cref{app:eqn:honeycombP_overcomplete_rotation,app:eqn:honeycombP_T}, we can rewrite the Hamiltonian $\mathcal{H}_0$ from \cref{app:eqn:honeycombP_ham} in terms of the creation operators on the $L$ lattice as 
\begin{equation}
	\mathcal{H}_{0} = - \frac{t_{\parallel}}{2 t^2} \sum_{\substack{\vec{R}_1, \vec{R}_2 \\ 1 \leq i,j \leq N_L}}  \left[T_{\vec{R}_1 - \vec{R}_2}\right]_{ij}  \hat{a}^\dagger_{\vec{R}_1,i} \hat{a}_{\vec{R}_2,j} - \sum_{\vec{R},i} \left( \frac{3 t_{\parallel}}{2}+ \mu \right) \hat{a}^\dagger_{\vec{R},i} \hat{a}_{\vec{R},i},
\end{equation}
thus proving our initial assertion that, up to rescaling and chemical potential terms, $\mathcal{H}_0$ is equivalent to the effective Hamiltonian derived from a BCL model Hamiltonian $\mathcal{H}$ ($N_{L} = 4$, $N_{\tilde{L}}=3$) with $A_{\vec{R}}= \epsilon \delta_{\vec{R},\vec{0}} \mathbb{1}$, $B_{\vec{R}} = \mathbb{0}$, and the $S_{\vec{R}}$ matrix from \cref{app:eqn:honeycombP_S}. We will now analyze the spectrum of $\mathcal{H}$ within the $P6/mmm1'$ group (SSG 191.234 in the notation of BCS)~\citeBCS{}, and explain the origin and topology of the three flat bands. 

The band structure of $\mathcal{H}$ is shown in \cref{app:fig:honeycombP:b} for $\epsilon \neq 0$ and features three distinct flat bands at different energies. As expected, there is exactly $N_{L}-N_{\tilde{L}} = 1$ flat band pinned at energy $\epsilon$. Using the BRs corresponding to the two sublattices (see \cref{app:tab:char_p6mmm} of \cref{app:sec:example:char} for the character table of the SSG)~\citeBCS{} ,
\begin{align}
	\mathcal{BR}_{L} &= \left(\mathrm{E}'\right)_{2c} \uparrow \mathcal{G} = \left(\Gamma_{5}^{+} \oplus \Gamma_{6}^{-} \right) + \left(\mathrm{K}_1 \oplus \mathrm{K}_4 \oplus \mathrm{K}_5 \right) + \left(\mathrm{M}_{1}^{+} \oplus \mathrm{M}_{2}^{+} \oplus \mathrm{M}_{3}^{-} \oplus \mathrm{M}_{4}^{-} \right), \\
	\mathcal{BR}_{\tilde{L}} &= \left(\mathrm{A}_{g}\right)_{3f} \uparrow \mathcal{G} = \left(\Gamma_{1}^{+} \oplus \Gamma _{5}^{+} \right) + \left(\mathrm{K}_1 \oplus \mathrm{K}_5 \right) + \left(\mathrm{M}_{1}^{+} \oplus \mathrm{M}_{3}^{-} \oplus \mathrm{M}_{4}^{-} \right), 
\end{align}
where the two-dimensional (co)irrep $\left(\mathrm{E}'\right)_{2c}$ corresponds to a $p_x$ and $p_y$ orbitals at the $2c$ Wyckoff position, we can determine the (co)irrep of the flat band from \cref{app:eqn:br_subtraction} to be 
\begin{equation}
	\label{app:eqn:honeycombP_BR}
	\mathcal{B}_{\mathrm{FB}} = \left( \Gamma_{6}^{-} \boxminus \Gamma _{1}^{+} \right) + \left(\mathrm{K}_4 \right) + \left( \mathrm{M}_{2}^{+} \right).
\end{equation}
At the $\Gamma$ point, the presence of a formal (co)irrep difference signals a band touching point between the flat and dispersive bands. 

In addition to the flat band stemming from the kernel of $S^{\dagger}_{\vec{k}}$, \cref{app:fig:honeycombP:b} shows the presence of two other flat bands whose wave functions are of the form in \cref{app:eqn:eig_dispersive_mu}. They originate from the fact that $S_{\vec{k}}$ has a $\vec{k}$-independent nonzero singular value. The additional flat bands of $\mathcal{H}$ would become dispersive should arbitrary intra-sublattice hopping be introduced in the $\tilde{L}$ sublattice. To better understand their origin, we can employ \cref{app:eqn:Ttilde_deff} and construct the effective Hamiltonian defined on the \emph{smaller} sublattice $\tilde{L}$ (which has $N_{\tilde{L}} = 3$)
\begin{equation}
	\label{app:eqn:honeycombP_Ttilde}
	\left[\tilde{T}_{\vec{R}} \right]_{ij} = t^2 \left[ \delta_{\vec{R}, \lambda\left( \hat{\vec{e}}_j -\hat{\vec{e}}_i \right)} + \delta_{\vec{R},\vec{0}} +4 \delta_{\vec{R},\vec{0}} \delta_{ij} \right].
\end{equation}
Up to rescaling and chemical potential terms, the effective Hamiltonian from \cref{app:eqn:honeycombP_Ttilde} is the Hamiltonian for $s$ orbitals on Kagom\'e lattice with nearest-neighbor real hopping. As such, $\tilde{T}_{\vec{k}}$ features a gapless flat band at energy $ 4t^2 $ whose (co)irreps are given by~\citeBCS{} 
\begin{equation}
	\label{app:eqn:honeycombP_otherBR}
	\mathcal{B}_{\mathrm{FB}'} = \left[\left(\mathrm{A}_{g}\right)_{3f} \uparrow \mathcal{G}\right] \boxminus \left[ \left(\mathrm{A}'_1\right)_{2c} \uparrow \mathcal{G} \right] = \left( \Gamma_{5}^{+} \boxminus \Gamma _{4}^{-} \right) + \left(\mathrm{K}_1 \right) + \left( \mathrm{M}_{3}^{-} \right).
\end{equation}
This implies that the $S_{\vec{k}}$ matrix has a $\vec{k}$-independent singular value $2 t$, which gives rise to flat bands of $H_{\vec{k}}$ with energies $\epsilon \pm 2 t$, in line with \cref{app:eqn:eig_dispersive_mu}. As illustrated in \cref{app:tab:relation_between_eigenstates} of \cref{app:sec:unitary_sym}, the (co)irreps of each of the bands of $H_{\vec{k}}$ with energy $\epsilon \pm 2 t$, must be the same as the (co)irreps of the band of $\tilde{T}_{\vec{k}}$ with energy $4t^2$, the latter of which were given in \cref{app:eqn:honeycombP_otherBR}. This is confirmed in the band structure presented in \cref{app:fig:honeycombP:b}.

In summary, $H_{\vec{k}}$ features three gapless flat bands with different energies, as shown in \cref{app:fig:honeycombP:b}: one flat band at energy $\epsilon$ (stemming from the kernel of $S^{\dagger}_{\vec{k}}$), and two flat bands with energies $\epsilon \pm 2t$ (stemming from a constant singular value of the $S_{\vec{k}}$ matrix). The (co)irreps of the flat band at $\epsilon$ can be directly computed using our BR subtraction scheme, as was done in \cref{app:eqn:honeycombP_BR}. The flat bands at $\epsilon \pm 2t$ can be related to the gapless flat band arising in $\tilde{T}_{\vec{k}}$ (which is the Hamiltonian of a Kagom\'e lattice of $s$ orbitals and nearest neighbor hopping) at energy $4t^2$. Their (co)irreps can be worked out indirectly from the \cref{app:eqn:honeycombP_otherBR}, which also diagnoses the band touching point enforced at $\Gamma$.

\subsection{Flat bands with globally stable band touching points}\label{app:sec:example:C2T}

\begin{figure}[!t]
\captionsetup[subfloat]{farskip=0pt}\sbox\nsubbox{
		\resizebox{\textwidth}{!}
		{\includegraphics[height=6cm]{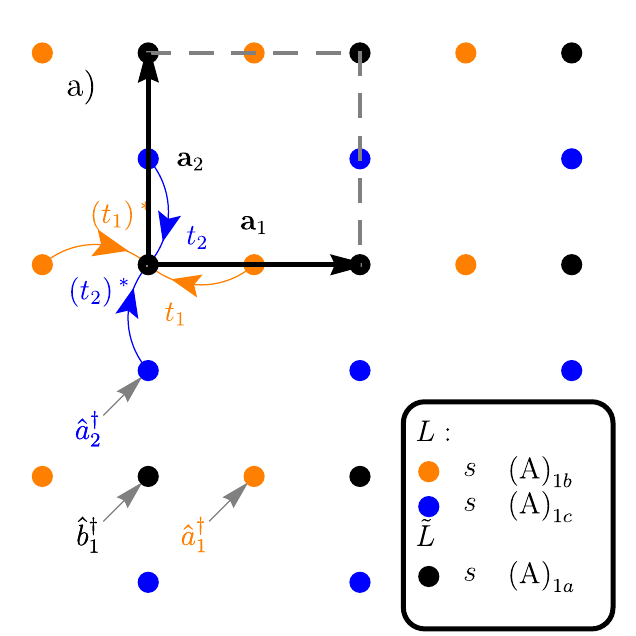}\includegraphics[height=6cm]{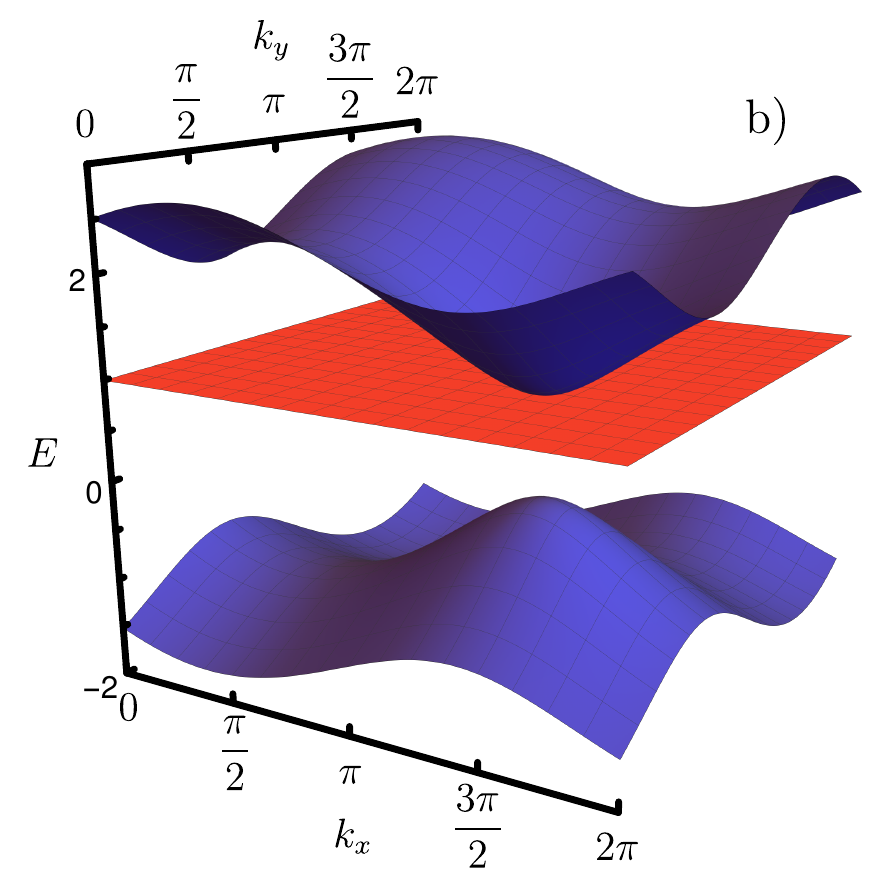}}}\setlength{\nsubht}{\ht\nsubbox}\centering\subfloat{\label{app:fig:C2TBTP_spectrum:a}\includegraphics[height=\nsubht]{6b.pdf}}\subfloat{\label{app:fig:C2TBTP_spectrum:b}\includegraphics[height=\nsubht]{6a.pdf}}
\caption{Flat bands in a BCL model with globally stable band touching points. We take the intra-sublattice hopping matrices to be $A_{\vec{k}}=\epsilon \mathbb{1}$, $B_{\vec{k}} = \mathbb{0}$, while the $S_{\vec{k}}$ matrix is given in \cref{app:eqn:c2zt_S}. In panel a), we illustrate the lattice structure of the model. The meaning of the symbols is the same as in \cref{app:fig:TBG_tight_binding:a}. By applying two-dimensional translation operations along $\vec{a}_1$ and $\vec{a}_2$, the entire $S_{\vec{k}}$ matrix can be recovered from the four hopping processes shown. Panel b) shows the band structure of the model inside the two-dimensional Brillouin zone, which features one gapless perfectly flat band (shown in red) and two dispersive bands (shown in blue). Here, we consider $t_{1} = 0.647 - 0.470 i$, $t_{2} = -0.728 + 0.529 i$, and $\epsilon=1$. The quadratic band touching point between the flat and the dispersive bands is globally enforced by the $C_{2z} \mathcal{T}$ symmetry and cannot be removed even when arbitrarily long-ranged, $C_{2z} \mathcal{T}$-symmetric inter-sublattice hopping processes (with finite support) are included.}
\label{app:fig:C2TBTP_spectrum}
\end{figure}

In the main paper, we argued that if the co(irreps) of $\mathcal{BR}_{\tilde{L}}$ form a subset of the (co)irreps of $\mathcal{BR}_{L}$, then there can be no locally stable band touching points between the flat and dispersive BCL bands at a given momentum point. In this appendix, we show that one can still have \emph{globally} stable band touching points, even if this is the case and even if all compatibility relations are automatically satisfied: while their positions can be shifted in $\vec{k}$-space by adding perturbations to the $S_{\vec{k}}$ matrix, the flat bands cannot be completely gapped from the dispersive ones.

As a simple example, we consider a modified (spinless) Lieb lattice Hamiltonian in which we break all symmetries with the exception of $C_{2z} \mathcal{T}$. As shown in \cref{app:fig:C2TBTP_spectrum:a}, we take the $L$ lattice to be formed of $s$ orbitals at the $1b$ and $1c$ Wyckoff positions of a two-dimensional square lattice, while the $\tilde{L}$ lattice contains $s$ orbitals at the $1a$ position. We now define the BCL Hamiltonian of the model to be given by $\mathcal{H}$ from \cref{app:eqn:gen_bcl_ham_rs} with $A_{\vec{k}}= \epsilon \mathbb{1} $, $B_{\vec{k}} = \mathbb{0}$, and 
\begin{equation}
	\label{app:eqn:c2zt_S}
	S_{\vec{k}} = \begin{pmatrix}
		2 \abs{t_1} \cos\left( \frac{ k_x}{2}+\theta_1 \right) \\
		2 \abs{t_2} \cos\left( \frac{ k_y}{2}+\theta_2 \right) 
	\end{pmatrix},
\end{equation}
where $t_1 = \abs{t_1}e^{i \theta_1}$ and $t_2 = \abs{t_2}e^{i \theta_2}$ represent arbitrary nearest-neighbor hopping parameters (as shown in \cref{app:fig:C2TBTP_spectrum:a}). The basis for the Hamiltonian is given in \cref{app:eqn:general_bas_spinor}. Regardless of the choice of hopping amplitudes, there will always be a point in the Brillouin zone where the rank of $S_{\vec{k}}$ is zero: $\left(k_x,k_y \right)=\left(\pi - 2 \theta_1, \pi - 2\theta_2 \right)$. As discussed in \cref{app:sec:breaking_chiral}, this indicates the existence of a quadratic band touching point between the flat and dispersive bands (even if $B_{\vec{k}} \neq \mathbb{0}$). On the other hand, the symmetry group of the system is given by the $P2'$ group (SSG 3.3 in the notation of BCS)~\citeBCS{}, which only features a single spinless (co)irrep at every single momentum point inside the Brillouin zone. As such, we trivially have $\mathcal{BR}_{\tilde{L}} \subset \mathcal{BR}_{L}$ and no \emph{locally} stable band touching points between the flat and dispersive bands. 

One can show that the globally stable band touching point survives the addition of arbitrarily long-ranged inter-sublattice hopping (but with finite support), as long as the $C_{2z}\mathcal{T}$ symmetry of $\mathcal{H}$ is not broken. A heuristic argument for this can be given in real space: in order to produce gapped flat bands, one would need to hybridize the $s$ orbitals at the $1b$ and $1c$ Wyckoff positions from sublattice $L$ (see \cref{app:fig:C2TBTP_spectrum:a}) into an $s$ and a $p$ orbital at the $1a$ position. This new $s$ orbital could then gap the $s$ orbital from the $\tilde{L}$ lattice giving rise to a gapped flat band. However, the $s$ orbitals from the $L$ sublattice have their positions pinned by $C_{2z}\mathcal{T}$ symmetry, and so no continuous $C_{2z} \mathcal{T}$-preserving deformations of the Hamiltonian can be made that could bring the $s$ orbitals in the $L$ sublattice at the $1a$ position.  

For a more mathematical proof, we consider the action of the (anti-unitary) $C_{2z}\mathcal{T}$ symmetry on the BCL fermions
\begin{equation}
	\begin{split}
		\left(C_{2z} \mathcal{T} \right) \hat{a}^\dagger_{\vec{k},i} \left(C_{2z} \mathcal{T} \right)^{-1} &= \hat{a}^\dagger_{\vec{k},i}, \quad \text{for} \quad i=1,2, \\
		\left(C_{2z} \mathcal{T} \right) \hat{b}^\dagger_{\vec{k},1} \left(C_{2z} \mathcal{T} \right)^{-1} &= \hat{b}^\dagger_{\vec{k},1}.
	\end{split}
\end{equation}
Since $C_{2z}\mathcal{T}$ is an anti-unitary symmetry of $\mathcal{H}$ (\ie $\left[ C_{2z}\mathcal{T},\mathcal{H}\right] = 0$), it follows that
\begin{equation}
    \label{app:eqn:c2zt_real_matrix_H}
    H_{\vec{k}} = H^{*}_{\vec{k}},
\end{equation}
which implies that 
\begin{equation}
    \label{app:eqn:c2zt_real_matrix_S}
    S_{\vec{k}} = S^{*}_{\vec{k}}.
\end{equation}

The BCL fermion operators also obey the following embedding relations
\begin{equation}
	\label{app:eqn:embedding_c2zt}
	\hat{a}_{\vec{k} + \vec{G},i} = \sum_{j=1}^{N_{L}} \left[V_{\vec{G}}\right]_{ij} \hat{a}_{\vec{k},j}, \qquad
	\hat{b}_{\vec{k} + \vec{G},i} = \sum_{j=1}^{N_{\tilde{L}}} \left[\tilde{V}_{\vec{G}}\right]_{ij} \hat{b}_{\vec{k},j},
\end{equation}
where $\vec{G}$ is any reciprocal lattice vector of the BCL, while the embedding matrices $V_{\vec{G}}$ and $\tilde{V}_{\vec{G}}$ are given by
\begin{equation}
	\left[ V_{\vec{G}} \right]_{mn} = \delta_{mn} e^{- i\vec{G} \cdot \vec{r}^{\hat{a}}_m} ,\qquad
	\left[ \tilde{V}_{\vec{G}} \right]_{mn} = \delta_{mn} e^{- i\vec{G} \cdot \vec{r}^{\hat{b}}_m}.
\end{equation}
with the displacement vectors $\vec{r}^{\hat{a}}_m$ and $\vec{r}^{\hat{b}}_m$ being defined in the text surrounding \cref{app:eqn:momentum_ops} as the displacement of the BCL fermions from the unit cell origins. \Cref{app:eqn:embedding_c2zt} imposes a gauge condition on the inter-sublattice hopping matrix
\begin{equation}
	S_{\vec{k}+\vec{G}} = V^{\dagger}_{\vec{G}} S_{\vec{k}} \tilde{V}_{\vec{G}},
\end{equation}
which requires that 
\begin{equation}
	\label{app:eqn:embeddingPeriodicity}
	S_{\vec{k}+ \vec{G}_{1}} = \begin{pmatrix}
		-\left[ S_{\vec{k}} \right]_{11} \\
		 \left[ S_{\vec{k}} \right]_{21}
	\end{pmatrix} \qquad \text{and} \qquad
	S_{\vec{k}+ \vec{G}_{2}} = \begin{pmatrix}
		\left[ S_{\vec{k}} \right]_{11} \\
		 -\left[ S_{\vec{k}} \right]_{21}
	\end{pmatrix},	 
\end{equation}
where $\vec{G}_1 = \left(2 \pi, 0 \right)$ and $\vec{G}_2 = \left(0, 2 \pi \right)$ are the primitive reciprocal lattice vectors of the BCL. We will now prove that the reality of $S_{\vec{k}}$ from \cref{app:eqn:c2zt_real_matrix_S} together with the gauge condition in \cref{app:eqn:embeddingPeriodicity} are enough to guarantee the existence of a band touching point, for any translation invariant inter-sublattice hopping matrix obeying $C_{2z}\mathcal{T}$. 

\begin{figure}[!t]
\captionsetup[subfloat]{farskip=0pt}\sbox\nsubbox{
		\resizebox{0.666\textwidth}{!}
		{\includegraphics[height=6cm]{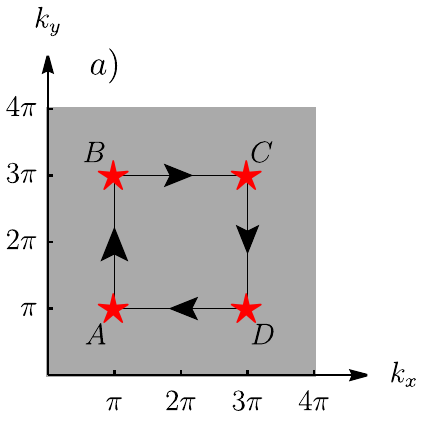}\includegraphics[height=6cm]{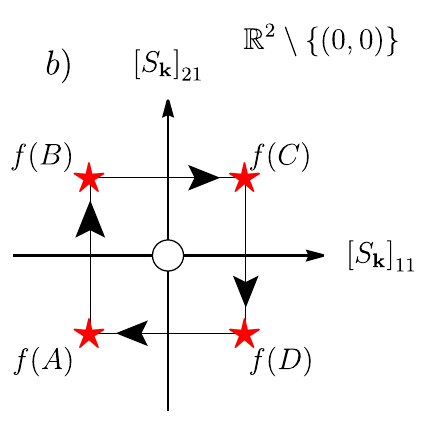}}}\setlength{\nsubht}{\ht\nsubbox}\centering\subfloat{\label{app:fig:C2TBTP_explain:a}\includegraphics[height=\nsubht]{7b.pdf}}\subfloat{\label{app:fig:C2TBTP_explain:b}\includegraphics[height=\nsubht]{7a.pdf}}
\caption{Mapping a closed path from the Brillouin zone to the $S_{\vec{k}}$-matrix space. Panel a) shows a path starting at a point $A = \left( k'_x, k'_y \right)$ and then encircling the Brillouin zone along the path $A \rightarrow B \rightarrow C \rightarrow D \rightarrow A$, where $B=\left( k'_x, k'_y +2 \pi \right)$, $C=\left( k'_x + 2 \pi, k'_y + 2 \pi \right)$, and $D=\left( k'_x + 2 \pi, k'_y \right)$. Assuming that $S_{\vec{k}}$ has no zeroes, a mapping $f : \left[0, 4\pi  \right) \times \left[0, 4\pi  \right) \rightarrow \left(\mathbb{R}^2 \setminus \left\lbrace\left(0,0 \right) \right\rbrace \right)$ must exist between the Brillouin zone and the $S_{\vec{k}}$-matrix space. We find that the corresponding path in the $S_{\vec{k}}$-matrix space (shown in panel b) must be topologically non-trivial (as it encircles the origin an odd number of times). Because the path in the Brillouin zone is homotopic to a point while its image in the $S_{\vec{k}}$-matrix space is not, the mapping $f$ cannot exist as defined, implying that $S_{\vec{k}}$ has at least one zero.}
\label{app:fig:C2TBTP_explain}
\end{figure}

Mathematically, the inter-sublattice hopping matrix $S_{\vec{k}}$ is a mapping from the two-dimensional Brillouin zone $\mathbb{T}^2$ to $\mathbb{R}^2$ (with elements $\left[ S_{\vec{k}} \right]_{11}$ and $\left[ S_{\vec{k}} \right]_{21}$). Moreover, since we only consider intersublattice hopping with finite support, this mapping is necessarily smooth. The embedding factors from \cref{app:eqn:embedding_c2zt} make the mapping $4 \pi$-periodic, so we expand the Brillouin zone (\ie the domain of $S_{\vec{k}}$), such that the mapping is strictly periodic. As such, we define a function $f : \left[0, 4\pi  \right) \times \left[0, 4\pi  \right) \rightarrow \mathbb{R}^2$ which obeys $f \left( k_x,k_y \right) = S_{\vec{k}}$. If $\mathcal{H}$ features no band touching points between the dispersive and flat bands, then the image of $f$ in the target space $\mathbb{R}^2$ does not contain the origin and its definition can be changed to $f : \left[0, 4\pi  \right) \times \left[0, 4\pi  \right) \rightarrow \left(\mathbb{R}^2 \setminus \left\lbrace\left(0,0 \right) \right\rbrace \right)$ (as if $S^T= \left(0,0 \right)$, a band touching point is enforced). We now chose a point $A = \left(k'_x,k'_y \right)$ in the Brillouin zone and denote its translation by $2\pi$ in the $\vec{G}_1$, $\vec{G}_1 + \vec{G}_2$, $\vec{G}_2$ directions by $B$, $C$, and $D$, respectively (see \cref{app:fig:C2TBTP_explain:a}). As required by the embedding constraint from \cref{app:eqn:embeddingPeriodicity}, 
\begin{alignat}{4}
	& A:&& f(k'_x, k'_y) = (r,s), \qquad && B:&& f(k'_x, k'_y + 2 \pi) = (r,-s), \nonumber \\
	& C:&& f(k'_x + 2 \pi, k'_y + 2 \pi) = (-r,-s), \qquad && D:&& f(k'_x + 2 \pi, k'_y) = (-r,s),  
\end{alignat}
where $r,s \in \mathbb{R}$. The image of a closed path around the edges of the unexpanded Brillouin zone (shown in \cref{app:fig:C2TBTP_explain:a}), $f(A) \rightarrow f(B) \rightarrow f(C) \rightarrow f(D) \rightarrow f(A)$ (shown in \cref{app:fig:C2TBTP_explain:b}) is forced to be topologically nontrivial: as the path $A \rightarrow B \rightarrow C \rightarrow D \rightarrow A$ is traversed, in its target space, $f$ winds from $(r,s) \rightarrow (r,-s) \rightarrow (-r,-s) \rightarrow (-r,s) \rightarrow (r,s)$. Up to homotopy, a path connecting two points in $\mathbb{R}^2 \setminus \left\lbrace\left(0,0 \right) \right\rbrace$ can be deformed to a straight line plus $nI$, with $I$ a clockwise winding about the origin and $n \in \mathbb{Z}$. Because of the ``reflection symmetry'' enforced by \cref{app:eqn:embeddingPeriodicity} that sends $\left[ S_{\vec{k}} \right]_{21} \rightarrow -\left[ S_{\vec{k}} \right]_{21}$, any windings $nI$ along the path $f(A) \rightarrow f(B)$ are canceled by the reflection $f(C) \rightarrow f(D)$, which gives $-nI$ and cancels the windings.  The same argument holds for the paths $f(B) \rightarrow f(C)$ and $f(D) \rightarrow f(A)$. Thus, the map $f$ winds once around the origin in the target space: the path $(r,s) \rightarrow (r,-s) \rightarrow (-r,-s) \rightarrow (-r,s) \rightarrow (r,s)$ encircles the origin once and any additional windings along the paths are canceled by the ``reflection symmetry'' in the $\mathbb{R}^2 \setminus \left\lbrace\left(0,0 \right) \right\rbrace$ target space. This is a contradiction: in momentum space, the path $A \rightarrow B \rightarrow C \rightarrow D \rightarrow A$ is homotopic to a point, but in $S_{\vec{k}}$-matrix space, the target path is not. Hence there must be a point inside the Brillouin zone where the $S_{\vec{k}}$ matrix crosses the origin: a band touching point.

\subsection{Flat bands protected by particle-hole symmetry}\label{app:sec:example:PH_flatBands}
\begin{figure}[!t]
\captionsetup[subfloat]{farskip=0pt}\sbox\nsubbox{
		\resizebox{\textwidth}{!}
		{\includegraphics[height=6cm]{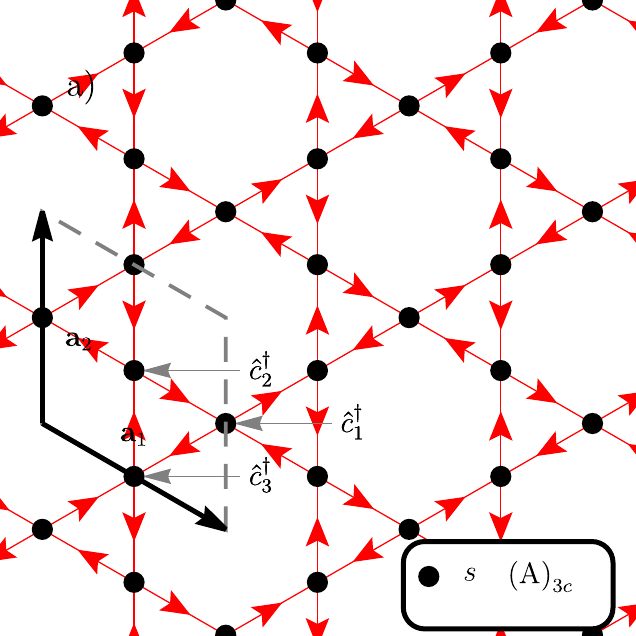}\includegraphics[height=6cm]{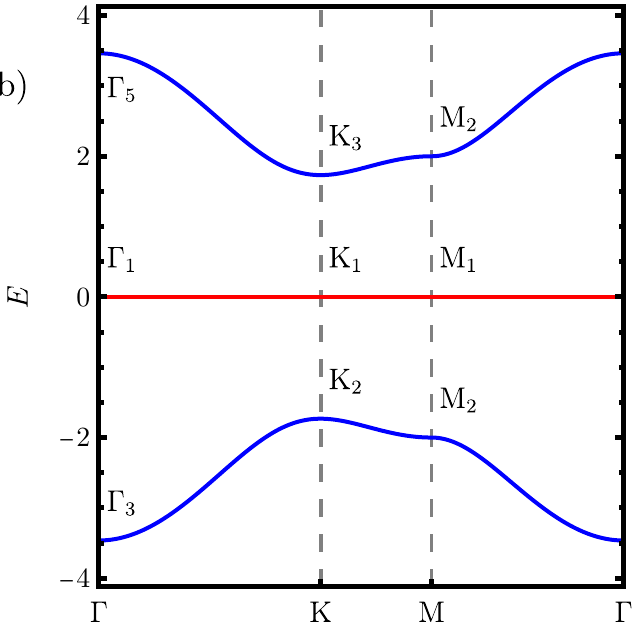}}}\setlength{\nsubht}{\ht\nsubbox}\centering\subfloat{\label{app:fig:C2P_original:a}\includegraphics[height=\nsubht]{8b.pdf}}\subfloat{\label{app:fig:C2P_original:b}\includegraphics[height=\nsubht]{8a.pdf}}
\caption{Non-degenerate flat bands protected by $C_{2z} \mathcal{P}$. In panel a), we illustrate the lattice structure of the model. The unit cell, spanned by the lattice vectors $\vec{a}_{1}$ and $\vec{a}_{2}$, is denoted by a dashed gray square. We represent the three spinless $s$ orbitals within each unit cell with filled dots. The orbital corresponding to each fermionic operator $\hat{c}^\dagger_{i}$, with $1 \leq i \leq 3$ (where we suppress the unit cell index $\vec{R}$) is shown by a gray arrow. The red arrow denotes nearest neighbor imaginary hopping with amplitude $i t_{0}$. The symmetry group of the Hamiltonian is $P6$ (SSG 168.109 in the notation of BCS)~\citeBCS{}. Panel b) shows the band structure of the model for $t_0=1$. At each high-symmetry momentum point, we also indicate the (co)irreps of the corresponding bands. Even though the dispersive bands have opposite energies, the model cannot be a chiral BCL, since the (co)irreps of the dispersive bands do not match: observe the $K$-point, where the upper band carries irrep $K_3$ while the lower carries $K_2$. The model features a single flat band pinned at zero energy (shown in red) as a consequence of the $C_{2z} \mathcal{P}$ symmetry of the Hamiltonian.}
\label{app:fig:C2P_original}
\end{figure}
Most flat band models reported in the literature can be directly related to our BCL construction. In this appendix, we present an example~\cite{GRE10,ELS19} of a Hamiltonian with a gapped, non-degenerate, perfectly flat band that is neither a BCL, nor an effective sublattice Hamiltonian. We prove this by appealing to the existence of the Chern number for the dispersive bands. Nevertheless, after explaining the origin of the flat band in this model, we construct an \emph{equivalent} BCL Hamiltonian in the same SSG, featuring a gapped flat band with a wave function identical to the one in the original model.

We begin by outlining the Hamiltonian of the model. Initially proposed by Ref.~\cite{GRE10}, a number of variations of the original model have been constructed in Refs.~\cite{TAN11,CHE14,ELS19,WEI20,GAO20}. Here, we will follow the conventions of Ref.~\cite{GRE10}. The tight-binding Hamiltonian $\mathcal{H}_0$ is defined on a Kagom\'e lattice and includes nearest-neighbor imaginary hopping, such that each triangular plaquette is threaded by a flux $\phi = 3\pi/2$, as seen in \cref{app:fig:C2P_original:a}. We introduce the fermion operators $\hat{c}^\dagger_{\vec{R},i}$ for $i=1,2,3$ corresponding to a spinless $s$ orbital at the $3c$ Wyckoff position in the unit cell $\vec{R}$ of a two-dimensional hexagonal lattice. The position vector of fermion $\hat{c}^\dagger_{\vec{R},i}$ within unit cell $\vec{R}$ is given by $\vec{r}^{\hat{c}}_i$, where
\begin{equation}
	\vec{r}^{\hat{c}}_1 = \vec{a}_1 + \frac{1}{2} \vec{a}_2, \quad
	\vec{r}^{\hat{c}}_2 = \frac{1}{2} \vec{a}_1 + \frac{1}{2} \vec{a}_2, \quad
	\vec{r}^{\hat{c}}_3 = \frac{1}{2} \vec{a}_1.
\end{equation}  
Following the same convention as in \cref{app:eqn:momentum_ops} of \cref{app:sec:bipartite_crystalline_lattices:notation}, we define the momentum space operators
\begin{equation}
	\label{app:eqn:mom_space_c}
	\hat{c}^\dagger_{\vec{k},i} = \frac{1}{\sqrt{N}}\sum_{\vec{R}}\hat{c}^\dagger_{\vec{R},i} e^{-i \vec{k} \cdot \left( \vec{R} + \vec{r}^{\hat{c}}_i \right)},
\end{equation}
allowing us to write the Hamiltonian of the model as 
\begin{equation}
	\mathcal{H}_{0} = \sum_{\substack{\vec{k} \\ i,j}} \left[h_{\vec{k}}\right]_{ij} \hat{c}^\dagger_{\vec{k},i}\hat{c}_{\vec{k},j},
\end{equation}
where the momentum space Hamiltonian matrix is given by 
\begin{equation}
	\label{app:eqn:h_c2p_matrix}
	h_{\vec{k}} = t_0 \begin{pmatrix}
 0 & 2 i \cos \left(\frac{1}{4} \left(\sqrt{3} k_x-k_y\right)\right) & -2 i \cos \left(\frac{1}{4} \left(\sqrt{3} k_x+k_y\right)\right) \\
 -2 i \cos \left(\frac{1}{4} \left(\sqrt{3} k_x-k_y\right)\right) & 0 & 2 i \cos \left(\frac{k_y}{2}\right) \\
 2 i \cos \left(\frac{1}{4} \left(\sqrt{3} k_x+k_y\right)\right) & -2 i \cos \left(\frac{k_y}{2}\right) & 0 \\
\end{pmatrix} ,
\end{equation}
and $t_0$ is a real hopping parameter (see \cref{app:fig:C2P_original:a}).

In \cref{app:fig:C2P_original:b}, we plot the band structure of the $\mathcal{H}_{0}$ together with the corresponding (co)irreps. The model has a six-fold rotation symmetry $C_{6z}$, but breaks time-reversal symmetry $\mathcal{T}$. Therefore, we analyze the band structure of $\mathcal{H}_{0}$ within the $P6$ group (SSG 168.109 in the notation of BCS)~\citeBCS{}. As seen in \cref{app:fig:C2P_original:b}, the model features a gapped, non-degenerate, perfectly flat band pinned at zero energy. Coupled with the fact that the two dispersive bands come in pairs with opposite energies, one might mistakenly attribute the origin of bands to chiral symmetry and wrongly conclude that $\mathcal{H}_{0}$ is a chiral BCL Hamiltonian. The dispersive bands of $\mathcal{H}_{0}$ are however, strongly topological and, in fact, carry a Chern number $\mathcal{C}=\pm 1$~\cite{GRE10}. Had $h_{\vec{k}}$ been a BCL Hamiltonian matrix, the corresponding effective Hamiltonians $T_{\vec{k}}$ and $\tilde{T}_{\vec{k}}$ would have existed, where $\tilde{T}_{\vec{k}}$ would have featured a single strongly-topological band (as a consequence of the relation between the spectra of $H_{\vec{k}}$ and $\tilde{T}_{\vec{k}}$ discussed in \cref{app:tab:relation_between_eigenstates}). However, that is impossible as both $T_{\vec{k}}$ and $\tilde{T}_{\vec{k}}$ are constructed from atomic orbitals. 

The flat bands of $\mathcal{H}_{0}$ arise not from chiral symmetry, but from the combined operation of particle-hole symmetry and two-fold rotation (as well as an odd number of bands). To see this, we define the unitary particle-hole operator $\mathcal{P}$ which maps creation and annihilation operators into one another
\begin{equation}
	\mathcal{P} \hat{c}^\dagger_{\vec{R},i} \mathcal{P}^{-1} = \hat{c}_{\vec{R},i}.
\end{equation}
Because $\mathcal{H}_{0}$ has only imaginary hopping between different orbitals and no onsite terms, it is particle-hole symmetric (\ie $\left[ \mathcal{P}, \mathcal{H}_{0} \right] = 0$). In momentum space, $\mathcal{P}$ maps momentum $\vec{k}$ to $-\vec{k}$. As such, if we combine particle-hole transformation with two-fold rotation, we can define a momentum-preserving symmetry operator $C_{2z} \mathcal{P}$, whose action on the operators defined in \cref{app:eqn:mom_space_c} is given simply by 
\begin{equation}
	\left(C_{2z} \mathcal{P} \right) \hat{c}^\dagger_{\vec{k},i} \left(C_{2z} \mathcal{P} \right)^{-1} = \hat{c}_{\vec{k},i}.
\end{equation} 
As $\left[C_{2z} \mathcal{P},\mathcal{H}_0 \right] = 0$, the Hamiltonian matrix from \cref{app:eqn:h_c2p_matrix} must be imaginary, \ie $h_{\vec{k}}=-h^{*}_{\vec{k}}$. Because $h_{\vec{k}}$ is also Hermitian, it follows that it must be skew-symmetric, a property which is reflected in \cref{app:eqn:h_c2p_matrix}. However, any odd-dimensional skew-symmetric matrix must have a zero mode~\footnote{Suppose $M$ is a $d$-dimensional skew-symmetric matrix. Because $A = -A^T$, it follows that $\det \left( A  \right) = \det \left( - A^T  \right) = \left( -1 \right)^d  \det \left( A^T  \right) = \left( -1 \right)^d \det \left( A \right) $. If $d$ is odd, we have that $\det \left( A \right) = 0$, and so $A$ must have at least a zero eigenvalue.}. Since $h_{\vec{k}}$ has a zero eigenvalue at all momenta, a flat band emerges whose energy is pinned at zero. In fact, \emph{any} quadratic Hamiltonian with translation invariance featuring $C_{2z} \mathcal{P}$ symmetry in two dimensions or $\mathcal{I} \mathcal{P}$ symmetry in three dimensions (where $\mathcal{I}$ denotes spatial inversion) and an \emph{odd} number of fermionic degrees of freedom, must have one perfectly flat band (protected by $C_{2z} \mathcal{P}$ or $\mathcal{I}\mathcal{P}$). We leave the complete classification of all such systems for a future work, though in \cref{app:sec:example:AZ_fb} we study the presence of flat bands in the ten Altland-Zirnbauer classes where the time-reversal and particle-hole symmetries are composed with inversion. We note, however, that $\mathcal{IP}$ is not a crystalline symmetry, so such flat band systems are not generic and we do not expect many examples in nature, unlike the sublattice-symmetry protected flat bands.

\begin{figure}[!t]
\captionsetup[subfloat]{farskip=0pt}\sbox\nsubbox{
		\resizebox{\textwidth}{!}
		{\includegraphics[height=6cm]{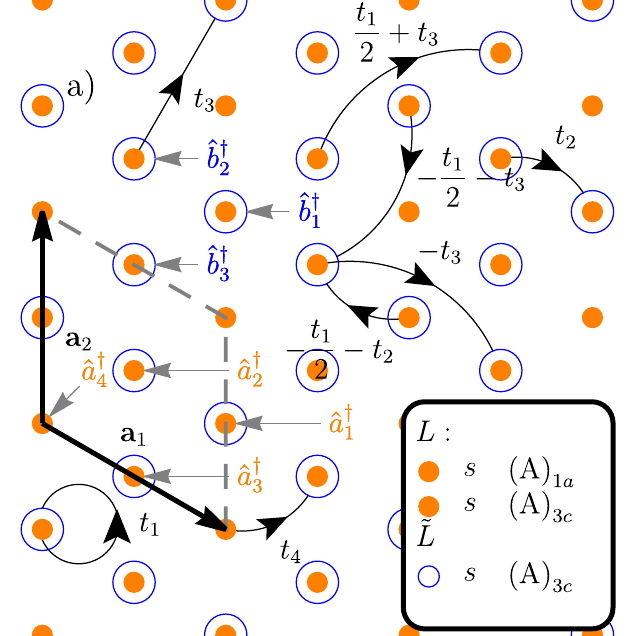}\includegraphics[height=6cm]{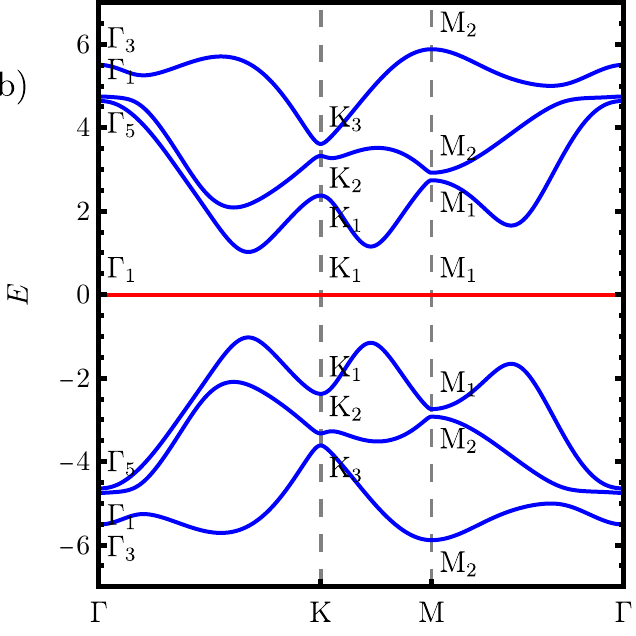}}}\setlength{\nsubht}{\ht\nsubbox}\centering\subfloat{\label{app:fig:C2P_equivalent:a}\includegraphics[height=\nsubht]{9b.pdf}}\subfloat{\label{app:fig:C2P_equivalent:b}\includegraphics[height=\nsubht]{9a.pdf}}
\caption{Recreating the $C_{2z} \mathcal{P}$-protected flat band within a chiral BCL Hamiltonian. We take the intra-sublattice hopping matrices to be $A_{\vec{k}}=\mathbb{0}$, $B_{\vec{k}} = \mathbb{0}$, while the $S_{\vec{k}}$ matrix is given in \cref{app:eqn:sMatrix_C2P}. In panel a), we illustrate the lattice structure of the model. The meaning of the symbols is the same as in \cref{app:fig:Lieb_molecular:a}. By applying the symmetry operations of the $P6$ group (SSG 168.109 in the notation of BCS)~\citeBCS{}, the entire tight-binding Hamiltonian of the model can be recovered from the eight hopping processes shown.
 Panel b) shows the band structure of the model, which features one gapped, perfectly flat band (shown in red) and six dispersive bands (shown in blue). Here, we consider $t_1= 0.00079 + 0.20087i$, $t_2 = 0.06502 + 0.81952i$, $t_3 = 0.98955 + 0.08976i $, and $t_4 = 0.96877 + 0.97070i$. At each high-symmetry momentum point, we also indicate the irreps of the corresponding bands.}
\label{app:fig:C2P_equivalent}
\end{figure}

In what follows, we will show explicitly that, although $\mathcal{H}_{0}$ is not a BCL Hamiltonian, one can construct an equivalent BCL Hamiltonian $\mathcal{H}$ within the same $P6$ group (SSG 168.109 in the notation of BCS)~\citeBCS{}, which features \emph{the same} gapped, non-degenerate flat band (in terms of wave function). A good place to start is the square of the original Hamiltonian matrix, because $h_{\vec{k}}^2$ corresponds to a Hamiltonian with one gapped flat band at zero energy and two degenerate dispersive bands with positive energy (thus resembling an effective sublattice Hamiltonian $T_{\vec{k}}$). One might therefore consider finding a $3 \times 2$ inter-sublattice hopping matrix $S_{\vec{k}}$ (corresponding to a certain $L \oplus \tilde{L}$ BCL), such that 
\begin{equation}
	\label{app:eqn:C2P_S_matrix_try}
	S_{\vec{k}} S^{\dagger}_{\vec{k}} = T_{\vec{k}} \stackrel{?}{=} h^2_{\vec{k}}. 
\end{equation}
$h_{\vec{k}}^2$ and $T_{\vec{k}}$ share identical orbital content, implying that sublattice $L$ of the BCL (on which $T_{\vec{k}}$ is defined) contains an $s$ orbital at the $3c$ Wyckoff position. By direct computation (see also \cref{app:tab:char_p6} of \cref{app:sec:example:char} for the character table of the SSG), the flat band of $h_{\vec{k}}$ has $\mathcal{B}_{\mathrm{FB}} = \left( \Gamma_1 \right) + \left( \mathrm{K}_1 \right) + \left( \mathrm{M}_1 \right)$, and so one can use \cref{app:eqn:br_subtraction} to determine the orbital content of $\tilde{L}$
\begin{equation}
	\mathcal{BR}_{\tilde{L}} = \mathcal{BR}_{L} \boxminus \mathcal{B}_{\mathrm{FB}} \stackrel{?}{=} \left( \Gamma_{3} \oplus \Gamma_{5} \right) + \left( \mathrm{K}_2 \oplus \mathrm{K}_2 \right) + \left( 2 \mathrm{M}_{2} \right) = \left(\mathrm{A}\right)_{3c} \uparrow \mathcal{G} \ominus \left( \mathrm{A} \right)_{1a} \uparrow \mathcal{G}.
\end{equation}
This leads to a contradiction, as $\mathcal{BR}_{\tilde{L}}$ would correspond to a set of topologically fragile bands, implying that a BCL with $N_{L} = 3$ and $N_{\tilde{L}}=2$ in the $P6$ group (SSG 168.109 in the notation of BCS)~\citeBCS{} cannot host a flat band with the required (co)irreps, let alone with the original wave function. On the other hand, by adding an $s$ orbital at the $1a$ position in both sublattices, one can find a BCL where both $\mathcal{BR}_{\tilde{L}}$ and $\mathcal{BR}_{L}$ are atomic limits and their subtraction produces the correct flat band (co)irreps. More precisely, we take the $L$ sublattice to contain $s$ orbitals at the $3c$ and $1a$ positions ($N_{L} = 4$), while the $\tilde{L}$ lattice contains $s$ orbitals at the $3c$ position ($N_{\tilde{L}} = 3$). The lattice structure and orbital content of the BCL are detailed in \cref{app:fig:C2P_equivalent:a}. 

Instead of adding trivial bands to $h^2_{\vec{k}}$ and then determining an $S_{\vec{k}}$ matrix through \cref{app:eqn:C2P_S_matrix_try}, we follow a slightly different approach. Denoting the flat band wave function of $h_{\vec{k}}$ by $\chi_{\vec{k}}$, with $h_{\vec{k}} \chi_{\vec{k}} = 0$, we find 
\begin{equation}
	\chi_{\vec{k}} \propto \begin{pmatrix}
 \cos \left(\frac{k_y}{2}\right) \\
 \cos \left(\frac{1}{4} \left(\sqrt{3} k_x+k_y\right)\right) \\
 \cos \left(\frac{1}{4} \left(\sqrt{3} k_x-k_y\right)\right) \\
\end{pmatrix} .
\end{equation}

We now restrict ourselves to the most general $4 \times 3$ inter-sublattice hopping matrix $S_{\vec{k}}$ which corresponds to a $P6$-symmetric BCL with the sublattice orbital content outlined shown in \cref{app:fig:C2P_equivalent:a}. The invariance under the $C_{6z}$ generating symmetry imposes the following constraint on the inter-sublattice hopping matrix
\begin{equation}
     \label{app:eqn:C2P_S_matrix_symmetry_symmetry}
    D^{\hat{a}} (C_{6z} ) S_{\vec{k}} D^{\hat{b} \dagger} (C_{6z} ) = S_{C_{6z}\vec{k}},
\end{equation}
where the representation matrices were defined in \cref{app:eqn:symmetry_action_examples} and are given explicitly by 
\begin{equation}
     D^{\hat{a}} (C_{6z} ) = \begin{pmatrix}
         0 & 0 & 1 & 0 \\
         1 & 0 & 0 & 0 \\
         0 & 1 & 0 & 0 \\
         0 & 0 & 0 & 1 \\
     \end{pmatrix} \quad \text{ and } \quad
     D^{\hat{b}} (C_{6z} ) = \begin{pmatrix}
         0 & 0 & 1 \\
         1 & 0 & 0 \\
         0 & 1 & 0 \\
     \end{pmatrix}.
\end{equation}
We limit the hopping processes described by $S_{\vec{k}}$ to include only hopping between atoms belonging to neighboring unit cells, \ie 
\begin{equation}
    \label{app:eqn:C2P_S_matrix_hopping_distance}
    S_\vec{R} = \mathbb{0} \qquad \text{if} \qquad \abs{\vec{R}} > \abs{\vec{a}_1} = \abs{\vec{a}_2},
\end{equation}
where $\vec{a}_1$ and $\vec{a}_2$ are the unit cell vectors (see \cref{app:fig:C2P_equivalent:a}). We then require that the BCL flat band corresponding to $S_{\vec{k}}$ has the same wave function as the flat band of $h_{\vec{k}}$. Because the $L$ sublattice (where the flat band wave function has support) contains one additional $s$ orbital at the $1a$ position compared to $h_{\vec{k}}$, this implies that 
\begin{equation}
    \label{app:eqn:C2P_S_matrix_wavf}
	S^{\dagger}_{\vec{k}} \begin{pmatrix}
		\chi_{\vec{k}} \\ 0
	\end{pmatrix} = 0.
\end{equation}  
In other words, the flat band is supported on the $s$ orbitals at the $3c$ Wyckoff position of sublattice $L$ with the wave function $\chi_{\vec{k}}$, but has no support on the $s$ orbital at the $1a$ position. The most general inter-sublattice hopping matrix obeying \cref{app:eqn:C2P_S_matrix_symmetry_symmetry,app:eqn:C2P_S_matrix_hopping_distance,app:eqn:C2P_S_matrix_wavf} is given by
\begin{align}
	\left[S_{\vec{k}}\right]_{:,1}=&\begin{pmatrix}
 4 \left(t_3\right){}^* \sin \left(\frac{\sqrt{3} k_x}{2}\right) \sin \left(\frac{k_y}{2}\right)+\left(t_1\right){}^* \left(\cos \left(\frac{1}{2} \left(\sqrt{3} k_x-k_y\right)\right)+1\right) \\
 e^{\frac{1}{4} i \left(\sqrt{3} k_x+3 k_y\right)} \left(e^{-\frac{1}{2} i \left(\sqrt{3} k_x+k_y\right)} \left(t_2+e^{i k_y} t_3\right){}^*+\left(t_2\right){}^* e^{-i k_y}+\left(t_3\right){}^*\right) \\
 -\left(\left(t_1\right){}^*+2 \left(t_3\right){}^*\right) \cos \left(\frac{1}{4} \left(\sqrt{3} k_x-3 k_y\right)\right)-\left(\left(t_1\right){}^*+2 \left(t_2\right){}^*\right) \cos \left(\frac{1}{4} \left(\sqrt{3} k_x+k_y\right)\right) \\
 2 \left(t_4\right){}^* \cos \left(\frac{k_y}{2}\right) \\
\end{pmatrix},\nonumber \\ 
\left[S_{\vec{k}}\right]_{:,2}=&\begin{pmatrix}
 -\left(\left(t_1\right){}^*+2 \left(t_2\right){}^*\right) \cos \left(\frac{1}{4} \left(\sqrt{3} k_x-k_y\right)\right)-\left(\left(t_1\right){}^*+2 \left(t_3\right){}^*\right) \cos \left(\frac{1}{4} \left(\sqrt{3} k_x+3 k_y\right)\right) \\
 2 \left(t_3\right){}^* \left(\cos \left(k_y\right)-\cos \left(\frac{1}{2} \left(\sqrt{3} k_x-k_y\right)\right)\right)+\left(t_1\right){}^* \left(\cos \left(k_y\right)+1\right) \\
 2 \left(\left(t_3\right){}^* \cos \left(\frac{\sqrt{3} k_x}{2}\right)+\left(t_2\right){}^* \cos \left(\frac{k_y}{2}\right)\right) \\
 2 \left(t_4\right){}^* \cos \left(\frac{1}{4} \left(\sqrt{3} k_x+k_y\right)\right) \\
\end{pmatrix},\nonumber \\ 
\left[S_{\vec{k}}\right]_{:,3}=&\begin{pmatrix}
 e^{-\frac{1}{4} i \left(\sqrt{3} k_x+k_y\right)} \left(e^{\frac{1}{2} i \left(\sqrt{3} k_x+k_y\right)} \left(t_2+e^{i k_y} t_3\right){}^*+\left(t_3\right){}^* e^{i k_y}+\left(t_2\right){}^*\right) \\
 -\left(\left(t_1\right){}^*+2 \left(t_3\right){}^*\right) \cos \left(\frac{\sqrt{3} k_x}{2}\right)-\left(\left(t_1\right){}^*+2 \left(t_2\right){}^*\right) \cos \left(\frac{k_y}{2}\right) \\
 \left(\left(t_1\right){}^*+2 \left(t_3\right){}^*\right) \cos \left(\frac{1}{2} \left(\sqrt{3} k_x+k_y\right)\right)-2 \left(t_3\right){}^* \cos \left(k_y\right)+\left(t_1\right){}^* \\
 2 \left(t_4\right){}^* \cos \left(\frac{1}{4} \left(\sqrt{3} k_x-k_y\right)\right) \\
\end{pmatrix} , \label{app:eqn:sMatrix_C2P}
\end{align}
where $\left[S_{\vec{k}} \right]_{:,i}$ denotes the $i$-th column of the $S_{\vec{k}}$ matrix, while $t_1$, $t_2$, $t_3$, $t_4$ are complex hopping parameters. The hopping pattern corresponding to $S_{\vec{k}}$ is shown schematically in \cref{app:fig:C2P_equivalent:a}. The corresponding chiral BCL Hamiltonian $\mathcal{H}$ is given by \cref{app:eqn:gen_bcl_ham_rs} with $A_{\vec{k}}=\mathbb{0}$ and $B_{\vec{k}}=\mathbb{0}$ and obeys all the symmetries of the $P6$ group (SSG 168.109 in the notation of BCS)~\citeBCS{}. As seen in \cref{app:fig:C2P_equivalent:b}, $\mathcal{H}$ features exactly one gapped flat band pinned at zero energy. Since the flat bands of $\mathcal{H}_0$ and $\mathcal{H}$ share the exact same wave functions, they also have identical (co)irreps. This proves that although the Hamiltonian themselves are not BCLs, the $C_{2z} \mathcal{P}$-protected flat bands reported in Refs.~\cite{GRE10,TAN11,CHE14,ELS19,WEI20,GAO20} are \emph{identical} to flat bands emerging in BCL Hamiltonians and are thus included in our classification.

\subsection{Flat bands in the ten Altland-Zirnbauer classes with inversion}\label{app:sec:example:AZ_fb}
We expand upon the flat band analysis of $C_{2z}\cal{P}$ symmetry in \cref{app:sec:example:PH_flatBands}, asking which of the ten Altland-Zirnbauer (AZ) classes host flat band physics~\cite{ZIR96, ALT97a, KIT09a, CHI16}. To generalize beyond two dimensions, we consider inversion symmetry $\cal{I}$ instead of $C_{2z}$. We modify the AZ classes by adding an inversion operator $\cal I$ in front of any time-reversal (${\cal T}$) or particle-hole (${\cal P}$) operators. The addition of $\mathcal{I}$ here ensures all symmetries preserve momentum: $\cal{IP}, \cal{IT}:  \vec{k} \rightarrow \vec{k}$, so any states pinned to zero energy are pinned throughout the entire Brillouin zone, yielding flat bands. The AZ classes with inversion symmetry have been studied previously by Ref.~\cite{SHI14}, though we study the problem in the presence of flat bands. We explicitly examine each of the ten AZ classes for the possibility of flat bands. There is another approach to classification: flat bands can be viewed as $n$-dimensional momentum-space defects in an $n$-dimensional BZ. The study of these defects (or gapless Fermi surfaces) has been carried out using general arguments of K-theory and Bott periodicity in Refs.~\cite{ZHA13,MAT13, ZHA14}: our results agree with theirs. Also of relevance are the study of AZ classes with spatial symmetries (though no flat bands) in Refs.~\cite{CHI13,MOR13,SHI14}. Note that the results of this section (and of the K-theory papers) only consider which of the ten AZ classes can \emph{support} flat bands, and \emph{not} the topology of the flat bands that can be diagnosed through (co)irreps (such as gapless points or fragile topological phases). 

In order to achieve flat bands, the symmetries must conspire to relate the positive energy spectrum of the first-quantized Hamiltonian to its negative one, pinning eigenstates at zero (the flat bands). We can therefore only rely on $\tilde{C}$ or $\cal{IP}$, as $\cal{IT}$ relates eigenstates at the same energy. Because all three possible symmetries preserve momentum, if a zero energy state is forced by the symmetry operators, it is forced for all $\vec{k}$ in the Brillouin zone, giving rise to flat bands.

We define the second-quantized Hamiltonian in momentum space as
\begin{align}
    {\cal H} = \sum_{\substack{\vec{k} \\ i,j}} \left[H_{\vec{k}}\right]_{ij} \hat{\psi}^\dagger_{\vec{k},i} \hat{\psi}_{\vec{k},j},
\end{align} 
with $\left[H_{\vec{k}}\right]_{ij}$ being the first-quantized Hamiltonian matrix and $\hat{\psi}_{\vec{k},i}$ annihilating a fermion of type $i$ at momentum $\vec{k}$. Here, we take $i$ and $j$ to be generic indices that incorporate the various fermionic degrees of freedom (\eg orbital, spin, etc.). The actions of anti-unitary symmetries are different for the second-quantized system and the first quantized system. We adopt the notation defined in Ref.~\cite{LUD15}, where the action of the three symmetries, $\cal{IT}$, $\cal{IP}$, and $\tilde{C}$, on the fermion operators is given by
\begin{align}
    {\cal IT} \hat{\psi}^\dagger_{\vec{k},i} {\cal (IT)}^{-1} &= \sum_{j} \left[ D({\cal IT}) \right]_{ji} \hat{\psi}^\dagger_{\vec{k},j}, \label{app:eqn:az_action_psi_T} \\
    {\cal IP} \hat{\psi}^\dagger_{\vec{k},i} {\cal (IP)}^{-1} &= \sum_j  \left[ D({\cal IP}) \right]^*_{ji} \hat{\psi}_{\vec{k},j}, \label{app:eqn:az_action_psi_P} \\
    \tilde{C} \hat{\psi}^\dagger_{\vec{k},i} {\tilde{C}}^{-1} &= \sum_{j} \left[ D({\tilde C}) \right]^*_{ji} \hat{\psi}_{\vec{k},j}, \label{app:eqn:az_action_psi_C}
\end{align}
with $D({\cal IT})$, $D({\cal IP})$, $D({\tilde C})$ denoting unitary matrices specifying the action of the three symmetry operators. The presence of $\cal{IT}$, $\cal{IP}$, or $\tilde{C}$ symmetry, respectively imposes a series of constraints on the first-quantized Hamiltonian matrix
\begin{align}
    D({\cal IT}) H^{*}_{\vec{k}} D^{\dagger} ({\cal IT}) &= H_{\vec{k}}, \label{app:eqn:az_action_H_T}\\
    D({\cal IP}) H_{\vec{k}} D^{\dagger} ({\cal IP}) &= -H^{*}_{\vec{k}}, \label{app:eqn:az_action_H_P}\\
    D(\tilde{C}) H_{\vec{k}} D^{\dagger} (\tilde{C}) &= -H_{\vec{k}}. \label{app:eqn:az_action_H_C}
\end{align}

Depending on the presence or absence of the three symmetries ${\cal IP}$, ${\cal IT}$, $\tilde{C}$, as well as on how the corresponding operators square, ten (modified) AZ classes can be defined (see \cref{tab:AZClassification}). We can always introduce relative phases in the operators so that $[{\cal{IT}}, {\cal{IP}}] = 0$, and if both are present for a certain system we define the chiral symmetry $\tilde{C} = e^{i\theta}{\cal{IT}}{\cal{IP}}$, with $\theta$ chosen so that $\tilde{C}^2 = 1$~\cite{LUD15}. The squares of the operators $({\cal IT})^2, ({\cal IP})^2, {\tilde C}^2$ all commute with $H_\vec{k}$.  Assuming that there are no further unitary symmetries, (\ie, $H_\vec{k}$ is a generic matrix that does not factor into a block-diagonal form), by Schur's lemma the squares of the operators are proportional to the identity~\cite{CHI16}:
\begin{align}
    D({\cal IT}) D({\cal IT})^* &= \pm 1 \label{app:eqn:az_matrix_sq_T}\\
    D({\cal IP}) D({\cal IP})^* &= \pm 1 \label{app:eqn:az_matrix_sq_P}\\
    e^{2i\theta} D({\cal IT}) D({\cal IP})^*  D({\cal IT}) D({\cal IP})^* &= 1 \label{app:eqn:az_matrix_sq_C}.
\end{align} 
\Cref{app:eqn:az_matrix_sq_C} can be recast as
\begin{equation}
    D({\cal IP}) D({\cal IT})^* = e^{2i\theta} D({\cal IT}) D({\cal IP})^*,
\end{equation}
which implies that
\begin{equation}
    ({\cal IP}) ({\cal IT}) = e^{2i\theta} ({\cal IT}) ({\cal IP}),
\end{equation} 
meaning that commuting ${\cal IP}$ past ${\cal IT}$ yields a phase. This phase can always be absorbed into a redefinition of the time-reversal operator $D({\cal IT'}) = e^{i\theta} D({\cal IT})$. This new operator commutes with ${\cal IP}$.

Using the constrains from \cref{app:eqn:az_action_H_T,app:eqn:az_action_H_P,app:eqn:az_action_H_C}, we now analyze the possibility of zero modes of $H_{\vec{k}}$ in the ten modified AZ classes defined in \cref{tab:AZClassification}.

\subsubsection{Classes A, AI, AII, AIII}\label{app:sec:example:AZ_fb:1}
The Wigner-Dyson classes A, AI, AII possess only time-reversal symmetry (if any at all) and no symmetries relating the positive and negative single-particle spectra of $H_{\vec{k}}$, and hence do not protect flat bands.  For class AIII, however, we have a chiral symmetry, the anti-unitary symmetry $\tilde{C}$ whose action on the single-particle Hamiltonian is given in \cref{app:eqn:az_action_H_C}. The BCL construction discussed in this paper restricted to systems without time-reversal symmetry falls under this classification. The AIII class allows any number of flat bands (equal to the difference of number of orbitals per unit cell between the two sublattices). Thus the flat band classification is $\mathbb{Z}$, counting the number of flat bands. In our construction, this invariant, $\nu = \Tr \left[ D(\tilde{C}) \right] \in \mathbb{Z}$, is precisely the difference in number of orbitals between sublattices $\nu = N_L - N_{\tilde L}$.

If a flat band system of $\nu$ and another of $\nu'$ are ``stacked'' atop one another, then the resulting system is labeled by $\nu + \nu'$.  This means that systems with $\nu = n$ and $\nu = -n$ are inequivalent; in the language of sublattices one has $N_L > N_{\tilde L}$ and the other $N_{\tilde L} > N_L$.  ``Stacking'' the two systems will allow the sublattices to fully gap each other out resulting in no flat bands, or $\nu = 0$.

\subsubsection{Classes BDI, CII}\label{app:sec:example:AZ_fb:2}
A similar sublattice construction (as performed for class AIII) holds also for the modified classes BDI and CII, as they also possess a chiral symmetry.  In addition, these classes also possess a time-reversal symmetry ${\cal{IT}}$ that \emph{commutes} with $\tilde{C}$ and squares to either $+1$ or $-1$:
\begin{alignat}{5}
& \text{BDI}: \quad && ({\cal{IT}})^2 = +1, \quad && ({\cal{IP}})^2 = +1, \quad && \tilde{C} = {\cal{IT}}{\cal{IP}}, \quad && \left[\tilde{C}, {\cal{IT}} \right] = 0, \\
& \text{CII}: \quad && ({\cal{IT}})^2 = -1, \quad && ({\cal{IP}})^2 = -1, \quad && \tilde{C} = {\cal{IT}}{\cal{IP}}, \quad && \left[\tilde{C}, {\cal{IT}}\right] = 0.
\label{}
\end{alignat}  

For class CII the anti-unitary symmetry $({\cal{IT}})^2 = -1$ protects a Kramers degeneracy for all momenta $\vec{k}$ (as the operator $\cal{IT}$ preserves momentum). Hence while class BDI is labeled by an invariant belonging to $\mathbb{Z}$, class CII is labeled by an element of $2\mathbb{Z}$. Class BDI (CII) is equivalent to a spinless (spinful) BCL Hamiltonian with time-reversal symmetry. In both classes, the corresponding invariant is $\nu = \Tr \left[ D(\tilde{C}) \right] = N_{L} - N_{\tilde{L}}$.  

\subsubsection{Classes C, D}\label{app:sec:example:AZ_fb:3}
We have shown in \cref{app:sec:example:PH_flatBands} that class D, with only $\mathcal{IP}$ symmetry having $({\cal{IP}})^2 = +1$, can host a flat band if there are an odd number of orbitals per unit cell. To see this, notice that  \cref{app:eqn:az_action_H_P} implies that $\det\left( H_{\vec{k}}\right) = \det\left( - H^{T}_{\vec{k}}\right) = (-1)^d \det\left( H_{\vec{k}}\right)$, where $d$ is the dimension of $H_{\vec{k}}$. If $H_{\vec{k}}$ is odd-dimensional (\ie $\mathcal{H}$ has an odd number of fermionic degrees of freedom per unit cell), it follows that its determinant must be zero, meaning that it has at least a zero eigenvalue: the flat band. There is no such restriction on even dimensional matrices, and hence the topological invariant for this class is given by $\mathbb{Z}_2$ (corresponding to the number of flat bands).

For class C, $\cal{IP}$ squares to $-1$, so any bands at zero energy must come in pairs (by Kramers theorem). However, two flat bands are not forced to be at zero energy: one can shift up to energy $+E$ and the other to $-E$. No flat bands are protected in class C.

\subsubsection{Classes CI,  DIII}\label{app:sec:example:AZ_fb:4}
There are two other classes that possess a chiral symmetry: CI and DIII. However, in these classes ${\cal IT}$ \emph{anti-commutes} with $\tilde{C}$:
\begin{alignat}{5}
& \text{CI}: \quad && ({\cal IT})^2 = +1, \quad && ({\cal IP})^2 = -1, \quad && \tilde{C} = i{\cal IT}{\cal IP}, \quad && \left\lbrace \tilde{C}, {\cal IT} \right\rbrace = 0, \\
& \text{DIII}: \quad && ({\cal IT})^2 = -1, \quad && ({\cal IP})^2 = +1, \quad && \tilde{C} = i{\cal IT}{\cal IP}, \quad && \left\lbrace \tilde{C}, {\cal IT} \right\rbrace = 0,
\end{alignat}  
which means it is impossible to create an unbalanced bipartite lattice: $L$ and $\tilde L$ are related by ${\cal IT}$ and must have the same dimension!  Because $({\cal IT}) \tilde{C} ({\cal IT})^{-1} = -\tilde{C}$,
\begin{equation}
    \Tr\left[ D^*(\tilde{C}) \right] = \Tr\left[ D(\tilde{C}) \right] = -\Tr\left[ D(\tilde{C}) \right] = 0
\end{equation} 
hence the sublattices have the same dimension. Thus no flat bands are protected by chiral symmetry.

However, because class DIII still has $\cal{IP}$ symmetry squaring to $+1$, it can still bind flat bands in a similar manner to class D. Consider two such flat bands that are Kramers pairs under $\mathcal{IT}$.  They transform into one another due to $\mathcal{IP}$, meaning that their energies are $+E$ and $-E$, but the Kramers degeneracy forces their energies to be the same, so $E = -E = 0$. A second Kramers pair will allow the bands to lift to finite energy, hence the classification is $Z_2$. Essentially, this topological flat band is a single flat band in class D plus its time-reversed partner.  An example and the topological protection of this type of flat band was realized in Ref.~\cite{HU18}, where the combination anti-unitary symmetries and a \emph{mirror} symmetry protected an flat band of Majorana Kramers pairs.  For class CI, while $\mathcal{IP}$ guarantees bands at $E$ and $-E$, $\mathcal{IT}$ no longer enforces a Kramers degeneracy, as it squares to $+1$.  These bands are not pinned to zero energy.  These results are summarized in \cref{tab:AZClassification}.   Note that this classification is identical to the classification of topological phases in 1D~\cite{QI08, SCH08, KIT09a, SCH09, RYU10, CHI16}.

\begin{table}
\begin{tabular}{c|c|c|c|c}
& $({\cal IT})^2$ & $({\cal IP})^2$ & ${\tilde{C}}^2$ & Flat Band Classification \\
\hline
A & \xmark & \xmark & \xmark & \xmark \\
AIII & \xmark & \xmark & $+1$ & $\mathbb{Z}$ \\
\hline
AI & $+1$ & \xmark & \xmark & \xmark \\
BDI & $+1$ & $+1$ & $+1$ & $\mathbb{Z}$ \\
D & \xmark & $+1$ & \xmark & $\mathbb{Z}_2$ \\
DIII & $-1$ & $+1$ & $+1$ & $\mathbb{Z}_2$ \\
AII & $-1$ & \xmark & \xmark & \xmark \\
CII & $-1$ & $-1$ & $+1$ & $2\mathbb{Z}$ \\
C & \xmark & $-1$ & \xmark & \xmark \\
CI & $+1$ & $-1$ & $+1$ & \xmark \\
\end{tabular}
\caption{The modified Altland-Zirnbauer classes that support flat bands: time-reversal and particle-hole symmetries come attached with an inversion symmetry to preserve momentum. For a given class, the absence of a symmetry is indicated by \xmark. In the presence of a given symmetry, the value of the operator squared is indicated. For each symmetry operators, the rightmost column indicates how many flat bands are supported: $\mathbb{Z}$ for any integer, $2\mathbb{Z}$ for twice the integers, $\mathbb{Z}_2$ for one (or a pair), and \xmark\, for none at all. The classification is identical to the topological classification of 1D Hamiltonians~\cite{SCH08, KIT09a, SCH09, RYU10, CHI16}, and agrees with the defect analysis of Refs.~\cite{ZHA13,MAT13}.}
\label{tab:AZClassification}
\end{table}

\subsection{Character tables}\label{app:sec:example:char}
\FloatBarrier
\begin{table}[!h]
	\begin{tabular}{c|ccccc|ccccc}
		& $E$ & $C_{2z}$ & $2C_{2x}$ & $2C_{2xy}$ & $2C_{4z}$ & $\bar{E}$ & $\bar{C}_{2z}$ & $2\bar{C}_{2x}$ & $2\bar{C}_{2xy}$ & $2\bar{C}_{4z}$ \\
		\hline\hline
		$\Gamma_1^+$ & $1$ & $1$ & $1$ & $1$ & $1$ & $1$ & $1$ & $1$ & $1$ & $1$ \\
		$\Gamma_2^+$ & $1$ & $1$ & $1$ & $-1$ & $-1$ & $1$ & $1$ & $1$ & $-1$ & $-1$ \\
		$\Gamma_5^-$ & $2$ & $-2$ & $0$ & $0$ & $0$ & $-2$ & $2$ & $0$ & $0$ & $0$ \\
		\end{tabular} \qquad
		\begin{tabular}{c|ccccc|ccccc}
		& $E$ & $C_{2z}$ & $2C_{2x}$ & $2C_{2xy}$ & $2C_{4z}$ & $\bar{E}$ & $\bar{C}_{2z}$ & $2\bar{C}_{2x}$ & $2\bar{C}_{2xy}$ & $2\bar{C}_{4z}$ \\
		\hline\hline
		$M_1^+$ & $1$ & $1$ & $1$ & $1$ & $1$ & $1$ & $1$ & $1$ & $1$ & $1$ \\
		$M_2^+$ & $1$ & $1$ & $1$ & $-1$ & $-1$ & $1$ & $1$ & $1$ & $-1$ & $-1$ \\
		$M_5^-$ & $2$ & $-2$ & $0$ & $0$ & $0$ & $-2$ & $2$ & $0$ & $0$ & $0$ \\
        \end{tabular} \\
		\begin{tabular}{c|cccc|cccc}
        & $E$ & $C_2$ & $C_{2y}$ & $C_{2x}$ & $\bar{E}$ & $\bar{C}_2$ & $\bar{C}_{2y}$ & $\bar{C}_{2x}$ \\
		\hline\hline
		$X_1^+$ & $1$ & $1$ & $1$ & $1$ & $1$ & $1$ & $1$ & $1$ \\
		$X_3^-$ & $1$ & $-1$ & $-1$ & $1$ & $-1$ & $1$ & $1$ & $-1$ \\
		$X_4^-$ & $1$ & $-1$ & $1$ & $-1$ & $-1$ & $1$ & $-1$ & $1$ \\
	\end{tabular}
	\caption{Character table of the (co)irreps for the little groups of the $P4/mmm1'$ group (SSG 123.340 in the notation of BCS)~\citeBCS{} at the high-symmetry momenta $\Gamma$, $\mathrm{M}$, and $\mathrm{X}$. We label each conjugacy class with a representative symmetry element. Here $C_{nx}$, $C_{ny}$, and $C_{nz}$ denote $n$-fold rotations around the $\hat{\vec{x}}$, $\hat{\vec{y}}$, and $\hat{\vec{z}}$ axes. Additionally, $C_{2xy}$ represents a rotation about the $x = \pm y$ axis, while $E$ is identity. An overhead bar indicates that the corresponding symmetry operation is combined with spatial inversion.}
	\label{app:tab:char_p4mmm}
\end{table}
\begin{table}[!h]
	\begin{tabular}{c|cccccc}
		& $E$ & $3C_{2y}$ & $C_{2z}$ & $2C_{3z}$ & $3C_{2x}$ & $2C_{6z}$ \\
		\hline\hline
		$\Gamma_1$ & $1$ & $1$ & $1$ & $1$ & $1$ & $1$ \\
		$\Gamma_2$ & $1$ & $-1$ & $1$ & $1$ & $-1$ & $1$ \\
		$\Gamma_5$ & $2$ & $0$ & $2$ & $-1$ & $0$ & $-1$ \\
		$\Gamma_6$ & $2$ & $0$ & $-2$ & $-1$ & $0$ & $1$ \\
		\end{tabular} \qquad
		\begin{tabular}{c|cccc}
		& $E$ & $C_{2z}$ & $C_{2y}$ & $C_{2x}$ \\
		\hline\hline
		$M_1$ & $1$ & $1$ & $1$ & $1$  \\
		$M_2$ & $1$ & $1$ & $-1$ & $-1$ \\
		$M_3$ & $1$ & $-1$ & $1$ & $-1$ \\
		$M_4$ & $1$ & $-1$ & $-1$ & $1$ \\
        \end{tabular} \qquad
		\begin{tabular}{c|ccc}
        & $E$ & $2C_{3z}$ & $3C_{2y}$ \\
		\hline\hline
		$K_1$ & $1$ & $1$ & $1$ \\
		$K_2$ & $1$ & $1$ & $-1$ \\
		$K_3$ & $2$ & $-1$ & $0$ \\
	\end{tabular}
	\caption{Character table of the (co)irreps for the little groups of the $P6/221'$ group (SSG 177.150 in the notation of BCS)~\citeBCS{} at the high-symmetry momenta $\Gamma$, $\mathrm{M}$, and $\mathrm{K}$. We label each conjugacy class with a representative symmetry element. Here $C_{nx}$, $C_{ny}$, and $C_{nz}$ denote $n$-fold rotations around the $\hat{\vec{x}}$, $\hat{\vec{y}}$, and $\hat{\vec{z}}$ axes. Additionally, $E$ represents identity.}
	\label{app:tab:char_p622}
\end{table}

\begin{table}[!h]
	\begin{tabular}{c|cccccc|cccccc}
		& $E$ & $3C_{2y}$ & $C_{2z}$ & $2C_{3z}$ & $3C_{2x}$ & $2C_{6z}$ & $\bar{E}$ & $3\bar{C}_{2y}$ & $\bar{C}_{2z}$ & $2\bar{C}_{3z}$ & $3\bar{C}_{2x}$ & $2\bar{C}_{6z}$ \\
		\hline\hline
		$\Gamma_1^+$ & $1$ & $1$ & $1$ & $1$ & $1$ & $1$ & $1$ & $1$ & $1$ & $1$ & $1$ & $1$ \\
		$\Gamma_5^+$ & $2$ & $0$ & $2$ & $-1$ & $0$ & $-1$ & $2$ & $0$ & $2$ & $-1$ & $0$ & $-1$ \\
		$\Gamma_6^-$ & $2$ & $0$ & $-2$ & $-1$ & $0$ & $1$ & $-2$ & $0$ & $2$ & $1$ & $0$ & $-1$ \\
		\end{tabular} \\
		\begin{tabular}{c|cccc|cccc}
		& $E$ & $C_{2z}$ & $C_{2y}$ & $C_{2x}$ & $\bar{E}$ & $\bar{C}_{2z}$ & $\bar{C}_{2y}$ & $\bar{C}_{2x}$ \\
		\hline\hline
		$M_1^+$ & $1$ & $1$ & $1$ & $1$ & $1$ & $1$ & $1$ & $1$ \\
		$M_2^+$ & $1$ & $1$ & $-1$ & $-1$ & $1$ & $1$ & $-1$ & $-1$ \\
		$M_3^-$ & $1$ & $-1$ & $1$ & $-1$ & $-1$ & $1$ & $-1$ & $1$ \\
		$M_4^-$ & $1$ & $-1$ & $-1$ & $1$ & $-1$ & $1$ & $1$ & $-1$ \\
        \end{tabular} \qquad
		\begin{tabular}{c|ccc|ccc}
        & $E$ & $2C_{3z}$ & $3C_{2y}$ & $\bar{C}_{2z}$ & $2\bar{C}_{6z}$ & $3\bar{C}_{2x}$ \\
		\hline\hline
		$K_1$ & $1$ & $1$ & $1$ & $1$ & $1$ & $1$ \\
		$K_4$ & $1$ & $1$ & $-1$ & $1$ & $1$ & $-1$ \\
		$K_5$ & $2$ & $-1$ & $0$ & $2$ & $-1$ & $0$ \\
	\end{tabular}
	\caption{Character table of the (co)irreps for the little groups of the $P6/mmm1'$ group (SSG 191.234 in the notation of BCS)~\citeBCS{} at the high-symmetry momenta $\Gamma$, $\mathrm{M}$, and $\mathrm{K}$. We label each conjugacy class with a representative symmetry element. Here $C_{nx}$, $C_{ny}$, and $C_{nz}$ denote $n$-fold rotations around the $\hat{\vec{x}}$, $\hat{\vec{y}}$, and $\hat{\vec{z}}$ axes, while $E$ is identity. An overhead bar indicates that the corresponding symmetry operation is combined with spatial inversion.}
	\label{app:tab:char_p6mmm}
\end{table}

\begin{table}[!h]
	\begin{tabular}{c|cccc}
		& $E$ & $C_{2z}$ & $C_{3z}$ & $C_{6z}$ \\
		\hline\hline
		$\Gamma_1$ & $1$ & $1$ & $1$ & $1$ \\
		$\Gamma_3$ & $1$ & $1$ & $\omega$ & $\omega^*$ \\
		$\Gamma_5$ & $1$ & $1$ & $\omega^*$ & $\omega$ \\
		\end{tabular} \qquad
		\begin{tabular}{c|cc}
		& $E$ & $C_{3z}$ \\
		\hline\hline
		$K_1$ & $1$ & $1$ \\
		$K_2$ & $1$ & $\omega$ \\
		$K_3$ & $1$ & $\omega^*$ \\
		\end{tabular} \qquad
		\begin{tabular}{c|cc}
		& $E$ & $C_{2z}$ \\
		\hline\hline
		$M_1$ & $1$ & $1$ \\
		$M_2$ & $1$ & $-1$ \\
	\end{tabular}
	\caption{Character table of the (co)irreps for the little groups of the $P6$ group (SSG 168.109 in the notation of BCS)~\citeBCS{} at the high-symmetry momenta $\Gamma$, $\mathrm{M}$, and $\mathrm{K}$. We label each conjugacy class with a representative symmetry element. Here $C_{nz}$ denotes an $n$-fold rotation around the $\hat{\vec{z}}$ axis.}
	\label{app:tab:char_p6}
\end{table}

\FloatBarrier

\section{Flat band classification}\label{app:sec:fb_clas}
In this appendix, we detail our flat band classification scheme, which employs Magnetic Topological Quantum Chemistry (MTQC)~\cite{BRA17, ELC20a}, the basics of which we will briefly review.  In short, MTQC studies band structures in terms of their symmetry properties. At the high-symmetry momenta, bands carry irreducible (co)representations [(co)irreps] of the SSG, which can be employed to diagnose their topology~\cite{BRA17,PO17,KRU17,WAT18,ELC20a}.

The great power of MTQC stems from enumerating all possible band representations -- topologically trivial bands which are induced from atomic limits (\ie orbitals located at different locations inside the unit cell, known as Wyckoff positions), in all 1651{} SSGs~\cite{BRA17,ELC20a}. On the other hand, the flat bands arising in BCLs were shown in \cref{app:sec:unitary_sym} to be formal differences of band representations. Therefore, MTQC serves as the primary tool for building a complete classification of BCL flat bands. First, we show that, although infinite, the set of \emph{all} (gapped and gapless) BCL flat bands is finitely generated.
After deriving an algorithm, we build the generating bases in all 1651{} SSGs. Next, we prove that the subset of \emph{gapped} BCL flat bands is also infinite, but finitely generated and outline the method for constructing the corresponding bases. Finally, we show that gapped BCL flat bands can realize \emph{any} topologically fragile state and provide a simple example illustrating the relation between gapped, gapless, and topologically fragile bands.

\subsection{Flat band symmetry data vectors}\label{app:sec:fb_clas:sym_dat_vec}
The symmetry properties of an electronic band are completely described by its decomposition into (co)irreps at high-symmetry momenta in the Brillouin zone~\cite{ZAK80,ZAK81,BRA17,KRU17,PO17,WAT18,ELC20a}. For a given gapped band or set of bands, the (co)irreps at two different momentum points are not independent, but instead have to satisfy certain compatibility relations~\cite{MIC99,MIC00,MIC01,BRA17,VER17,KRU17,ELC17,PO17,BRA18,WAT18,ELC20a}. The (co)irreps at the maximal momenta determine the (co)irreps across the entire Brillouin zone~\cite{BRA17,ELC20a}.
For any such band, as in Refs.~\cite{BRA17,ELC20a,CAN21}, we define the symmetry data vector $B$, which contains the multiplicities of all (co)irreps of a certain band at maximal momenta in the Brillouin zone
\begin{equation}
	\label{app:eqn:sym_dat_vect_def}
	B = \left( 
		m\left(\rho^1_{\mathcal{G}_{\vec{K}_1}} \right),
		m\left(\rho^2_{\mathcal{G}_{\vec{K}_1}} \right),
		\dots,
		m\left(\rho^1_{\mathcal{G}_{\vec{K}_2}} \right),
		m\left(\rho^2_{\mathcal{G}_{\vec{K}_2}} \right),
		\dots
	\right)^T,
\end{equation}
where $m\left(\rho^i_{\mathcal{G}_{\vec{K}_j}} \right)$ denotes the multiplicity of the (co)irrep $\rho^i_{\mathcal{G}_{\vec{K}_j}}$ of the little-group $\mathcal{G}_{\vec{K}_j}$ of the maximal momentum $\vec{K}_j$.

A central role in (Magnetic) Topological Quantum Chemistry is played by the (magnetic) elementary band representations (EBRs). An EBR is a special type of (atomic) SSG (co)representation that is induced from a certain (co)irrep of the site-symmetry group of a maximal Wyckoff position~\cite{ZAK80,ZAK81,MIC99,MIC00,MIC01,BRA17,CAN18a,ELC20a}. For each SSG $\mathcal{G}$, we define $EBR$ to be the matrix whose columns are the symmetry data vectors corresponding to all EBRs of $\mathcal{G}$, \ie
\begin{equation}
	\label{app:eqn:EBR_def}
	EBR = \left( 
		B^{\left(\rho_1\right)_{w_1}},
		B^{\left(\rho_2\right)_{w_1}},
		\dots,
		B^{\left(\rho_1\right)_{w_2}},
		B^{\left(\rho_2\right)_{w_2}},
		\dots
	\right),
\end{equation}
where $B^{\left(\rho_i\right)_{w_j}}$ is the symmetry data vector of the EBR induced from the (co)irrep $\rho_{i}$ of the Wyckoff position $w_j$. The EBR matrix has size $d_{\mathcal{G}} \times d_{EBR}$, where $d_{\mathcal{G}}$  counts the unique types of (co)irreps for \emph{all} maximal momenta, while $d_{EBR}$ is the total number of EBRs in the SSG $\mathcal{G}$, which are induced from the Wyckoff positions of maximal symmetry.  

In \cref{app:sec:unitary_sym}, we have shown that for a generalized BCL Hamiltonian (defined on the $L \oplus \tilde{L}$ lattice) and at a given momentum point, the (co)irreps of the flat bands (in the gapped case), or the formal (co)irrep difference of the flat band and corresponding band touching points (in the gapless case) are independent of the inter-sublattice hopping matrix  $S_{\vec{k}}$. Instead they depend exclusively on the orbital content of the two sublattices $L$ and $\tilde{L}$ according to the subtraction rule from \cref{app:eqn:br_subtraction}. Because both $\mathcal{BR}_{L}$ and $\mathcal{BR}_{\tilde{L}}$ are induced from the (co)irreps (orbitals) of the site-symmetry groups of the occupied Wyckoff positions in the two sublattices (i.e. both are `trivial' atomic limits), their symmetry data vectors, $B_L$ and $B_{\tilde{L}}$, are linear combinations of EBRs with nonnegative integer coefficients~\cite{BRA17,CAN18a,ELC20a}:
\begin{align}
	B_L &= EBR \cdot x_L, \\
	B_{\tilde{L}} &= EBR \cdot x_{\tilde{L}},
\end{align}
where $x_L, x_{\tilde{L}} \in \mathbb{N}^{d_{EBR}}$ are column vectors specifying the orbital content of the two sublattices  
\begin{align}
	x_L &= \left( 
		m_{L} \left[ \left(\rho_1\right)_{w_1} \right],
		m_{L} \left[ \left(\rho_2\right)_{w_1} \right],
		\dots
		m_{L} \left[ \left(\rho_1\right)_{w_2} \right],
		m_{L} \left[ \left(\rho_2\right)_{w_2} \right],
		\dots
	\right)^T, \label{app:eqn:sublat_orb_vector_L} \\
	x_{\tilde{L}} &= \left( 
		m_{\tilde{L}} \left[ \left(\rho_1\right)_{w_1} \right],
		m_{\tilde{L}} \left[ \left(\rho_2\right)_{w_1} \right],
		\dots
		m_{\tilde{L}} \left[ \left(\rho_1\right)_{w_2} \right],
		m_{\tilde{L}} \left[ \left(\rho_2\right)_{w_2} \right],
		\dots
	\right)^T.  \label{app:eqn:sublat_orb_vector_Ltilde}
\end{align}
We choose the convention in which the set of natural numbers $\mathbb{N}$ includes $0$. In \cref{app:eqn:sublat_orb_vector_L,app:eqn:sublat_orb_vector_Ltilde}, we have denoted by $m_{\mathcal{L}} \left[ \left(\rho_i\right)_{w_j}\right]$ the multiplicity of the (co)irrep $\rho_i$ of the Wyckoff position $w_j$ in sublattice $\mathcal{L}=L,\tilde{L}$. Without loss of generality, only the (co)irreps $ \left(\rho_i\right)_{w_j}$ that induce EBRs are considered because any band representation can be expressed as a linear combination of EBRs with positive coefficients. The number of fermionic degrees of freedom per unit cell within each sublattice is related to the components of $x_L$ and $x_{\tilde{L}}$ according to 
\begin{align}
	N_L &= \sum_{i,j} m_{L} \left[ \left(\rho_i\right)_{w_j}\right] d_{\left(\rho_i\right)_{w_j}} , \label{app:eqn:sublat_no_orb_L} \\
	N_{\tilde{L}} &= \sum_{i,j} m_{\tilde{L}} \left[ \left(\rho_i\right)_{w_j}\right] d_{\left(\rho_i\right)_{w_j}}, \label{app:eqn:sublat_no_orb_Ltilde}
\end{align}
where $d_{\left(\rho_i\right)_{w_j}}$ denotes the dimension of the (co)irrep $ \left(\rho_i\right)_{w_j}$. Using \cref{app:eqn:br_subtraction}, we find that the symmetry data vector $B_{\mathrm{FB}}$ corresponding to $\mathcal{B}_{\mathrm{FB}}$ [denoting the (co)irreps of the corresponding flat bands in the gapped case or the formal (co)irrep differences in the gapless case] is given by
\begin{equation}
	\label{app:eqn:sym_dat_vec_FB}
	B_{\mathrm{FB}} = B_L - B_{\tilde{L}} = EBR \left(x_L - x_{\tilde{L}} \right).
\end{equation}
It is worth noting that even though $B_{\mathrm{FB}}$ only contains the multiplicities of the (co)irreps at the maximal momenta, the (co)irreps at \emph{all} momenta in the Brillouin zone are completely specified through the compatibility relations. These compatibility relations are constraints relating the (co)irreps at two different momentum points~\cite{MIC99,MIC00,MIC01,BRA17,VER17,KRU17,ELC17,PO17,BRA18,WAT18,ELC20a}, which $B_{FB}$ automatically obeys as an integer linear combination of EBRs~\cite{BRA17,ELC20a}.
\subsection{Classifying all flat bands}\label{app:sec:fb_clas:all}

In this appendix, we  classify \emph{all} possible BCL flat bands, allowing for band touching points between the flat and dispersive bands. As such, $\mathcal{B}_{\mathrm{FB}}$ \emph{can} contain formal differences of (co)irreps (in the case of gapless flat bands), and correspondingly $B_{\mathrm{FB}}$ \emph{can} have both positive and negative entries. The case of gapped flat bands, for which all components of $B_{\mathrm{FB}}$ must be  non-negative (\ie $\left(B_{\mathrm{FB}}\right)_i \geq 0$ for $1 \leq i \leq d_{\mathcal{G}}$) will be discussed separately in \cref{app:sec:fb_clas:gapped}. The set of all flat bands and potential band touching points (co)irreps for a given SSG $\mathcal{G}$ (henceforth denoted by $\mathcal{F}_{\mathcal{G}}$) is given by all possible symmetry data vectors $B_{\mathrm{FB}}$ obtainable from \cref{app:eqn:sym_dat_vec_FB}, \ie
\begin{equation}
	\label{app:eqn:set_of_flat_bands_1.1}
	\mathcal{F}_{\mathcal{G}} = \left\lbrace B_{\mathrm{FB}} \in \mathbb{Z}^{d_{\mathcal{G}}} \mid B_{\mathrm{FB}} = EBR \left(x_L - x_{\tilde{L}} \right), \text{ with } x_L,x_{\tilde{L}} \in \mathbb{N}^{d_{EBR}} \text{ and } N_{L} > N_{\tilde{L}} \right\rbrace.
\end{equation}

We expect the set $\mathcal{F}_{\mathcal{G}}$ to be \emph{infinite}, but \emph{finitely generated}. We prove this statement by directly constructing the corresponding basis vectors. First, for a given symmetry data vector $B$ as defined in \cref{app:eqn:sym_dat_vect_def}, we introduce an \emph{augmented} symmetry data vector $\overline{B}$, whose components are given by
\begin{equation}
	\label{app:eqn:aug_sym_dat_vect_def}
	\overline{B} = \left(
		n,
		m\left(\rho^1_{\mathcal{G}_{\vec{K}_1}} \right),
		m\left(\rho^2_{\mathcal{G}_{\vec{K}_1}} \right),
		\dots,
		m\left(\rho^1_{\mathcal{G}_{\vec{K}_2}} \right),
		m\left(\rho^2_{\mathcal{G}_{\vec{K}_2}} \right),
		\dots
	\right)^T.
\end{equation}
Compared to $B$, the augmented symmetry data vector $\overline{B}$ contains one additional entry at the beginning ($n$) specifying the number of bands encoded by $B$.  This augmentation keeps track of the total number of bands, which eventually allows us to impose the constraint of a positive number of flat bands, that is, $N_L > N_{\tilde L}$. Letting $d_{\rho^i_{\mathcal{G}_{\vec{K}_j}}}$ be the dimension of the (co)irrep $\rho^i_{\mathcal{G}_{\vec{K}_j}}$, we find that $n$ obeys
\begin{equation}
	\label{app:eqn:number_of_flat_to_irrep_mult}
	n= \sum_{i} m\left(\rho^i_{\mathcal{G}_{\vec{K}_j}} \right) d_{\rho^i_{\mathcal{G}_{\vec{K}_j}}}
\end{equation}
for any maximal momentum $\vec{K}_j$ (note that $\vec{K}_j$ is not summed over). In analogy with \cref{app:eqn:EBR_def}, we also define the augmented EBR matrix 
\begin{equation}
	\label{app:eqn:aug_EBR_def}
	\overline{EBR} = \left( 
		\overline{B}^{\left(\rho_1\right)_{w_1}},
		\overline{B}^{\left(\rho_2\right)_{w_1}},
		\dots,
		\overline{B}^{\left(\rho_1\right)_{w_2}},
		\overline{B}^{\left(\rho_2\right)_{w_2}},
		\dots
	\right),
\end{equation}
of size $\left( d_{\mathcal{G}}+1 \right) \times d_{EBR} $, whose columns are the augmented symmetry data vectors of all EBRs of $\mathcal{G}$. In short, the augmented vector $\overline{B}^{\left(\rho_i\right)_{w_j}}$ is simply
\begin{equation}
	\overline{B}^{\left(\rho_i\right)_{w_j}} = \begin{pNiceArray}{C}[last-col]
		d_{\left(\rho_i\right)_{w_j}} & 1 \\
		B^{\left(\rho_i\right)_{w_j}} & d_{\mathcal{G}}
	\end{pNiceArray},
\end{equation}
where $B^{\left(\rho_i\right)_{w_j}}$ was defined in the text surrounding \cref{app:eqn:EBR_def}.

Due to the one-to-one mapping between symmetry data vectors and augmented symmetry data vectors, we are free to index the elements of $\mathcal{F}_{\mathcal{G}}$ according to their augmented symmetry data vectors and write
\begin{equation}
	\label{app:eqn:set_of_flat_bands_1.2}
	\mathcal{F}_{\mathcal{G}} = \left\lbrace \overline{B}_{\mathrm{FB}} \in \mathbb{Z}^{d_{\mathcal{G}}+1} \mid \overline{B}_{\mathrm{FB}} = \overline{EBR} \left(x_L - x_{\tilde{L}} \right), \text{ with } x_L,x_{\tilde{L}} \in \mathbb{N}^{d_{EBR}} \text{ and } N_{L} > N_{\tilde{L}} \right\rbrace,
\end{equation}
where $x_L, x_{\tilde L}$ are precisely the multiplicities of the orbitals that occur in the $L$, $\tilde L$ sublattices. 
From \cref{app:eqn:aug_EBR_def,app:eqn:set_of_flat_bands_1.2,app:eqn:sublat_orb_vector_L,app:eqn:sublat_orb_vector_Ltilde,app:eqn:sublat_no_orb_L,app:eqn:sublat_no_orb_Ltilde}, we find that the first entry of $\overline{B}_{\mathrm{FB}}$ obeys
\begin{equation}
	\left( \overline{B}_{\mathrm{FB}} \right)_1 = N_{L} - N_{\tilde{L}},
\end{equation}
being equal to the number of BCL flat bands. Thus, the constraint $N_L > N_{\tilde{L}}$ becomes equivalent to requiring that $\left( \overline{B}_{\mathrm{FB}} \right)_1 > 0$. Additionally, since $(x_L - x_{\tilde{L}})$ spans the integer lattice $\mathbb{Z}^{d_{EBR}}$ (for $x_L, x_{\tilde{L}} \in \mathbb{N}^{d_{EBR}}$), 
\begin{equation}
	\label{app:eqn:set_of_flat_bands_1.3}
	\mathcal{F}_{\mathcal{G}} = \left\lbrace \overline{B}_{\mathrm{FB}} \in \mathbb{Z}^{d_{\mathcal{G}}+1} \mid \overline{B}_{\mathrm{FB}} = \overline{EBR} \cdot x, \text{ with } x \in \mathbb{Z}^{d_{EBR}} \text{ and } \left( \overline{B}_{\mathrm{FB}} \right)_1 > 0 \right\rbrace.
\end{equation}
Note that the entries of $\overline{B}_{\mathrm{FB}}$ obey \cref{app:eqn:number_of_flat_to_irrep_mult} both at the gapped points [where the multiplicity of each (co)irrep is positive] and at the gapless points [where some of the (co)irrep multiplicities are negative]. 

As implied by \cref{app:eqn:set_of_flat_bands_1.3}, for any BCL flat band, the corresponding augmented symmetry data vector is given by
\begin{equation}
	\overline{B}_{\mathrm{FB}} = \overline{EBR} \cdot x,
\end{equation}
where $x \in \mathbb{Z}^{d_{EBR}}$ and $\left( \overline{B}_{\mathrm{FB}} \right)_1 > 0$. The EBRs are not generally linearly independent~\cite{BRA17,CAN18a,ELC20a} , implying that the rank $r$ of the EBR matrix (obeying $r \leq d_{EBR},d_{\mathcal{G}}$) is not necessarily maximal. The matrix $\overline{EBR}$ contains one additional row compared to $EBR$, which is a linear combination of the rows of the matrix $EBR$ and thus has the same rank $r$. As we seek a non-redundant basis for $\mathcal{F}_{\mathcal{G}}$, we proceed to write $\overline{EBR}$ using its Smith normal form~\cite{WIK21} as 
\begin{equation}
	\overline{EBR} = U \Lambda R,
\end{equation}
where $U$ ($R$) is a unimodular integer matrix of size $[d_{\mathcal{G}}+1] \times [d_{\mathcal{G}}+1]$ ($d_{EBR} \times d_{EBR}$), while $\Lambda$ is an integer diagonal matrix of size $\left(d_{\mathcal{G}}+1 \right) \times d_{EBR}$. Only the first $r$ elements on the main diagonal of $\Lambda$ are non-zero. Since $R$ is invertible among the integers,  the vector $y= Rx \in \mathbb{Z}^{d_{EBR}}$ spans the lattice of integers, and
\begin{equation}
	\label{app:eqn:set_of_flat_bands_1.4}
	\mathcal{F}_{\mathcal{G}} = \left\lbrace \overline{B}_{\mathrm{FB}} \in \mathbb{Z}^{d_{\mathcal{G}}+1} \mid \overline{B}_{\mathrm{FB}} = U \Lambda y, \text{ with } y \in \mathbb{Z}^{d_{EBR}} \text{ and } \left( \overline{B}_{\mathrm{FB}} \right)_1 > 0 \right\rbrace.
\end{equation}
Only the first $r$ diagonal entries of $\Lambda$ are non-zero and, consequently, only the first $r$ components of $y$ in \cref{app:eqn:set_of_flat_bands_1.4} determine the symmetry data vector $\overline{B}_{\mathrm{FB}}$. We can therefore ignore the remaining $d_{EBR} - r$ redundant components by defining $\tilde{U}$ to be the matrix containing the first $r$ columns of $U$~\footnote{The remaining $d_{EBR} -r$ columns of $U$ impose a set of compatibility relations on the augmented symmetry data vector from \cref{app:eqn:set_of_flat_bands_1.4}. Because $\overline{B}_{\mathrm{FB}}$ is a linear combination of EBRs with integer coefficients, it automatically satisfies these compatibility relations~\cite{ELC20}.} and $\tilde{\Lambda}$ to be the $r \times r$ diagonal matrix containing the upper-left $r \times r$ block of $\Lambda$, and $\tilde y$ the vector with the first $r$ entries of $y$. The set of all possible BCL flat bands is thus given by
\begin{equation}
	\label{app:eqn:set_of_flat_bands_2}
	\mathcal{F}_{\mathcal{G}} = \left\{ \overline{B}_{\mathrm{FB}} \in \mathbb{Z}^{d_{\mathcal{G}}+1} \mid \overline{B}_{\mathrm{FB}} = \tilde{U} \tilde{\Lambda} \tilde{y}, \text{ with } \tilde{y} \in \mathbb{Z}^{r} \text{ and } \left( \overline{B}_{\mathrm{FB}} \right)_1 > 0 \right\}.
\end{equation}  
An additional step eliminates the inequality constraint $\left( \overline{B}_{\mathrm{FB}} \right)_{1} > 0$. First, we define $D$ to be a $\left(d_{\mathcal{G}}+1\right) \times 1$ matrix with $D_{11} = 1$ and $D_{i1}=0$, for any $1<i \leq d_{\mathcal{G}}+1$. Enforcing $\left( \overline{B}_{\mathrm{FB}} \right)_{1} > 0$ is equivalent to finding the integer solutions of the inequality
\begin{equation}
	\label{app:eqn:inequality_constraint_fb}
    D^T \tilde{U} \tilde{\Lambda} \tilde{y} > 0.
\end{equation}
We now write the $1 \times r$ matrix $D^{T} \tilde{U} \tilde{\Lambda}$ using its Smith normal form~\cite{WIK21} as $D^{T} \tilde{U} \tilde{\Lambda} = W \mathcal{D} V$, where $W$ is a $1 \times 1$ unimodular matrix (meaning that $W_{11} = \pm 1$), $\mathcal{D}$ is a rank-$1$ $1 \times r$ matrix, and $V$ is a $r \times r$ unimodular matrix. The matrices $W$, $\mathcal{D}$, and $V$ can be chosen such that $W_{11} \mathcal{D}_{11} > 0$ (if $W_{11} \mathcal{D}_{11}$ is negative, we can redefine the matrices $W \rightarrow -W, V \rightarrow -V$). Once again, we employ the unimodularity of $V$ and trade $\tilde{y}$ for $z = V \tilde{y}$, where $z \in \mathbb{Z}^r$. The inequality constraint from \cref{app:eqn:inequality_constraint_fb} is equivalent to $z_1 > 0$. Thus all the possible values of $\tilde{y}$ are given by $\tilde{y} = V^{-1} z$, where $z \in \mathbb{Z}^r$ and $z_1>0$. We can now define a basis for the set $\mathcal{F}_{\mathcal{G}}$. Letting $e_i$ (for $1 \leq i \leq r$) denote the column vectors of the matrix, 
\begin{equation}
	\label{app:eqn:EBR_matrix_M}
	M \equiv \tilde{U} \tilde{\Lambda} V^{-1} = \left( 
		e_1,
		e_2,
		\dots,
		e_{r-1},
		e_{r}
	\right),
\end{equation}
the set $\mathcal{F}_{\mathcal{G}}$ is simply defined as
\begin{equation}
	\label{app:eqn:set_of_flat_bands_3}
	\mathcal{F}_{\mathcal{G}} = \left\{ \overline{B}_{\mathrm{FB}} \in \mathbb{Z}^{d_{\mathcal{G}}+1} \mid \overline{B}_{\mathrm{FB}} = \sum_{i=1}^{r} z_i e_i, \text{ with } z \in \mathbb{Z}^{r} \text{ and } z_1 > 0  \right\}.
\end{equation}
The set of all possible BCL flat bands is isomorphic to the direct product between a semi-group and a group $\mathcal{F}_{\mathcal{G}} \sim \mathbb{N}^{*} \otimes \mathbb{Z}^{r-1}$, where the ``semi-group entry'' ($z_1$) controls the number of flat bands, and $\mathbb{N}^{*}$ denotes the set of all natural numbers excluding $0$. It is worth noting that the $\left(d_{\mathcal{G}} + 1 \right) \times r$ matrix $M$ defined in \cref{app:eqn:EBR_matrix_M} is full-rank (having rank $r$). To see this, note that the columns of $\tilde{U}$ are linearly independent (as they originate from the unimodular and hence non-singular matrix $U$). Because $\tilde{\Lambda}$ and $V^{-1}$ are non-singular square matrices, the columns of $M$ are just independent linear combinations of the columns of $\tilde{U}$ and are therefore linearly independent. As a result, the matrix $M$ is full-rank.

In \cref{app:sec:fb_clas:table}, we provide the set of basis vectors $e_i$ (for $1 \leq i \leq r$) in all 1651{} SSG. For each basis vector, we only list the (co)irrep multiplicity for the maximal momentum points, which fully specify the (co)irreps of the flat bands (and potential band touching points) across the entire Brillouin zone through the compatibility relations~\cite{BRA17,ELC20a}.

\subsection{Constructing the set of all gapped flat bands}\label{app:sec:fb_clas:gapped}

In \cref{app:sec:fb_clas:all}, we have shown that the set of all (gapless and gapped) flat bands is infinite, but finitely generated. Here, we prove that the set of gapped flat bands  (\ie for which $\mathcal{B}_{\mathrm{FB}}$ contains no (co)irrep difference), henceforth denoted by $\mathcal{F^{\mathrm{G}}_{\mathcal{G}}}$, is also infinite, but finitely generated. Moreover, we will show that the problem of finding the corresponding generating basis is equivalent to finding the Hilbert basis of a pointed rational cone (see \cref{app:sec:fb_clas:maths} for a brief mathematical overview of the subject). The same technique was used in Ref.~\cite{SON20} as a key step towards obtaining all fragile bands in the 230 space groups. Owing to the computational difficulty of constructing the Hilbert bases for pointed rational cones~\cite{DUR02}, as well as the size of the resulting bases, we will not explicitly construct the gapped flat band bases in all 1651{} SSGs, but instead provide a series of examples in \cref{app:sec:fb_clas:gapped_examples}.

As shown in \cref{app:sec:gapless}, formal differences of (co)irreps lead to band touching points, or gapless points.  Thus, flat bands without locally-stable band touching points must have augmented symmetry data vectors with non-negative entries. Throughout this appendix we will refer to them as \emph{gapped} flat bands (even though they might contain globally stable band touching points that cannot be diagnosed from (co)irreps as exemplified in \cref{app:sec:example:C2T}). The set of all gapped flat bands is given by
\begin{align}
	\mathcal{F}^{\mathrm{G}}_{\mathcal{G}} &= \left\{ \overline{B}_{\mathrm{FB}} \in \mathbb{Z}^{d_{\mathcal{G}}+1} \mid \overline{B}_{\mathrm{FB}} = \sum_{i=1}^{r} z_i e_i, \text{ with } z \in \mathbb{Z}^{r}, z_1 > 0, \text{ and } \overline{B}_{\mathrm{FB}} \geq 0 \right\}  \nonumber \\
	&= \left\{ \overline{B}_{\mathrm{FB}} \in \mathbb{Z}^{d_{\mathcal{G}}+1} \mid \overline{B}_{\mathrm{FB}} = M z, \text{ with } z \in \left( \mathbb{Z}^{r} \setminus \left\lbrace \mathbb{0} \right\rbrace \right) \text{ and } \overline{B}_{\mathrm{FB}} \geq 0 \right\} , \label{app:eqn:set_of_gapped_flat_bands_1.1}
\end{align}
where the $\left( d_{\mathcal{G}}+1 \right) \times r$ matrix $M$ was defined in \cref{app:eqn:EBR_matrix_M}. Note that the constraint $\overline{B}_{\mathrm{FB}} \geq 0$ implies that $z_1 \geq 0$, with $z_1 = 0$ only if $\overline{B}_{\mathrm{FB}} = \mathbb{0}$ (and hence $z = \mathbb{0}$), justifying the expression on the second line of \cref{app:eqn:set_of_gapped_flat_bands_1.1}.

To show that the set $\mathcal{F}^{\mathrm{G}}_{\mathcal{G}}$ is finitely generated, we first need to solve the integer inequality
\begin{equation}
	\label{app:eqn:gapped_condition_ineq}
    M z \geq 0
\end{equation}
and determine the set of solutions 
\begin{equation}
	\label{app:eqn:gapped_z_set_def}
	\mathcal{Z} = \left\lbrace z \in \mathbb{Z}^{r} \mid M z \geq 0 \right\rbrace,
\end{equation}
In this notation, the set $\mathcal{F}^{\mathrm{G}}_{\mathcal{G}}$ can be rewritten simply as 
\begin{equation}
	\label{app:eqn:set_of_gapped_flat_bands_1.2}
	\mathcal{F}^{\mathrm{G}}_{\mathcal{G}} = \left\{ \overline{B}_{\mathrm{FB}} \in \mathbb{Z}^{d_{\mathcal{G}}+1} \mid \overline{B}_{\mathrm{FB}} = M z, \text{ with } z \in \left( \mathcal{Z} \setminus \left\lbrace \mathbb{0} \right\rbrace \right) \right\}.
\end{equation} 
In what follows, we will employ elements of polyhedral computation to show that the set $\mathcal{Z}$ is finitely generated. A similar procedure was used in Ref.~\cite{SON20} to find the fragile roots in all 230 space groups for systems with significant spin orbit coupling. We will quote the results of two theorems which are stated for completeness in \cref{app:sec:fb_clas:all} and refer the reader to Ref.~\cite{FUK14} for the detailed proofs. The set $\mathcal{Z}$ can be thought of as the intersection between a cone embedded in $r$ dimensions, written in the so-called H-representation (see Theorem 1 of \cref{app:sec:fb_clas:all})
\begin{equation}
	\label{app:eqn:P_in_H_rep}
	P = \left\lbrace z \in \mathbb{R}^{r} \mid M z \geq 0 \right\rbrace,
\end{equation}
and the $r$-dimensional integer lattice $\mathbb{Z}^{r}$, \ie
\begin{equation}
	\mathcal{Z} = P \cap \mathbb{Z}^{r}. 
\end{equation}
Because the matrix $M$ is full rank and contains only integer elements, the cone $P$ is also pointed and rational (see \cref{app:sec:fb_clas:all}). As a consequence of the Minkowski-Weyl Theorem (see \cref{app:sec:fb_clas:all}), $P$ is also a finitely generated cone, implying that there exists a matrix $Q  \in \mathbb{Z}^{r \times n}$ for some $n \in \mathbb{N}$, such that 
\begin{equation}
	\label{app:eqn:P_in_V_rep}
	P = \left\lbrace Q p \mid p \in \mathbb{R}^{n}_{+} \right\rbrace.
\end{equation}
Note that the matrix $Q$ only contains integer elements since $P$ is a rational cone. \Cref{app:eqn:P_in_V_rep} is known as the V-representation of the cone $P$ and a number of software packages can be employed to construct it from the H-representation of \cref{app:eqn:P_in_H_rep}. Here, we use the freely-available SageMath package~\cite{THE20}. 

As a consequence of Theorem 2 from \cref{app:sec:fb_clas:all}, the intersection between $P$ and the integer lattice $\mathbb{Z}^r$ is finitely generated. More precisely, $\mathcal{Z}$ admits a unique minimal Hilbert basis denoted by $\mathrm{Hil} \left(P\right)  = \left\lbrace b_1, b_2, \dots b_t \right\rbrace$ for some $t \in  \mathbb{N}$ and with $b_{i} \in Z^{r}$ for $1 \leq i \leq t$, such that
\begin{equation}
	\mathcal{Z} = P \cap \mathbb{Z}^{r} = \left\lbrace \sum_{i=1}^t p_i b_i \mid p_i \in \mathbb{Z} \text{ and } p_i \geq 0 \right\rbrace.
\end{equation}
The Hilbert basis $\mathrm{Hil} \left(P\right) $ is minimal in the sense that none of the basis vectors $b_{i} \in Z^{r}$ for $1 \leq i \leq t$ can be written as linear combinations of one another with positive integer coefficients. A number of algorithms and packages exist for constructing $\mathrm{Hil} \left(P\right) $ from the V-representation of $P$ from \cref{app:eqn:P_in_V_rep}. Throughout this paper, we employ the SageMath package~\cite{THE20} to obtain $\mathrm{Hil} \left(P\right) $ from the V-representation of $P$.

Once the set $\mathcal{Z}$ has been explicitly constructed with the aid of $\mathrm{Hil} \left(P\right) $, we can use \cref{app:eqn:set_of_gapped_flat_bands_1.2} and show that the set of all gapped flat bands is indeed finitely generated
\begin{equation}
	\label{app:eqn:set_of_gapped_flat_bands_2}
	\mathcal{F}^{\mathrm{G}}_{\mathcal{G}} = \left\lbrace \overline{B}_{\mathrm{FB}} \in \mathbb{Z}^{d_{\mathcal{G}}+1} \mid \overline{B}_{\mathrm{FB}} = \sum_{i=1}^t p_i g_i, \text{ with } p_i \in \mathbb{Z}, p_i \geq 0, \text{ and } \sum_{i=1}^{t} p_i \neq 0 \right\rbrace,
\end{equation}
where the gapped flat band basis vectors $g_i \in \mathbb{Z}^{d_{\mathcal{G}} +1 }$ are defined in terms of $\mathrm{Hil} \left(P\right) $ as
\begin{equation}
	\label{app:eqn:gapped_flat_bands_basis_def}
	g_i = M b_i,
\end{equation}
for all $1 \leq i \leq t$. Unlike the set of all (gapped and gapless) flat bands, the set of gapped flat bands is isomorphic with a semi-group, \ie $\mathcal{F}^{\mathrm{G}}_{\mathcal{G}} \sim \mathbb{N}^{t} \setminus \left\lbrace \mathbb{0} \right\rbrace $. 

Note that in \cref{app:eqn:set_of_gapped_flat_bands_2} $\overline{B}_{\mathrm{FB}}$ contains only the (co)irrep multiplicities at maximal momenta. Nevertheless, because $\overline{B}_{\mathrm{FB}}$ is essentially a linear combination of EBRs with integer coefficients, it automatically satisfies the compatibility relations~\cite{ELC20}. As such, the multiplicities of \emph{all} the (co)irreps across the Brillouin zone are fully determined by the entries of $\overline{B}_{\mathrm{FB}}$ and the corresponding multiplicities are necessarily positive (implying that no locally-stable band touching points exist). 

From a computational perspective, obtaining the Hilbert basis for $P$ is a hard problem, as it allows to easily solve the associated integer linear programming problem (which is known to be NP-complete)~\cite{DUR02}. Moreover, as a step for obtaining the fragile band roots, Ref.~\cite{SON20} has found that the Hilbert basis necessary to enumerate all symmetry data vectors $B$ that can be written as integer linear combinations of EBRs and satisfy $B \geq 0$ can contain $t \approx 10^{5}$ elements. As such, a complete enumeration of all the gapped flat band basis vectors in all SSGs is beyond the scope of the present work. Nevertheless, using flat band bases for gapped and gapless flat bands defined in \cref{app:eqn:set_of_flat_bands_3} which are fully tabulated in \cref{app:sec:fb_clas:table}, one can obtain the gapped flat band basis vectors using the algorithm we described.

As an example of our algorithm, in the following \cref{app:sec:fb_clas:gapped_examples}, we explicitly provide the gapped flat band bases for two SSG. An additional example, explaining the relation between the set of all BCL flat bands, the set of gapped BCL flat bands and the set of gapped BCL flat bands with fragile topology is given in \cref{app:sec:fb_clas:relation_example}.

\subsection{Examples of gapped flat band bases}\label{app:sec:fb_clas:gapped_examples}

In this appendix, we exemplify our algorithm for determining the set of all possible gapped flat bands. A similar procedure was employed by Ref.~\cite{SON20} as a key step towards determining the fragile roots in the 230 space groups for systems with significant spin-orbit coupling. As in \cref{app:sec:fb_clas:gapped}, we define gapped flat bands as bands without any locally-stable band touching points [\ie which are characterized by (co)irreps at all momenta in the Brillouin zone, rather than formal (co)irrep differences]. Here, we focus on the $P31'$ group (SSG 143.2 in the notation of BCS)~\citeBCS{} and systems without significant spin-orbit coupling in \cref{app:sec:fb_clas:gapped_examples:1}. In \cref{app:sec:fb_clas:gapped_examples:2}, we provide an additional example in the $P6/221'$ group (SSG 177.150 in the notation of BCS) for systems with significant spin-orbit coupling. Another example will be discussed in \cref{app:sec:fb_clas:relation_example} to outline the relation between gapped, gapless, and topologically fragile BCL flat bands.

\subsubsection{The gapped flat band basis in SSG 143.2}\label{app:sec:fb_clas:gapped_examples:1}

For the $P31'$ group (SSG 143.2 in the notation of BCS)~\citeBCS{} and systems without significant spin-orbit coupling, the matrix $M$ introduced in \cref{app:eqn:EBR_matrix_M} can be constructed from the (gapped and gapless) flat band basis vectors $e_i$ provided in \cref{app:sec:fb_clas:table} and reads 
\begin{equation}
	\label{app:eqn:Mmatrix_example:1}
	M = \begin{pNiceArray}{CCCC}[last-col]
 1 & 0 & 0 & 0 & n \\
 1 & -2 & 0 & 0 & \text{}^{\text{}}\Gamma _1^{\text{}} \\
 0 & 1 & 0 & 0 & \text{}^{\text{}}\Gamma _2^{\text{}}\text{}^{\text{}}\Gamma _3^{\text{}} \\
 1 & -2 & 0 & 0 & \text{}^{\text{}}\text{A}_1^{\text{}} \\
 0 & 1 & 0 & 0 & \text{}^{\text{}}\text{A}_2^{\text{}}\text{}^{\text{}}\text{A}_3^{\text{}} \\
 1 & -2 & -1 & -1 & \text{}^{\text{}}\text{H}_1^{\text{}} \\
 0 & 1 & 1 & 0 & \text{}^{\text{}}\text{H}_2^{\text{}} \\
 0 & 1 & 0 & 1 & \text{}^{\text{}}\text{H}_3^{\text{}} \\
 1 & -2 & -1 & -1 & \text{}^{\text{}}\text{HA}_1^{\text{}} \\
 0 & 1 & 1 & 0 & \text{}^{\text{}}\text{HA}_2^{\text{}} \\
 0 & 1 & 0 & 1 & \text{}^{\text{}}\text{HA}_3^{\text{}} \\
 1 & -2 & -1 & -1 & \text{}^{\text{}}\text{K}_1^{\text{}} \\
 0 & 1 & 1 & 0 & \text{}^{\text{}}\text{K}_2^{\text{}} \\
 0 & 1 & 0 & 1 & \text{}^{\text{}}\text{K}_3^{\text{}} \\
 1 & -2 & -1 & -1 & \text{}^{\text{}}\text{KA}_1^{\text{}} \\
 0 & 1 & 1 & 0 & \text{}^{\text{}}\text{KA}_2^{\text{}} \\
 0 & 1 & 0 & 1 & \text{}^{\text{}}\text{KA}_3^{\text{}} \\
 1 & 0 & 0 & 0 & \text{}^{\text{}}\text{L}_1^{\text{}} \\
 1 & 0 & 0 & 0 & \text{}^{\text{}}\text{M}_1^{\text{}} \\
\end{pNiceArray} .
\end{equation}
The (co)irreps corresponding to each row of $M$ in \cref{app:eqn:Mmatrix_example:1} have been explicitly listed as a separate column outside the matrix. Additionally, the first row of $M$ denotes the number of bands encoded by each column and has therefore been labeled by $n$, in line with \cref{app:eqn:aug_sym_dat_vect_def}. According to \cref{app:eqn:EBR_matrix_M,app:eqn:set_of_flat_bands_3}, the gapped and gapless flat band augmented symmetry data vectors $\overline{B}_{\mathrm{FB}}$ are given by integer linear combinations of the columns of $M$.

Following the algorithm outlined in \cref{app:sec:fb_clas:gapped}, the next step involves explicitly finding the set $\mathcal{Z}$ containing the solutions of \cref{app:eqn:gapped_condition_ineq}. For this we employ the SageMath package~\cite{THE20}, to determine the set of Hilbert basis vectors $\mathrm{Hil} \left(P\right) $ of the rational pointed cone $P$ defined in \cref{app:eqn:P_in_H_rep}. The Hilbert basis vectors of $\mathcal{Z}$ are given by the columns of the following matrix
\begin{equation}
	\label{app:eqn:HilbVecs_example:1}
	\left(b_1, b_2, \dots, b_9 \right) = \left(
\begin{array}{ccccccccc}
 1 & 1 & 1 & 2 & 2 & 2 & 2 & 2 & 2 \\
 0 & 0 & 0 & 1 & 1 & 1 & 1 & 1 & 1 \\
 0 & 0 & 1 & -1 & -1 & -1 & 0 & 0 & 1 \\
 0 & 1 & 0 & -1 & 0 & 1 & -1 & 0 & -1 \\
\end{array}
\right) .
\end{equation}
Notice that the Hilbert basis vectors from \cref{app:eqn:HilbVecs_example:1} are \emph{not} linearly independent over the set of real numbers, but \emph{are} linearly independent overe the set of positive integers.  Using \cref{app:eqn:gapped_flat_bands_basis_def}, the Hilbert basis vectors can be employed to find the gapped flat band basis vectors: they are given as the columns of the following matrix
\begin{equation}
	\label{app:eqn:gapBasis_example:1}
	\left(g_1, g_2, \dots, g_9 \right)=\begin{pNiceArray}{CCCCCCCCC}[last-col]
 1 & 1 & 1 & 2 & 2 & 2 & 2 & 2 & 2 & n \\
 1 & 1 & 1 & 0 & 0 & 0 & 0 & 0 & 0 & \text{}^{\text{}}\Gamma _1^{\text{}} \\
 0 & 0 & 0 & 1 & 1 & 1 & 1 & 1 & 1 & \text{}^{\text{}}\Gamma _2^{\text{}}\text{}^{\text{}}\Gamma _3^{\text{}} \\
 1 & 1 & 1 & 0 & 0 & 0 & 0 & 0 & 0 & \text{}^{\text{}}\text{A}_1^{\text{}} \\
 0 & 0 & 0 & 1 & 1 & 1 & 1 & 1 & 1 & \text{}^{\text{}}\text{A}_2^{\text{}}\text{}^{\text{}}\text{A}_3^{\text{}} \\
 1 & 0 & 0 & 2 & 1 & 0 & 1 & 0 & 0 & \text{}^{\text{}}\text{H}_1^{\text{}} \\
 0 & 0 & 1 & 0 & 0 & 0 & 1 & 1 & 2 & \text{}^{\text{}}\text{H}_2^{\text{}} \\
 0 & 1 & 0 & 0 & 1 & 2 & 0 & 1 & 0 & \text{}^{\text{}}\text{H}_3^{\text{}} \\
 1 & 0 & 0 & 2 & 1 & 0 & 1 & 0 & 0 & \text{}^{\text{}}\text{HA}_1^{\text{}} \\
 0 & 0 & 1 & 0 & 0 & 0 & 1 & 1 & 2 & \text{}^{\text{}}\text{HA}_2^{\text{}} \\
 0 & 1 & 0 & 0 & 1 & 2 & 0 & 1 & 0 & \text{}^{\text{}}\text{HA}_3^{\text{}} \\
 1 & 0 & 0 & 2 & 1 & 0 & 1 & 0 & 0 & \text{}^{\text{}}\text{K}_1^{\text{}} \\
 0 & 0 & 1 & 0 & 0 & 0 & 1 & 1 & 2 & \text{}^{\text{}}\text{K}_2^{\text{}} \\
 0 & 1 & 0 & 0 & 1 & 2 & 0 & 1 & 0 & \text{}^{\text{}}\text{K}_3^{\text{}} \\
 1 & 0 & 0 & 2 & 1 & 0 & 1 & 0 & 0 & \text{}^{\text{}}\text{KA}_1^{\text{}} \\
 0 & 0 & 1 & 0 & 0 & 0 & 1 & 1 & 2 & \text{}^{\text{}}\text{KA}_2^{\text{}} \\
 0 & 1 & 0 & 0 & 1 & 2 & 0 & 1 & 0 & \text{}^{\text{}}\text{KA}_3^{\text{}} \\
 1 & 1 & 1 & 2 & 2 & 2 & 2 & 2 & 2 & \text{}^{\text{}}\text{L}_1^{\text{}} \\
 1 & 1 & 1 & 2 & 2 & 2 & 2 & 2 & 2 & \text{}^{\text{}}\text{M}_1^{\text{}} \\
\end{pNiceArray} .
\end{equation}
As implied by \cref{app:eqn:set_of_gapped_flat_bands_2}, the set of all possible BCL gapped flat bands in the $P31'$ group (SSG 143.2 in the notation of BCS)~\citeBCS{} are given by positive integer linear combinations of the $g_i$ basis vectors given in \cref{app:eqn:gapBasis_example:1}. 

\subsubsection{The gapped flat band basis in SSG 177.150}\label{app:sec:fb_clas:gapped_examples:2}

For the $P6/221'$ group (SSG 177.150 in the notation of BCS)~\citeBCS{} and systems with significant spin-orbit coupling, the matrix $M$ reads 
\begin{equation}
	\label{app:eqn:Mmatrix_example:2}
	M = \begin{pNiceArray}{CCCC}[last-col]
 2 & 0 & 0 & 0 & n \\
 1 & -1 & -1 & 0 & \text{}^{\text{}}\overline{\Gamma }_7^{\text{}} \\
 0 & 1 & 0 & 0 & \text{}^{\text{}}\overline{\Gamma }_8^{\text{}} \\
 0 & 0 & 1 & 0 & \text{}^{\text{}}\overline{\Gamma }_9^{\text{}} \\
 1 & -1 & -1 & 0 & \text{}^{\text{}}\overline{\text{A}}_7^{\text{}} \\
 0 & 1 & 0 & 0 & \text{}^{\text{}}\overline{\text{A}}_8^{\text{}} \\
 0 & 0 & 1 & 0 & \text{}^{\text{}}\overline{\text{A}}_9^{\text{}} \\
 1 & -1 & -1 & -1 & \text{}^{\text{}}\overline{\text{H}}_4^{\text{}} \\
 1 & -1 & -1 & -1 & \text{}^{\text{}}\overline{\text{H}}_5^{\text{}} \\
 0 & 1 & 1 & 1 & \text{}^{\text{}}\overline{\text{H}}_6^{\text{}} \\
 1 & -1 & -1 & -1 & \text{}^{\text{}}\overline{\text{K}}_4^{\text{}} \\
 1 & -1 & -1 & -1 & \text{}^{\text{}}\overline{\text{K}}_5^{\text{}} \\
 0 & 1 & 1 & 1 & \text{}^{\text{}}\overline{\text{K}}_6^{\text{}} \\
 1 & 0 & 0 & 0 & \text{}^{\text{}}\overline{\text{L}}_5^{\text{}} \\
 1 & 0 & 0 & 0 & \text{}^{\text{}}\overline{\text{M}}_5^{\text{}} \\
\end{pNiceArray} .
\end{equation}
As in \cref{app:sec:fb_clas:gapped_examples:1}, we employ the SageMath package~\cite{THE20}, to determine the set of Hilbert basis vectors $\mathrm{Hil} \left(P\right) $ of the rational pointed cone $P$ defined in \cref{app:eqn:P_in_H_rep}. The Hilbert basis vectors of $\mathcal{Z}$ are given by the columns of the following matrix
\begin{equation}
	\label{app:eqn:HilbVecs_example:2}
	\left(b_1, b_2, \dots, b_6 \right) = \left(
\begin{array}{cccccc}
 1 & 1 & 1 & 1 & 1 & 1 \\
 0 & 0 & 0 & 0 & 1 & 1 \\
 0 & 0 & 1 & 1 & 0 & 0 \\
 0 & 1 & -1 & 0 & -1 & 0 \\
\end{array}
\right) .
\end{equation}
The gapped flat band basis vectors can then be found using \cref{app:eqn:gapped_flat_bands_basis_def}: they are given as the columns of the following matrix
\begin{equation}
	\label{app:eqn:gapBasis_example:2}
	\left(g_1, g_2, \dots, g_6 \right)=\begin{pNiceArray}{CCCCCC}[last-col]
 2 & 2 & 2 & 2 & 2 & 2 & n \\
 1 & 1 & 0 & 0 & 0 & 0 & \text{}^{\text{}}\overline{\Gamma }_7^{\text{}} \\
 0 & 0 & 0 & 0 & 1 & 1 & \text{}^{\text{}}\overline{\Gamma }_8^{\text{}} \\
 0 & 0 & 1 & 1 & 0 & 0 & \text{}^{\text{}}\overline{\Gamma }_9^{\text{}} \\
 1 & 1 & 0 & 0 & 0 & 0 & \text{}^{\text{}}\overline{\text{A}}_7^{\text{}} \\
 0 & 0 & 0 & 0 & 1 & 1 & \text{}^{\text{}}\overline{\text{A}}_8^{\text{}} \\
 0 & 0 & 1 & 1 & 0 & 0 & \text{}^{\text{}}\overline{\text{A}}_9^{\text{}} \\
 1 & 0 & 1 & 0 & 1 & 0 & \text{}^{\text{}}\overline{\text{H}}_4^{\text{}} \\
 1 & 0 & 1 & 0 & 1 & 0 & \text{}^{\text{}}\overline{\text{H}}_5^{\text{}} \\
 0 & 1 & 0 & 1 & 0 & 1 & \text{}^{\text{}}\overline{\text{H}}_6^{\text{}} \\
 1 & 0 & 1 & 0 & 1 & 0 & \text{}^{\text{}}\overline{\text{K}}_4^{\text{}} \\
 1 & 0 & 1 & 0 & 1 & 0 & \text{}^{\text{}}\overline{\text{K}}_5^{\text{}} \\
 0 & 1 & 0 & 1 & 0 & 1 & \text{}^{\text{}}\overline{\text{K}}_6^{\text{}} \\
 1 & 1 & 1 & 1 & 1 & 1 & \text{}^{\text{}}\overline{\text{L}}_5^{\text{}} \\
 1 & 1 & 1 & 1 & 1 & 1 & \text{}^{\text{}}\overline{\text{M}}_5^{\text{}} \\
\end{pNiceArray} .
\end{equation}
As implied by \cref{app:eqn:set_of_gapped_flat_bands_2}, the set of all possible BCL gapped flat bands in the $P6/221'$ group (SSG 177.150 in the notation of BCS)~\citeBCS{} are given by positive integer linear combinations of the $g_i$ basis vectors given in \cref{app:eqn:gapBasis_example:2}. 

\subsection{Flat bands as platforms for fragile topology}\label{app:sec:fb_clas:fragile}

Our flat band subtraction prescription \cref{app:eqn:br_subtraction} offers an effortless technique of not only obtaining flat bands, but flat bands hosting fragile topology. (Magnetic) Topological Quantum Chemistry defines topologically trivial bands, as bands whose (co)irreps can be written as linear combinations of EBRs with positive integer coefficients~\cite{BRA17,ELC20a}. Fragile topological bands are a special type of topologically nontrivial bands that can only be written as integer linear combinations of EBRs with some of the coefficients \emph{necessarily} being strictly negative~\cite{PO18c,CAN18,SON20,SON20c}. Unlike strong (or stable) topological states, fragile topological bands can be trivialized by coupling them to certain topologically trivial bands. The importance of fragile topology cannot be overstated, as it underpins one of the most intriguing examples of strongly correlated physics, twisted bilayer graphene. In this appendix, we show explicitly that \emph{any} bands displaying fragile topology diagnosed by (co)irreps can also be realized as a BCL flat band, thus offering a pathway towards strongly correlated phases of matter emerging in systems with non-trivial topology.

To show that any fragile topological phase diagnosable from (co)irreps can be realized in gapped BCL flat bands, we consider the symmetry data vector $B'$ characterizing a fragile topological phase in the SSG $\mathcal{G}$. By definition, $B'\geq 0 $ and $B'$ can be written as an integer linear combination of EBRs for which some of the coefficients \emph{must} be strictly negative. As such, there exist $x'_1, x'_2 \in \mathbb{N}^{d_{EBR}}$, such that 
\begin{equation}
	B' = EBR\left(x'_1 - x'_2 \right),
\end{equation}
with the $EBR$ matrix being defined in \cref{app:eqn:EBR_def} and $x'_2$ necessarily having at least one nonzero entry. Now consider constructing an $L \oplus \tilde{L}$ BCL Hamiltonian such that the orbital content of each sublattice, encoded by the vectors $x_L$ and $x_{\tilde{L}}$ from \cref{app:eqn:sublat_orb_vector_L,app:eqn:sublat_orb_vector_Ltilde}, is given by 
\begin{equation}
	x_L = x'_1 \quad \text{and} \quad x_{\tilde{L}} = x'_2.
\end{equation}
Then, from \cref{app:eqn:sym_dat_vec_FB}, the flat band symmetry data vectors of the BCL Hamiltonian will be given by $B_{\mathrm{FB}} = B'$. Because, $B_{\mathrm{FB}} \geq 0$, the flat bands contain no locally-stable band touching points across the Brillouin zone and are generically gapped. Moreover, because the symmetry data vector $B'$ corresponds to a fragile topological band, the $L \oplus \tilde{L}$ BCL flat band will also display fragile topology indicated by (co)irreps. This completes the proof that \emph{any} (co)irrep-indicated topologically fragile band can be realized as a BCL flat band.

The classification of all topologically fragile BCL flat bands is thus entirely equivalent to a classification of all topologically fragile bands. The latter has been worked out by Ref.~\cite{SON20} in all 230 space groups for systems with significant spin-orbit coupling using elements of polyhedron computation. In \cref{app:sec:fb_clas:relation_example} we will provide a simple example in the $I 2_1 3$ group (SSG 199.12 in the notation of BCS)~\citeBCS{} for systems with significant spin-orbit coupling illustrating the relation between the set of gapped, gapless and topologically fragile flat bands.

\subsection{Relation between the gapped, gapless, and topologically fragile flat bands}\label{app:sec:fb_clas:relation_example}

\begin{figure}[!t]
\includegraphics[width=0.333\textwidth]{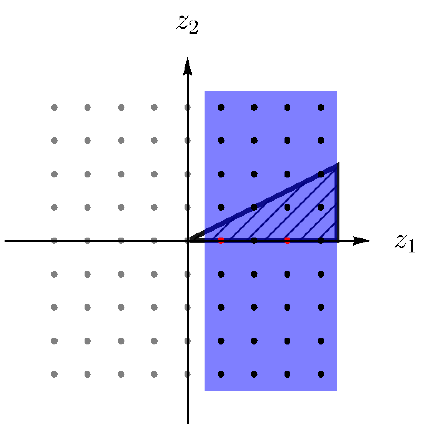}
\caption{The relation between gapped, gapless, and topologically fragile flat bands in the the $I 2_1 3'$ group (SSG 199.13 in the notation of BCS)~\citeBCS{} for systems with significant spin-orbit coupling. The symmetry data vectors corresponding to integer linear combinations of EBRs are parameterized by a two component integer vector $z\in \mathbb{Z}^2$ and thus form the $\mathbb{Z}^2$ lattice (shown by the black and red dots). In the same parameterization, the set of all (gapped and gapless) BCL flat bands is given by the $z_1>0$ subset of the $\mathbb{Z}^2$ lattice (the black and red dots situated inside the blue region). Out of all flat bands, the gapped ones are represented by the black and red dots that are also inside the hatched two-dimensional cone. Employing the fragile criteria derived in Ref.~\cite{SON20}, we find that some of the gapped flat bands are topologically fragile (represented as the red dots).}
\label{app:fig:relationFlatBandSets}
\end{figure}

To better understand the relation between the set of all flat bands $\mathcal{F}_{\mathcal{G}}$ defined in \cref{app:sec:fb_clas:all}, the set of gapped flat bands $\mathcal{F}^{G}_{\mathcal{G}}$ introduced in \cref{app:sec:fb_clas:gapped}, and the topologically fragile states discussed in \cref{app:sec:fb_clas:fragile}, we present a concrete example in the $I 2_1 3'$ group (SSG 199.13 in the notation of BCS)~\citeBCS{} for systems with significant spin-orbit coupling. The same SSG was used by Ref.~\cite{SON20} to outline the derivation of fragile criteria for generic (\ie not necessarily flat) bands.

For the $I 2_1 3'$ group (SSG 199.13 in the notation of BCS)~\citeBCS{} and systems with significant spin-orbit coupling, the matrix $M$ introduced in \cref{app:eqn:EBR_matrix_M} whose columns form the flat band basis vectors $e_i$ provided in \cref{app:sec:fb_clas:table} reads
\begin{equation}
	\label{app:eqn:Mmatrix_example:relation}
	M = \begin{pNiceArray}{CC}[last-col]
 4 & 0 & n \\
 0 & 2 & \text{}^{\text{}}\overline{\Gamma }_5^{\text{}} \\
 1 & -1 & \text{}^{\text{}}\overline{\Gamma }_6^{\text{}}\text{}^{\text{}}\overline{\Gamma }_7^{\text{}} \\
 0 & 2 & \text{}^{\text{}}\overline{\text{H}}_5^{\text{}} \\
 1 & -1 & \text{}^{\text{}}\overline{\text{H}}_6^{\text{}}\text{}^{\text{}}\overline{\text{H}}_7^{\text{}} \\
 2 & 0 & \text{}^{\text{}}\overline{\text{N}}_3^{\text{}}\text{}^{\text{}}\overline{\text{N}}_4^{\text{}} \\
 1 & -2 & \text{}^{\text{}}\overline{\text{P}}_4^{\text{}} \\
 0 & 1 & \text{}^{\text{}}\overline{\text{P}}_5^{\text{}} \\
 0 & 1 & \text{}^{\text{}}\overline{\text{P}}_6^{\text{}} \\
 1 & 0 & \text{}^{\text{}}\overline{\text{P}}_7^{\text{}} \\
 1 & -2 & \text{}^{\text{}}\overline{\text{PA}}_4^{\text{}} \\
 0 & 1 & \text{}^{\text{}}\overline{\text{PA}}_5^{\text{}} \\
 0 & 1 & \text{}^{\text{}}\overline{\text{PA}}_6^{\text{}} \\
 1 & 0 & \text{}^{\text{}}\overline{\text{PA}}_7^{\text{}} \\
\end{pNiceArray} .
\end{equation}
Correspondingly, the set of all BCL flat bands in the $I 2_1 3'$ group is given by
\begin{equation}
	\label{app:eqn:set_of_flat_bands_example:relation}
	\mathcal{F}_{I 2_1 3'} = \left\{ \overline{B}_{\mathrm{FB}} \in \mathbb{Z}^{14} \mid \overline{B}_{\mathrm{FB}} = M z, \text{ with } z \in \mathbb{Z}^{2} \text{ and } z_1 > 0  \right\}.
\end{equation}
The augmented symmetry data vectors corresponding to the flat bands (and any band touching points) is completely specified by the two integer parameters $z_1$ and $z_2$. As implied by \cref{app:eqn:set_of_flat_bands_example:relation} and shown in \cref{app:fig:relationFlatBandSets}, the set of all possible BCL flat bands in the $I 2_1 3'$ group for systems with significant spin-orbit coupling is given by the $\mathbb{Z}^2$ lattice points in the positive $z_1>0$ half-plane, up to a multiplication by the $M$ matrix from \cref{app:eqn:Mmatrix_example:relation}.

According to \cref{app:eqn:set_of_gapped_flat_bands_1.2}, the set of gapped [as indicated from (co)irreps] BCL flat bands is given by the set
\begin{equation}	\label{app:eqn:set_of_gapped_flat_bands_example:relation}
	\mathcal{F}^{\mathrm{G}}_{I 2_1 3'} = \left\{ \overline{B}_{\mathrm{FB}} \in \mathbb{Z}^{14} \mid \overline{B}_{\mathrm{FB}} = M z, \text{ with } z \in \left( \mathcal{Z} \setminus \left\lbrace \mathbb{0} \right\rbrace \right) \right\},
\end{equation}
where the set $\mathcal{Z}$ is defined in \cref{app:eqn:gapped_z_set_def}. Essentially, $\mathcal{Z}$ is the set of all two-dimensional integer vectors $z$ that give rise to augmented symmetry data vectors containing only positive entries. As shown in \cref{app:sec:fb_clas:gapped}, $\mathcal{Z} = P \cap \mathbb{Z}^{2}$, where the pointed rational cone $P$ is defined in the V-representation as
\begin{equation}
	P = \left\lbrace \begin{pmatrix}
		2 & 1 \\ 1 & 0 \end{pmatrix} p \mid p \in \mathbb{R}^{2}_+ 
	\right\rbrace.
\end{equation}
\Cref{app:fig:relationFlatBandSets} illustrates that, up to a multiplication by the $M$ matrix, the gapped flat bands are given by those points in the $\mathbb{Z}^{2}$ lattice that are also inside the two-dimensional pointed rational cone $P$. Moreover, one can see that, as expected, $\mathcal{F}^{\mathrm{G}}_{I 2_1 3'} \subset \mathcal{F}_{I 2_1 3'}$.

Ref.~\cite{SON20} has derived criteria for diagnosing fragile topological phases in all 230 space groups with significant spin-orbit coupling. These criteria can be defined in terms of a non-redundant parameterization of all symmetry data vectors that can be written as integer linear combinations of EBRs (and thus form candidates for (co)irrep indicated fragile phases). Since the matrix $M$ also provides a non-redundant parameterization of all augmented symmetry data vectors representing integer linear combinations of EBRs, we can can always find a invertible map between the parameterization introduced in Ref.~\cite{SON20} and our parameterization in terms of the $z$-vectors. When applied to our parameterization and for the specific case of the $I 2_1 3'$ group, Ref.~\cite{SON20} has shown that a band with the augmented symmetry data vector $\overline{B}_{\mathrm{FB}} = Mz \geq 0$ for which $z_1$ is odd and $z_2 = 0$ is necessarily fragile. This is indicated in \cref{app:fig:relationFlatBandSets}, where some of the gapped flat bands are marked as topologically fragile.

\subsection{Mathematical theorems}\label{app:sec:fb_clas:maths}

In this appendix, we provide two results from the theory of polyhedral computation. The two theorems are crucial for proving that the set of all gapped flat bands is finitely generated in \cref{app:sec:fb_clas:gapped}. They were also employed extensively by Ref.~\cite{SON20} in deriving the fragile roots in all 230 space groups for systems with significant spin-orbit coupling. They are listed here without proof, but a proof can be found in Ref.~\cite{FUK14}.
\begin{itemize}
	\item \emph{Theorem 1: The Minkowski-Weyl Theorem for Cones}. For $P \subseteq \mathbb{R}^{d}$, the following two statements are equivalent:
	\begin{enumerate}
		\item (H-representation) $P$ is a polyhedral cone, \ie there exists $A \in \mathbb{R}^{m \times d}$ for some $m \in \mathbb{N}$, such that $P = \left\lbrace x \in \mathbb{R}^{d} \mid A x \geq 0 \right\rbrace$.
		\item (V-representation) $P$ is a finitely generated cone, \ie there exist a matrix $R \in \mathbb{R}^{d \times n}$ for some $n \in \mathbb{N}$, such that $P = \left\lbrace R p \mid p \in \mathbb{R}^{n}_+ \right\rbrace$.
	\end{enumerate}
	A cone is called pointed if $x,-x \in P$ implies that $x=\mathbb{0}$. When written in the H-representation, a cone is pointed if $A$ is full rank. A cone is called rational if the matrix $A$ (in the H-representation) or the matrix $R$ (in the V-representation) contains only integer elements.
	\item \emph{Theorem 2:} Every rational pointed cone $P$ in $d$ dimensions admits a unique minimal integral Hilbert basis denoted by $\mathrm{Hil} \left(P\right)  = \left\lbrace b_1, b_2, \dots b_t \right\rbrace$ for some $t \in  \mathbb{N}$ and with $b_{i} \in Z^{d}$ for $1 \leq i \leq t$ such that 
	\begin{equation}
		P \cap \mathbb{Z}^{d} = \left\lbrace \sum_{i=1}^t p_i b_i \mid p_i \in \mathbb{Z} \text{ and } p_i \geq 0 \right\rbrace.
	\end{equation}
	Moreover, for any $b_j \in \mathrm{Hil} \left(P\right) $, the equation $b_j = \sum_{i=1}^{t} \alpha_i b_i$ with $\alpha_i \in \mathbb{Z}$ and $\alpha_i \geq 0$ for $1 \leq i \leq t$ has only the trivial solutions $\alpha_i= \delta_{ij}$ (\ie none of the vectors in $\mathrm{Hil} \left(P\right) $ can be written as a linear combination of the other vectors with  non-negative integer coefficients). 
	There are a number of packages for finding the Hilbert basis from the V-representation of a pointed cone. Throughout this paper, we employ the SageMath package~\cite{THE20}.
	
\end{itemize}

\newpage
\newcommand{\PreserveBackslash}[1]{\let\temp=\\#1\let\\=\temp}
\newcolumntype{C}[1]{>{\PreserveBackslash\centering}p{\dimexpr(\linewidth -1.8 pt) / #1 \relax}}
\setlength\tabcolsep{0 pt}
\setlength{\LTpost}{0pt}
\setlength{\LTpre}{0pt}

\subsection{Tabulated flat band bases for all SSG}\label{app:sec:fb_clas:table}

Refs.~\cite{BRA17,ELC20a} have recently tabulated the EBRs in all 1651{} SSG. Using the data available on the BCS~\citeBCS{}, we have applied the algorithm described in \cref{app:sec:fb_clas:all} to all SSGs, obtaining the corresponding flat band bases. The results are listed in \cref{app:sec:fb_clas:table:soc} for systems with significant spin-orbit coupling and in \cref{app:sec:fb_clas:table:no_soc} for systems without significant spin-orbit coupling. 

The table corresponding to each SSG $\mathcal{G}$ starts with a row denoting the SSG number in the notation of BCS~\citeBCS{}, \eg \textbf{SSG 168.109}. The following row lists the basis of the augmented symmetry data vectors in $\mathcal{G}$. This row always begins with $n$ (the number of bands in $\overline{B}$) and then lists all the (co)irreps corresponding to the maximal momenta of the SSG $\mathcal{G}$. The next $r$ rows (where $r$ is the rank of the EBR matrix) denote the $e_i$ basis vectors ($1 \leq i \leq r$) from \cref{app:eqn:set_of_flat_bands_3}, which generate the augmented symmetry vectors of all the possible BCL flat bands in the SSG $\mathcal{G}$. When listing the $e_i$ basis vectors, we always denote negative numbers with an overline (\eg $-1 = \overline{1}$).

\newpage
\subsubsection{Systems with significant spin-orbit coupling}\label{app:sec:fb_clas:table:soc} 
\newpage
\foreach \x in {1,...,157}
{\includepdf[pages={\x}]{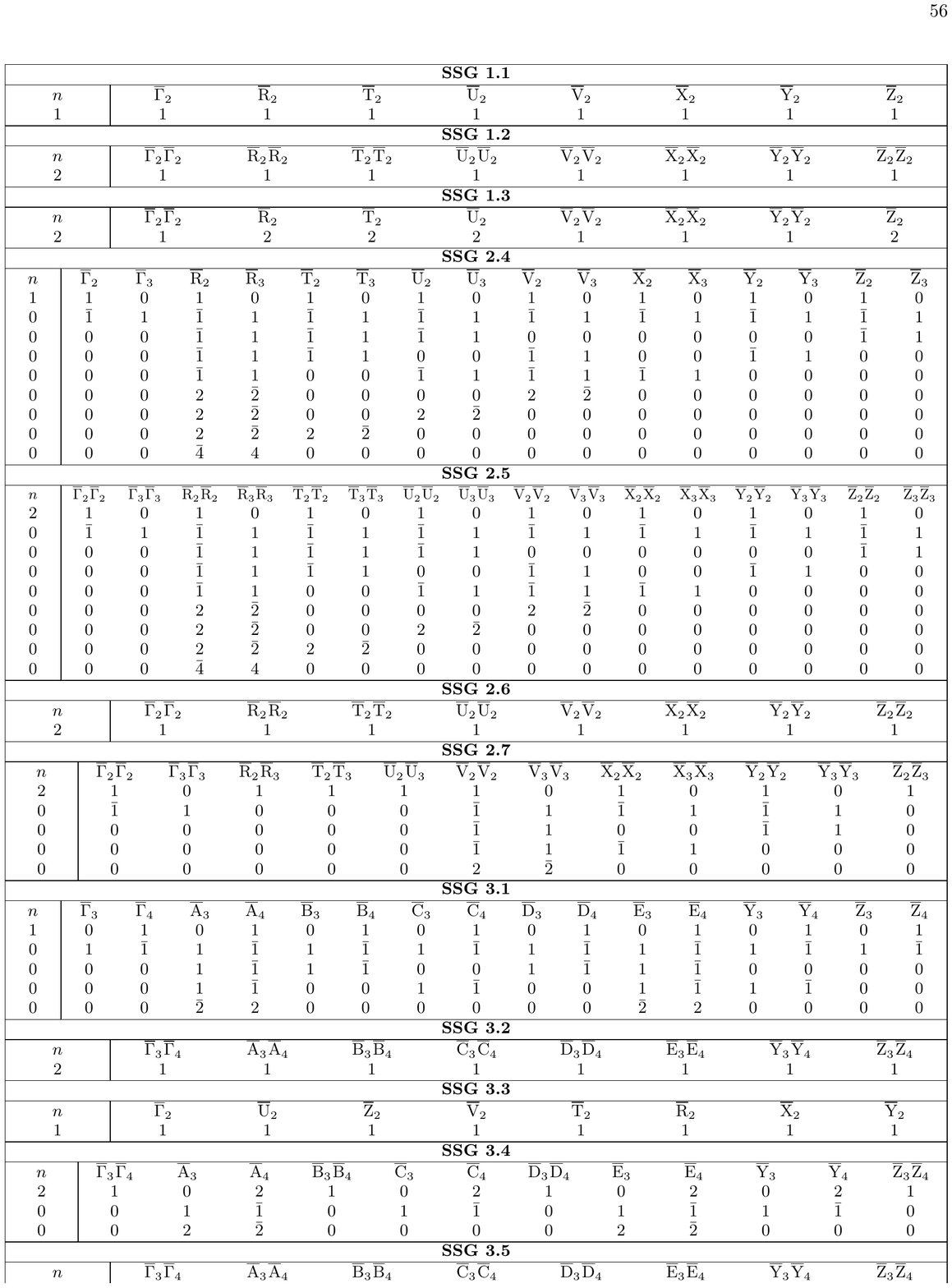} 
}
\subsubsection{Systems without significant spin-orbit coupling}\label{app:sec:fb_clas:table:no_soc}
\newpage
\foreach \x in {1,...,213}
{\includepdf[pages={\x}]{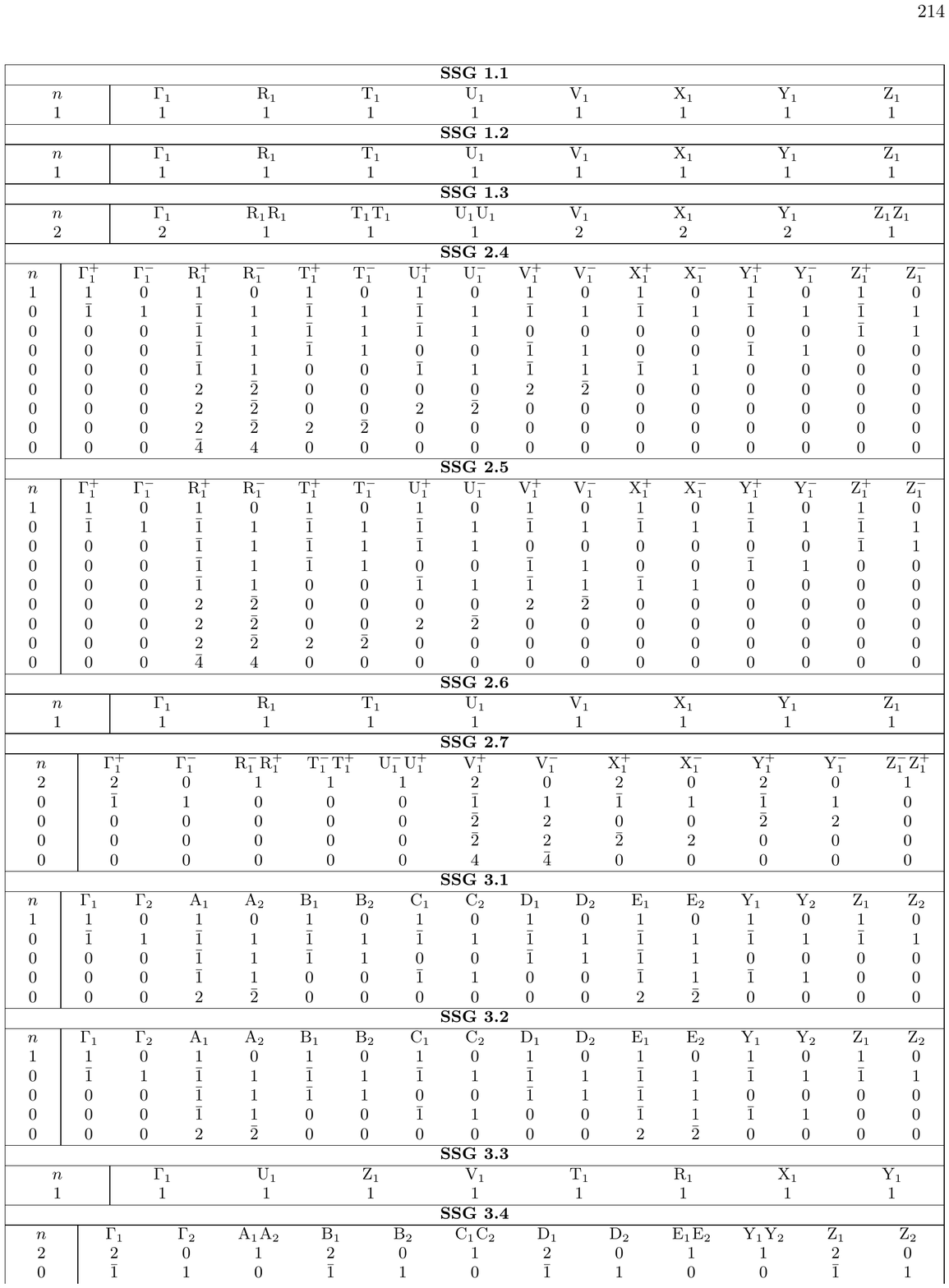} 
}

\end{document}